%% file: these.tex
\documentclass[12pt,compat2,openright,twoside]{report}
\usepackage[a4paper,hmargin=2.5cm,textheight=23cm]{geometry}
\usepackage[francais]{babel}
\usepackage[latin1]{inputenc}
\usepackage[T1]{fontenc}
\usepackage{amssymb,amsfonts,amsmath}
\usepackage{makeidx}
\usepackage{graphicx}
\usepackage{color}

\catcode`\ =10 
\def«{\og\ignorespaces} \def»{\fg}
\numberwithin{equation}{section}

\let\vraibfseries\bfseries
\renewcommand{\bfseries}{\boldmath\vraibfseries}

\renewcommand{\a}{\alpha}
\renewcommand{\b}{\beta}
\newcommand{\g}{\gamma}
\newcommand{\eps}{\varepsilon}
\newcommand{\s}{\sigma}
\renewcommand{\t}{\tau}
\newcommand{\w}{\omega}

\newcommand{\I}{\mathbb{I}}
\newcommand{\R}{\mathbb{R}}
\newcommand{\RP}{\mathbb{RP}}
\newcommand{\Z}{\mathbb{Z}}

\newcommand{\Lg}{\mathfrak{g}}

\newcommand{\A}{\mathcal{A}}
\newcommand{\F}{\mathcal{F}}
\newcommand{\K}{\mathcal{K}}
\newcommand{\M}{\mathcal{M}}
\newcommand{\N}{\mathcal{N}}
\newcommand{\T}{\mathcal{T}}

\newcommand{\<}{\langle}
\renewcommand{\>}{\rangle}
\renewcommand{\d}{\partial}
\newcommand*{\ordnorm}[1]{\,\mathopen:\,#1\mathclose:\,}

\DeclareMathOperator{\diag}{diag}
\DeclareMathOperator{\im}{Im}

\DeclareMathOperator{\sign}{sign}
\DeclareMathOperator{\OO}{O}
\DeclareMathOperator{\SL}{SL}
\DeclareMathOperator{\SO}{SO}
\DeclareMathOperator{\SU}{SU}
\DeclareMathOperator{\Tr}{Tr}
\DeclareMathOperator{\U}{U}
\DeclareMathOperator{\USp}{USp}

\newcommand{\nposodd}{\substack{n>0 \\ \text{impair}}}
\newcommand{\D}{\text{D}}
\newcommand{\parfrac}{\genfrac{(}{)}{}{}}

\makeindex

\newcommand*{\citemoi}[1]{\cite{#1} (annexe~#1 de cette thèse)}

\begin{document}
\renewcommand{\O}{\mathcal{O}} 
\pagestyle{empty}
\begin{center}
\textsc{\large Laboratoire de Physique Théorique et Hautes Énergies}
\vskip .3cm
\centerline{\textsc{\large Laboratoire de Physique Théorique de l'École
                           Normale Supérieure}}
\vskip 1.5cm
\centerline{\textbf{\Large TH{\`E}SE DE DOCTORAT DE L'UNIVERSITÉ PARIS VI}}
\vskip .8cm
{\large Spécialité : \textbf{\textsc{Physique théorique}}}
\vskip 1.2cm
présentée par
\vskip .6cm
\textbf{\Large Nicolas \textsc{Couchoud}}
\vskip 1cm
pour obtenir le grade de
\vskip .6cm
\textbf{\Large Docteur de l'Université Paris VI}
\vskip 1.5cm
Sujet :
\vskip 1cm
\textbf{\textit{\LARGE D-branes et orientifolds dans des espaces}}
\vskip .3cm
\textbf{\textit{\LARGE courbes ou dépendant du temps}}
\end{center}
\vskip 3cm
\noindent
Soutenue le 1\ier\ octobre 2004 devant le jury compos{\'e} de~:
\vskip 1cm
\noindent\begin{tabular}{rlll}
&      M. & Constantin \textsc{Bachas},  & directeur de thèse, \\
&M\up{me} & Ilka \textsc{Brunner}, \\
&      M. & Vladimir \textsc{Dotsenko},  & président du jury,\\
&      M. & Marios \textsc{Petropoulos}, & rapporteur, \\
&      M. & Bert \textsc{Schellekens},   & rapporteur, \\
&      M. & Volker \textsc{Schomerus}, \\
et &   M. & Paul \textsc{Windey},        & directeur de thèse.
\end{tabular}
\cleardoublepage

\linespread{1.3} \selectfont

\begin{center}
\textit{\large Remerciements}
\end{center}

Tout d'abord, je remercie Laurent Baulieu d'une part, Jean Iliopoulos et
Eugène Cremmer d'autre part, de m'avoir accueilli dans les laboratoires
qu'ils dirigent ou ont dirigé afin que je puisse y entreprendre mon travail
de thèse.

Ensuite, je suis infiniment reconnaissant à Costas Bachas et Paul Windey
d'avoir accepté de diriger ma thèse et de m'avoir guidé dans cette matière
ardue, mais passionnante qu'est la théorie des cordes, et d'avoir bien
voulu relire cette thèse.

Je voudrais également remercier tous ceux avec qui j'ai été amené à
discuter au cours de cette thèse, notamment Pedro Bordalo, Eric Gimon,
Boris Pioline, Sylvain Ribault et Christoph Schweigert, et évoquer le rôle
important de l'encouragement mutuel de mes collègues de bureau et moi-même
dans nos travaux respectifs.

Je suis aussi reconnaissant aux rapporteurs, Marios Petropoulos et Bert
Schellekens, d'avoir accepté la charge de lire ma thèse, et au jury d'avoir
bien voulu accorder de l'intérêt à ma thèse, parfois en venant de loin pour
cela.

Enfin, je remercie ma famille et mes amis de m'avoir accordé leur soutien
dans cette entreprise.
 
\cleardoublepage
\pagestyle{plain}
\tableofcontents

\makeatletter
\renewcommand*{\@schapter}[1]{{\c@secnumdepth=\m@ne\@chapter[#1]{#1}}}
\makeatother

\include{intro}
\include{cordes}
\include{WZW}
\include{AdS}
\include{pulse}
\include{concl}

\appendix
\chapter{Orientifolds of the 3-sphere}
\begin{flushright}
\textsc{\large J. High Energy Phys. 12 (2001) 003}
\end{flushright}
\cleardoublepage
\addtocounter{page}{20}

\chapter{D-branes and orientifolds of SO(3)}
\begin{flushright}
\textsc{\large J. High Energy Phys. 03 (2002) 026}
\end{flushright}
\cleardoublepage
\addtocounter{page}{12}

\chapter{Anti-de Sitter branes with Neveu-Schwarz and Ramond-Ramond
  backgrounds}
\begin{flushright}
\textsc{\large J. High Energy Phys. 03 (2003) 007}
\end{flushright}
\cleardoublepage
\addtocounter{page}{8}

\newcommand*{\hepth}[1]{\texttt{hep-th/#1}}
\newcommand*{\ap}[2]{\textit{Ann. Phys.} \textbf{#1} (#2) }
\newcommand*{\cmp}[2]{\textit{Commun. Math. Phys.} \textbf{#1} (#2) }
\newcommand*{\forp}[2]{\textit{Fortsch. Phys.} \textbf{#1} (#2) }
\newcommand*{\jgp}[2]{\textit{J. Geom. Phys} \textbf{#1} (#2) }
\newcommand*{\jhep}[2]{\textit{J. High Energy Phys.} \textbf{#1} (#2) }
\newcommand*{\jmp}[2]{\textit{J. Math. Phys.} \textbf{#1} (#2) }
\newcommand*{\mpla}[2]{\textit{Mod. Phys. Lett.} \textbf{A~#1} (#2) }
\newcommand*{\npb}[2]{\textit{Nucl. Phys.} \textbf{B~#1} (#2) }
\newcommand*{\plb}[2]{\textit{Phys. Lett.} \textbf{B~#1} (#2) }
\newcommand*{\prep}[2]{\textit{Phys. Rept.} \textbf{#1} (#2) }
\newcommand*{\prd}[2]{\textit{Phys. Rev.} \textbf{D~#1} (#2) }
\newcommand*{\prl}[2]{\textit{Phys. Rev. Lett.} \textbf{#1} (#2) }

\newcommand*{\bibsec}[1]{\subsection*{\hskip-\leftmargin#1}}
\linespread{1}\selectfont

\newpage
\addcontentsline{toc}{chapter}{Index}
\printindex
\end{document}

%% file: intro.tex
\chapter*{Introduction}
\subsubsection{Généralités sur la théorie des cordes}
Le monde tel que nous le connaissons est décrit par le Modèle Standard de
la physique des particules (interactions électrofaible et forte) et la
Relativité Générale (interaction gravitationnelle), qui rendent compte des
données expérimentales actuelles des échelles subatomiques aux échelles
cosmologiques. Cette compréhension de l'univers est cependant loin d'être
satisfaisante : la Relativité Générale est en effet une théorie classique,
et il faudrait, pour obtenir un tout cohérent, une description quantique de
la gravitation, ce qui est un problème très difficile, et non encore
pleinement résolu.

Il y a eu une tentative de quantification de la gravitation à partir de ses
principes fondamentaux, à savoir l'invariance par difféomorphismes, nommée
\emph{gravité quantique de boucles}. On ne sait malheureusement pas faire
grand-chose avec : notamment, elle n'a pas de développement perturbatif, ce
qui rend difficile des prédictions physiques.

La théorie des cordes, à l'inverse, n'est pas fondée sur les principes
fondamentaux de la Relativité Générale. En fait, lorsqu'elle est née au
début des années soixante-dix, la théorie des cordes avait vocation à
décrire les interactions fortes. Mais d'une part, ses prédictions en tant
que théorie des interactions fortes étaient incompatibles avec
l'expérience, d'autre part il est apparu qu'elle contient nécessairement
une excitation de spin~2 et de masse nulle, naturellement identifiée au
graviton.

Le principe de la théorie des cordes est que les différentes formes de
matière et de rayonnement sont des excitations d'une corde, au lieu d'être
des particules ponctuelles. L'intérêt de considérer des objets étendus est
que cela permet de ne pas avoir les divergences ultraviolettes que l'on
obtient habituellement en théorie des champs, en particulier dans le cas de
la gravitation, qui est non renormalisable.

La théorie des cordes, telle qu'elle est formulée actuellement, est une
théorie perturbative, en ce sens qu'on y considère des excitations autour
d'un vide fixé, alors qu'on s'attend à ce qu'il soit déterminé
dynamiquement. Il découle que, comme en théorie quantique des champs, on ne
sait guère calculer de quantités physiques qu'à couplage faible. Il ne
s'agit donc pas d'une théorie complète, mais du développement perturbatif
d'une théorie à l'heure actuelle inconnue. On peut toutefois formuler des
conjectures relatives aux propriétés à couplage fort de cette théorie
inconnue en utilisant la supersymétrie présente dans la théorie
perturbative et les théorèmes de non-renormalisation qui en découlent. On
trouve alors notamment des dualités couplage fort/couplage faible, telles
que la S-dualité que nous utiliserons dans le chapitre~3. (On notera que
cela suppose que la supersymétrie reste présente dans la théorie non
perturbative. Sans cette hypothèse, on ne pourrait rien faire.)

Même en restant dans le domaine perturbatif, on ne sait résoudre la théorie
des cordes que dans peu de champs de fond : l'espace de Minkowski et ses
compactifications toroïdales, bien sûr, et aussi les variétés de groupes
compacts. Ces dernières sont un moyen d'étudier les espaces courbes, et
nous avons utilisé les techniques qui leur sont spécifiques dans le
chapitre 2. Les cas plus généraux restent assez largement inconnus, en
particulier les fonds dépendant du temps, d'une grande importance en
cosmologie, notamment pour voir si la théorie des cordes permet de résoudre
la singularité initiale du Big Bang.

Un développement essentiel dans les années quatre-vingt-dix est
la découverte de l'importance des D-branes. Il s'agit d'objets étendus, de
dimensions variées, sur lesquels les extrémités des cordes ouvertes
s'accrochent. Elles jouent un rôle essentiel dans les dualités mentionnées
ci-dessus : par exemple, la S-dualité de la théorie dite de type IIB
échange les cordes fondamentales --- celles de la théorie perturbative ---
et les D-branes de dimension spatiale un. Elles joueront un rôle essentiel
dans toute cette thèse.

La cohérence de la théorie requiert qu'elle soit formulée en dimension dix,
ce qui en fait six de trop si on veut décrire notre univers avec. Le
premier moyen envisagé pour obtenir une théorie effective à quatre
dimensions est de \emph{compactifier} les six dimensions supplémentaires,
c'est-à-dire considérer les cordes sur le produit cartésien de l'espace de
Minkowski à quatre dimensions et d'une variété compacte de dimension six.
Les D-branes permettent d'autres possibilités, puisque les cordes ouvertes
qui y sont accrochées sont contraintes de rester à leur voisinage ; nous
évoquerons cela plus en détail dans le chapitre~3.

\subsubsection{Plan de cette thèse}
Le premier chapitre vise à introduire la théorie des cordes, en mettant
l'accent sur ce qui sera nécessaire dans la suite pour présenter nos
travaux. Il commence par la quantification de la corde bosonique, qui est
une théorie des cordes sans fermions avec laquelle nous travaillerons dans
la plus grande partie de cette thèse par souci de simplicité, et explique
comment on fait interagir les cordes bosoniques fermées. Ensuite nous
introduisons les cordes ouvertes, et montrons que les D-branes apparaissent
naturellement dans la théorie. Nous étudions ensuite les cordes non
orientées et les orientifolds, qui sont un autre type de défaut
topologique. Enfin, nous décrivons brièvement la théorie des supercordes.

Le deuxième chapitre étudie les cordes sur un groupe compact, dont la
structure algébrique permet de résoudre la théorie des cordes (dans le même
sens qu'on sait la résoudre sur l'espace de Minkowski). Nous commençons par
y rappeler des résultats connus sur les cordes et les D-branes dans de tels
espaces, et notamment le fait remarquable que, dans ces géométries, un
calcul semi-classique des fluctuations des D-branes permet de trouver leur
spectre exact tel qu'on sait le calculer en théorie conforme, en nous
focalisant sur les cas les plus simples, à savoir la 3-sphère (groupe
$\SU(2)$) et l'espace projectif $\RP^3$ (groupe $\SO(3)$). Nous décrivons
ensuite nos résultats sur les orientifolds dans ces espaces, à savoir leur
position et leurs interactions avec les cordes ouvertes et fermées.

Le troisième chapitre s'intéresse aux D-branes dans des espaces
anti-de~Sitter en présence de champs Ramond-Ramond. La motivation de cette
étude est liée au fait que, grâce à la possibilité de localiser la gravité
sur une D-brane, on peut envisager un \emph{univers branaire}, c'est-à-dire
où ce que nous observons est contenu dans une D-brane à 4 dimensions ; nous
commençons donc par rappeler divers faits à propos de ce type de
configurations. Ensuite, nous rappelons des résultats connus sur les
D-branes anti-de~Sitter dans des fonds Neveu-Schwarz et nos résultats sur
les fonds Ramond-Ramond, utilisant la S-dualité reliant ces deux types de
fonds.

Le quatrième chapitre étudie les cordes ouvertes sur une D-brane parcourue
par une onde plane. L'intérêt d'une telle étude est qu'il s'agit d'un fond
dépendant du temps suffisamment simple pour arriver à calculer
explicitement certaines amplitudes, et présente une analogie avec certains
modèles-jouets de singularité cosmologique. Nous présentons les résultats
précédemment connus, concernant les amplitudes à deux cordes, et les
nôtres, hélas très partiels, concernant les amplitudes à trois cordes.

Nous terminons, de façon originale, par une conclusion, où nous rappelons
les principaux résultats de cette thèse et donnons quelques perspectives de
recherche.

Les résultats obtenus au cours de cette thèse, à l'exception de ceux du
chapitre~4, ont donné lieu à des publications jointes en annexe.

%% file: cordes.tex
\chapter{Éléments de théorie des cordes}
Nous allons introduire ici quelques notions fondamentales de théorie
des cordes, en mettant l'accent sur ce qui sera utile dans la suite de
cette thèse. Ce sujet étant raisonnablement bien traité dans les
ouvrages tels que \cite{GSW} et \cite{Pol}, nous ne donnerons
généralement pas de références, et le lecteur pourra se référer aux
bibliographies desdits ouvrages.

\section{Cordes bosoniques fermées}
\subsection{L'action}
Il est assez aisé d'écrire une action pour une corde relativiste en
s'inspirant de l'action d'une particule. L'action d'une particule
relativiste dans l'espace-temps  plat s'écrivant comme son temps propre :
\[
S = -m \int d\t \sqrt{-\frac{dX^{\mu}}{d\t}\frac{dX_{\mu}}{d\t}}
\]
où $\t$ est un paramètre le long de la ligne d'univers de la
particule, et où la métrique est de signature $(-+\ldots+)$, il est assez
naturel d'écrire pour une corde la généralisation suivante, connue
sous le nom d'\emph{action de Nambu-Goto}\index{Nambu-Goto, action de},
où l'action est la surface propre de la surface d'univers de la corde :
\[
S = -\frac{1}{2\pi\a'} \int d\t d\s \sqrt{-\det h}
\quad\text{avec}\quad
h_{\a\b} = \d_{\a}X^{\mu}\d_{\b}X_{\mu}.
\]
Les coordonnées de la surface d'univers sont ici $\t\equiv\s^0$ et
$\s\equiv\s^1$, et $1/(2\pi\a')$ est la tension de la corde ; $h$ n'est
rien d'autre que la métrique induite sur la surface d'univers par celle de
l'espace-temps cible. Dans le cas d'une corde fermée (une boucle), la
coordonnée $\s$ est périodique, et le plus souvent, la période choisie est
$2\pi$. Dans le cas d'une corde ouverte (un segment), $\s$ parcourt
l'intervalle $[0;\pi]$.

Cette action étant peu aisée à manier, en particulier lorsqu'il
s'agit de quantifier la corde, on lui préfère généralement celle de
Polyakov\index{Polyakov, action de} :
\begin{equation}
\label{AcPol}
S= -\frac{1}{4\pi\a'} \int d\t d\s \sqrt{-\det \g}\,
   \g^{\a\b} \d_{\a}X^{\mu} \d_{\b}X_{\mu}
\end{equation}
qui comprend, en plus du plongement de la corde dans l'espace-temps
cible $X^{\mu}$, une métrique \emph{variable} $\g_{\a\b}$. Les
équations du mouvement impliquant que la métrique $\g$ est égale à la
métrique induite à un facteur scalaire près, cette action est
classiquement équivalente à celle de Nambu-Goto.

Elle admet comme symétries locales les difféomorphismes de la surface
d'univers et la symétrie de Weyl\index{Weyl@Weyl, symétrie de}, qui multiplie
la métrique par un scalaire quelconque et laisse le reste invariant. Les
difféomorphismes permettent de mettre la métrique sous la forme
$\g_{\a\b}(\t,\s) = \Lambda(\t,\s) \eta_{\a\b}$ ; la symétrie de Weyl a
alors pour effet que l'action ne dépend pas de $\Lambda$, et on aboutit
alors à une théorie des champs bidimensionnelle avec $D$ champs scalaires
(du point de vue de la surface d'univers) libres, avec comme équations du
mouvement $\Box X^{\mu}=0$ et, dans le cas de la corde ouverte, la
\textit{condition aux bords de Neumann}
\index{Neumann|see{condition aux bords}}
\index{condition aux bords!Neumann}
$\d_{\s}X^{\mu}=0$ (nous reviendrons plus tard sur ce point). Il
faut ajouter à ces équations une contrainte découlant de la variation de la
métrique dans l'action \eqref{AcPol} : le tenseur énergie-impulsion
\begin{equation}
\label{contrainte}
T_{\a\b} = \frac{1}{\a'}(\d_\a X^\mu \d_\b X_\mu - \text{trace})
\end{equation}
doit s'annuler. 

Le fait que sa trace soit nulle sans imposer la contrainte est une
conséquence de la symétrie de Weyl. Cette dernière est affectée par une
anomalie, qui se manifeste précisément par la non-annulation de cette trace
au niveau quantique ; dans le cas d'un espace-temps plat, on montre que
cette anomalie s'annule uniquement en dimension 26. Nous nous placerons
dans ce cas dans la suite.

\subsection{L'invariance conforme et le tenseur énergie-impulsion}
Après fixation de la métrique, il reste encore des symétries, qui sont les
combinaisons de transformations de coordonnées et de transformations de
Weyl qui laissent la métrique invariante. Faire une telle transformation
revient à faire un changement de coordonnées tel que le changement de
métrique associé est une transformation de Weyl, sans faire ladite
transformation de Weyl, autrement dit une transformation conforme.
\index{invariance conforme}

L'annulation de la trace du tenseur énergie-impulsion est une propriété
générale des théories invariantes conformes (CFT). Dans les coordonnées
$\s^\pm=\t\pm\s$, la trace est $T_{+-}$, donc $T_{\a\b}$ a deux composantes
non nulles, $T\equiv T_{++}$ et $\bar{T}\equiv T_{--}$. La conservation de
l'énergie-impulsion s'écrit alors
\[
\d_- T = \d_+ \bar{T} = 0,
\]
c'est-à-dire que $T$ est une fonction de $\s^+$ uniquement, et $\bar{T}$
de $\s^-$ uniquement. Dans la suite, nous parlerons principalement de $T$,
étant entendu que $\bar{T}$ a des propriétés similaires.

$T$ admet le développement en modes suivant :
\index{tenseur énergie-impulsion}
\[
T = \sum_n L_n e^{-in(\t+\s)}
\]
avec $L_{-n}=L_n^\dagger$. On montre alors que les $L_n$ admettent les
relations de commutation suivantes (algèbre de Virasoro) :
\index{Virasoro@Virasoro, algèbre de}
\begin{equation}
\label{virasoro}
[L_m, L_n] = (m-n)L_{m+n} + \delta_{m+n} \frac{c}{12} (m^3-m)
\end{equation}
où $c$ est appelée charge centrale de l'algèbre. Il existe de même une
charge $\bar{c}$ pour l'algèbre des $\bar{L}$ ; dans la suite, nous
considérerons des théories où $\bar{c}=c$.

\subsection{Le spectre}
\index{quantification!covariante}
\index{spectre!corde bosonique fermée}
Nous décrivons ici l'« ancienne quantification covariante » de la corde
bosonique, où on impose la contrainte \eqref{contrainte} comme pour la
quantification de Gupta-Bleuler de l'électrodynamique. Nous n'évoquerons
pas ici la « nouvelle » quantification, dite BRST, basée sur les fantômes
de Fadeev et Popov. La quantification non covariante dans la jauge du cône
de lumière, similaire à la quantification de l'électrodynamique dans la
jauge de Coulomb, sera traitée dans le chapitre~4.

Dans le cas de la corde fermée, les champs $X^\mu$ peuvent être développés
en modes comme suit :
\index{developpement@développement en modes!corde fermée}
\begin{equation}
\label{DevModesF}
X^\mu(\t,\s) = x^\mu + \a'p^\mu\t + i \sqrt{\frac{\a'}{2}} \sum_{n\neq0}
 \left( \frac{\a^\mu_n}{n}e^{-in\s^+} + \frac{\bar{\a}^\mu_n}{n}e^{-in\s^-}
 \right).
\end{equation}
La réalité de $X^\mu$ impose $\a^\mu_{-n}=(\a^\mu_n)^\dagger$. Les
relations de commutation de ces modes s'écrivent :
\[
[x^\mu,p^\nu] = i\eta^{\mu\nu}, \quad
[\a^\mu_m,\bar{\a}^\nu_n] = 0, \quad
[\a^\mu_m,\a^\nu_n] = [\bar{\a}^\mu_m,\bar{\a}^\nu_n] 
                    = m\delta_{m+n}\eta^{\mu\nu}.
\]
Les états s'obtiennent alors en faisant agir $\a^\mu_{-n}$ et
$\bar{\a}^\mu_{-n}$, $n>0$, sur un vide annulé par $\a^\mu_{n}$ et
$\bar{\a}^\mu_{n}$. On a alors, comme lors de la quantification de
l'électrodynamique, un espace de Hilbert avec un produit scalaire non
défini positif à cause du mauvais signe des relations de commutation des
$\a^0$. Il faut alors encore imposer la contrainte \eqref{contrainte}.

Les modes du tenseur énergie-impulsion s'écrivent :
\index{tenseur énergie-impulsion}
\begin{equation}
\label{Ln}
L_n = \frac{1}{2} \sum_m \ordnorm{\a_n \cdot \a_{m-n}}
\end{equation}
avec $\a^\mu_0=\bar{\a}^\mu_0=\sqrt{\a'/2}\,p^\mu$. L'ordre normal
(opérateurs de création à gauche des opérateurs d'annihilation) est
nécessaire pour $L_0$ parce que $\a_n$ et $\a_{-n}$ ne commutent pas. Il
n'est pas très difficile de vérifier alors que les relations
\eqref{virasoro} sont vérifiées, avec $c$ égal à la dimension de
l'espace-temps.

Pour imposer la contrainte, on exige des états physiques qu'ils vérifient
$L_n|\psi\>=0$ pour $n>0$ et $(L_0-a)|\psi\>=0$, le $a$ étant nécessaire à
cause des problèmes d'ordre des opérateurs. Il en découle l'annulation de
$T$ pris entre deux états physiques. On montre que la cohérence de la
théorie impose que $a=1$, en plus de $D=26$ que nous avons évoqué avant.
(En fait, si on ne considère que des cordes sans interactions, d'autres
possibilités sont cohérentes, notamment $D\leq25$, $a\leq1$. Mais on ne
peut mettre d'interactions cohérentes que dans le cas $D=26$, $a=1$).

L'annulation de $L_0-a$ et $\bar{L}_0-a$ donne les équations suivantes :
\[
m^2 = \frac{2}{\a'} (N+\bar{N}-2)
\]
\[
N = \bar{N}
\]
où $N=\sum_{n>0}\a_{-n}\cdot\a_n$ est le nombre d'oscillateurs gauches
(étant entendu que $\a_{-n}$ en crée $n$) et $\bar{N}$ le nombre
d'oscillateurs droits.

L'état fondamental de ce spectre est un tachyon. On peut vérifier que, pour
un tel état, $L_n$ s'annule automatiquement pour $n>0$. La présence d'un
tachyon est gênante, car cela signifie qu'on ne développe pas la théorie
autour d'un vide stable. Heureusement, il a le bon goût de disparaître dans
la théorie des supercordes que nous évoquerons plus loin.

Le premier état excité, obtenu avec $N=\bar{N}=1$, s'écrit de façon
générale $\zeta_{\mu\nu}\a^\mu_{-1}\bar{\a}^\nu_{-1}|k\>$, où $|k\>$ est un
état du vide d'impulsion $k$. Sachant que, pour un tel état, $L_n$ s'annule
automatiquement pour $n>1$, on n'a à vérifier que l'annulation de $L_1$. Il
est aisé de voir que cela implique une condition de polarisation
$k^\mu\zeta_{\mu\nu}=k^\nu\zeta_{\mu\nu}=0$. L'espace des états ainsi
obtenu a un produit scalaire positif, mais pas défini : les états où
$\zeta_{\mu\nu}$ est de la forme $k_{\mu}a_{\nu}$ ou $a_{\mu}k_{\nu}$
découplent (ont un produit scalaire nul avec tout état physique), et il
faut donc quotienter l'espace des états par ces états (ce qui revient à une
invariance de jauge) pour obtenir les états physiques. Ces excitations se
décomposent en un scalaire $\Phi$, appelé dilaton, une 2-forme $B_{\mu\nu}$
et un tenseur symétrique de trace nulle qui est le graviton $G_{\mu\nu}$.

\index{Polyakov, action de}
\index{couplage!à un champ de fond|see{champ de fond}}
\index{champ de fond!$B_{\mu\nu}$}
L'action de Polyakov \eqref{AcPol} peut être généralisée pour inclure des
champs de fond correspondant à une superposition cohérente d'excitations de
masse nulle. L'action s'écrit alors
\begin{equation}
\label{ModSigma}
\begin{split}
S &= -\frac{1}{4\pi\a'} \int d^2\s \sqrt{-\det \g}\,
      \g^{\a\b} \d_{\a}X^{\mu} \d_{\b}X^{\nu} G_{\mu\nu}(X)
    \\ &\quad
    + \frac{1}{2\pi\a'} \int \hat{B}(X)
    - \frac{1}{4\pi} \int d^2\s \sqrt{-\det \g}\, \Phi(X) R
\end{split}
\end{equation}
où $\hat{B}$ est la 2-forme induite par $B$ sur la surface d'univers et $R$
est la courbure de la métrique sur la surface d'univers. L'invariance de
Weyl impose des contraintes sur les champs de fond : ils doivent vérifier
des équations que l'on peut dériver de l'action effective suivante, à
l'ordre le plus bas en $\a'$ (c'est-à-dire dans la limite de basse
énergie) :
\index{action effective!corde fermée}
\[
S_{\mathrm{eff}} = \frac{1}{2\kappa_0^2} \int d^Dx \sqrt{-\det G} e^{-2\Phi}
  \left[ R + 4\d_\mu\Phi\d^\mu\Phi - \frac{1}{12} H_{\mu\nu\rho}H^{\mu\nu\rho}
  \right]
\]
où $H=dB$. On y reconnaît notamment l'action d'Einstein-Hilbert, ce qui
montre que la théorie des cordes contient la gravitation. La constante
$\kappa_0$ n'a pas de signification physique, puisqu'on peut changer sa
normalisation par une translation de $\Phi$. Ce qui, en revanche, a une
signification, est $\kappa=\kappa_0e^{\Phi_0}$, où $\Phi_0$ est la valeur
moyenne de $\Phi$ : cela s'interprète comme le couplage gravitationnel. On
peut donner au premier terme de l'action la forme habituelle (sans le
préfacteur $e^{-2\Phi}$) en écrivant l'action en termes de la
\emph{métrique d'Einstein} $G_E$, qui s'exprime en termes de la
\emph{métrique de corde fermée} $G$ par $G_E=e^{-4\Phi/(D-2)}G$.
\index{metrique@métrique!d'Einstein}

\subsection{Passage à une surface d'univers euclidienne}
\label{eucl}
La surface d'univers de la corde telle que nous l'avons vue jusqu'ici est
minkowskienne, car il s'agit de la trajectoire d'une corde dans un
espace-temps cible minkowskien. Dans la suite, il apparaîtra néanmoins
utile de considérer une surface d'univers euclidienne, car cela simplifie
considérablement certains calculs d'amplitudes, notamment parce que les
groupes conformes euclidien et minkowskien sont assez différents. Il
convient donc de justifier que ce passage du minkowskien vers l'euclidien
est légitime.

Pour cela, nous partons de l'intégrale de chemin, qui s'écrit
$\int[de\,dX]\exp(iS)$, où $S$ est l'action du modèle sigma
\eqref{ModSigma}, et $e$ le vielbein, en termes de laquelle la métrique
s'écrit $\g_{\a\b}=-e^0_\a e^0_\b+e^1_\a e^1_\b$. Puisque l'intégrale de
chemin est un produit (infini) d'intégrales ordinaires, on peut déformer
les contours d'intégration. On les déforme donc de façon à donner à
$e^0_\a$ des valeurs imaginaires pures, c'est-à-dire qu'on pose
$e^0_\a=-i\tilde{e}^0_\a$. La métrique devient alors euclidienne, et
l'intégrale de chemin devient $\int[d\tilde{e}\,dX]\exp(-S_E)$, où l'action
euclidienne $S_E$ s'écrit
\index{Polyakov, action de!euclidienne}
\[
\begin{split}
S_E &= \frac{1}{4\pi\a'} \int d^2\s \sqrt{\det \g}\,
        \g^{\a\b} \d_{\a}X^{\mu} \d_{\b}X^{\nu} G_{\mu\nu}(X)
      \\ &\quad
      - \frac{i}{2\pi\a'} \int B(X)
      + \frac{1}{4\pi} \int d^2\s \sqrt{\det \g}\, \Phi(X) R.
\end{split}
\]
Ainsi, il découle que les amplitudes de la théorie des cordes données par
l'action minkowskienne \eqref{ModSigma} sont les mêmes que celles de
l'action euclidienne $S_E$. C'est cette dernière que nous utiliserons le
plus souvent pour étudier les interactions.

Il est important de noter que cette procédure est applicable à l'action
avant toute fixation de jauge. Lorsque, par exemple à la suite du choix
d'une jauge non covariante, telle que la jauge du cône de lumière que nous
utiliserons au chapitre~4, il apparaît dans l'action des termes dépendant
du temps $\t$ de la surface d'univers, alors cela ne marchera plus, et
c'est l'action minkowskienne qu'il faudra utiliser, comme nous le
vérifierons explicitement en comparant avec un formalisme covariant.

\subsection{Interactions et fonction de partition}
\begin{figure}
\begin{center}
\includegraphics[scale=0.7]{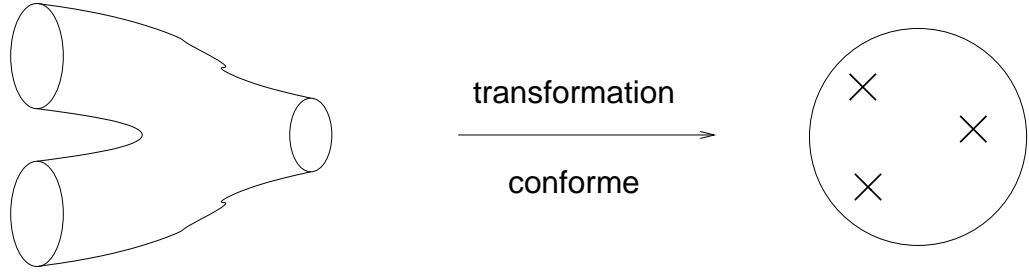}
\end{center}
\caption{Exemple de surface d'univers intervenant dans une interaction à
 trois cordes.}
\label{inter}
\end{figure}

\index{interactions!corde fermée}
En physique des particules, les interactions s'obtiennent par une sommation
sur tous les diagrammes de Feynman possibles, lesdits diagrammes étant
constitués de lignes se rejoignant en des vertex. La généralisation de ceci
aux cordes consiste à sommer sur des surfaces, orientables dans le cas de
cordes orientées, où chaque corde entrante est un tuyau venant de l'infini.
Par une transformation conforme appropriée, dans le cas euclidien ---~et
pas dans le cas minkowskien, d'où l'intérêt de travailler avec une action
euclidienne~--- ceci se ramène à une surface compacte privée d'autant de
points qu'il y a de cordes, de genre égal au nombre de boucles, et les
amplitudes s'obtiennent alors comme des moyennes de produits
d'\textit{opérateurs de vertex} insérés en chaque point (fig. \ref{inter}).
\index{operateur@opérateur de vertex}
Pour un état sans excitation (tachyon), l'opérateur de vertex est une
fonction de $X$ uniquement (pas de ses dérivées) donnée par le fait qu'il
se transforme de la même façon que l'état correspondant par les
transformations de Poincaré, donc $\ordnorm{e^{ik\cdot X}}$ pour un état
d'impulsion $k$. Les excitations s'obtiennent en multipliant cela par
$\d_+^nX^\mu$ pour chaque opérateur de création $\a_{-n}^\mu$ : ainsi, les
états de masse nulle ont un opérateur de vertex proportionnel à
$\zeta_{\mu\nu} \ordnorm{\d_+X^\mu\d_-X^\nu e^{ik\cdot X}}$.

Dans le cas d'une amplitude à l'ordre des arbres, le diagramme est une
sphère, et les amplitudes s'obtiennent simplement en prenant la moyenne du
produit des opérateurs de vertex, intégrée sur l'ensemble des positions
possibles des vertex. Lorsqu'il y a des boucles, les choses sont plus
compliquées, car les métriques possibles sur une surface de genre non nul
ne sont pas équivalentes par transformation conforme, donc il faut sommer
sur les classes d'équivalence des différentes métriques, qui forment un
espace à deux paramètres réels pour une boucle (tore), et $6n-6$ paramètres
pour $n$ boucles avec $n\geq2$. Nous nous limiterons ici au tore.

\index{fonction de partition!tore}
\index{tore|see{fonction de partition}}
\index{invariance modulaire}
Dans le cas du tore, on peut, par une transformation conforme, se ramener
au cas d'un tore plat défini comme le plan complexe muni de sa métrique
habituelle, quotienté par un réseau engendré par deux complexes $z_1$ et
$z_2$. Par des rotations et dilatations appropriées, on peut se ramener à
$z_1=1$, de sorte que seul le rapport $\tau=z_2/z_1$ caractérise la
structure conforme du tore ; en outre, on pourra toujours prendre
$\im(\t)>0$, étant entendu que si cela n'est pas le cas, on peut y remédier
en échangeant $z_1$ et $z_2$. La base $(z_1,z_2)$ pour le réseau n'est pas
unique, puisqu'on peut changer de base par l'action d'une matrice
unimodulaire à coefficients entiers (groupe $\SL(2,\Z)$), l'action qui en
résulte sur $\t$ étant $\t\rightarrow\frac{a\t+b}{c\t+d}$, avec $a$, $b$,
$c$ et $d$ entiers. Il sera essentiel, dans la suite, de s'assurer de
l'invariance par ces transformations, dites modulaires, des amplitudes sur
le tore. $\SL(2,\Z)$ étant généré par :
\[
T : \t \rightarrow \t+1, \quad S : \t \rightarrow -\frac{1}{\t}
\]
il suffit de vérifier l'invariance par ces deux transformations.

L'amplitude du tore qui nous intéressera particulièrement est celle du
vide, c'est-à-dire sans opérateur de vertex. Elle s'obtient, pour un tore
de module $\t=\t_1+i\t_2$, en considérant une corde fermée qu'on fait
évoluer pendant le temps euclidien $2\pi\t_2$, puis que l'on translate de
$2\pi\t_1$ le long de sa coordonnée spatiale $\s$ avant de recoller les
bouts. Cela donne la trace
\[
Z(\t)=\Tr\bigl[\exp(2\pi i\t_1P - 2\pi\t_2H)\bigr]
\]
où la trace est prise sur le spectre de la théorie. Cette amplitude,
faisant intervenir une trace pondérée par le spectre du hamiltonien, n'est
autre que la fonction de partition de la corde. L'impulsion est l'intégrale
de $T_{\t\s}$ sur la corde et vaut donc $L_0-\bar{L}_0$. Le hamiltonien
vaudrait naïvement $L_0+\bar{L}_0$, mais les corrections quantiques
conduisent à y ajouter un terme $-c/12$, où $c$ est la charge centrale de
l'algèbre de Virasoro, de sorte que l'amplitude s'écrit
\[
Z(\t) = \Tr\bigl[q^{L_0-c/24}\bar{q}^{\bar{L}_0-c/24}\bigr]
\]
avec $q=e^{2\pi i\t}$%
\footnote{Dans la mesure où $q$ n'est pas un réel positif et $L_0-c/24$
n'est pas entier, la notation $q^{L_0-c/24}$ est mal définie, et on aurait
dû lui préférer $\exp[2\pi i\t(L_0-c/24)]$. C'est, hélas, la notation
traditionelle.}. Sachant que le spectre se décompose en paires de
représentations de l'algèbre de Virasoro (une à gauche et une à droite), ou
d'une éventuelle algèbre de symétrie étendue (dont nous verrons un exemple
au chapitre~\ref{WZW}) et que pour une paire donnée les excitations gauche
et droite sont indépendantes (la condition $L_0=\bar{L}_0$ est une
condition de couche de masse que nous n'avons pas à imposer ici), la
fonction de partition se décompose en
\[
Z(\t) = \sum_{R,\bar{R}} n_{R,\bar{R}} \chi_R(q) \chi_{\bar{R}}^*(\bar{q})
\]
où $n_{R,\bar{R}}$ est la multiplicité de la représentation $(R,\bar{R})$
dans le spectre, et
\[
\chi_R(q) = \Tr_R(q^{L_0-c/24}).
\]
La somme doit être remplacée par une intégrale en cas de spectre continu.
L'invariance modulaire de la fonction de partition a une conséquence
importante sur les transformations modulaires des caractères $\chi_R$ : en
effet, le fait que ces transformations laissent invariante une forme
sesquilinéaire des caractères entraîne qu'elles agissent linéairement sur
l'ensemble des caractères, ce qui nous sera extrêmement utile au chapitre~%
\ref{WZW}.

\index{couplage!constante de}
On peut s'étonner de l'absence apparente de constante de couplage dans les
amplitudes. En fait, elle apparaît sous la forme d'un terme $\Phi_0\chi$
dans l'action, où
\begin{equation}
\label{Euler}
\chi = \frac{1}{4\pi} \int d^2\s \sqrt{\det\g} R
\end{equation}
est un invariant topologique, appelé \emph{caractéristique d'Euler} de la
surface d'univers, qui, pour une surface orientée sans bord, vaut $2-2n$,
où $n$ est le genre de la surface, c'est-à-dire pour nous le nombre de
boucles. L'intégrale de chemin fait alors apparaître un facteur
$e^{\Phi_0(2n-2)}$, ce qui conduit à interpréter $g=e^{\Phi_0}$ comme la
constante de couplage des cordes. Il en découle que la constante de
couplage n'est pas un paramètre de la théorie, comme dans le cas des
théories des champs habituelles, mais une propriété du champ de fond autour
duquel on développe. Plus généralement, il n'y a pas en théorie des cordes
de paramètre libre adimensionné, contrairement au modèle standard qui en a
une vingtaine, mais un ensemble de vides, les couplages dépendant du vide
considéré. (Dans la mesure où on ne sait pour l'instant pas comment choisir
le vide, cela ne fait que déplacer le problème.)

\subsection{Compactification et T-dualité}
\index{compactification toroïdale}
Comme nous l'avons vu, la théorie des cordes bosoniques n'est cohérente
qu'en 26 dimensions, ce qui fait 22 de trop. Nous verrons plus loin que la
théorie des supercordes requiert 10 dimensions, donc 6 de trop. Un moyen de
régler ce problème est de \emph{compactifier} les dimensions excédentaires,
c'est-à-dire considérer des cordes sur le produit cartésien de $\R^4$ par
une variété compacte. Nous nous limiterons ici au cas le plus simple,
c'est-à-dire le tore ; nous n'allons même ici compactifier qu'une seule
dimension, laissant le cas de plusieurs dimensions aux références. D'autres
compactifications plus compliquées seront évoquées dans le chapitre~%
\ref{WZW}.

Considérons donc une corde fermée dans un espace où la dimension $X$ est
compactifiée, c'est-à-dire que $X\sim X+2\pi R$. Cela a deux conséquences :
\begin{itemize}
\item Une conséquence que l'on rencontre aussi en théorie des champs est la
  quantification de l'impulsion selon la direction compactifiée : $p=n/R$,
  avec $n$ entier.
\item Une conséquence spécifique aux cordes est l'existence d'états
  enroulés autour de la dimension compactifiée, c'est-à-dire avec une
  relation de périodicité modifiée : $X(\s+2\pi) = X(\s) + 2\pi Rw$, où $w$
  est un entier comptant le nombre de tours que fait la corde autour de la
  direction compacte.
\end{itemize}

\begin{figure}
\begin{center}
\includegraphics{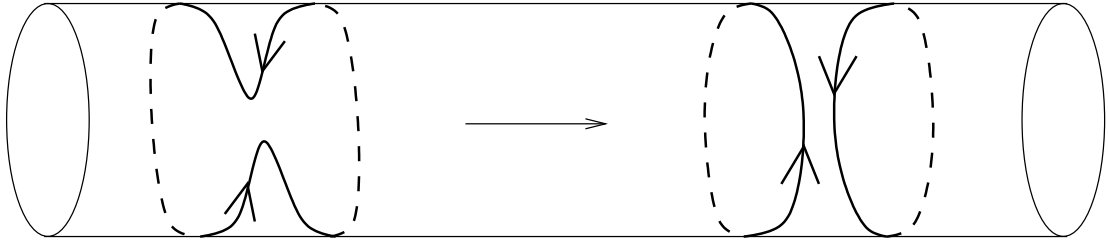}
\end{center}
\caption{Processus par lequel une corde enroulée donne deux cordes
enroulées de nombres d'enroulement opposés.}
\label{enroule}
\end{figure}

On peut se demander s'il est nécessaire d'inclure les états enroulés dans
le spectre. La réponse est oui : d'une part il est aisé d'imaginer un
processus par lequel une corde non enroulée se déforme et arrive à
interagir avec elle-même pour donner deux cordes de nombres d'enroulement
$+1$ et $-1$ (fig. \ref{enroule}), d'autre part ils sont nécessaires à
l'invariance modulaire.

Ceci est une propriété assez générale : lorsqu'on quotiente une théorie des
cordes par un sous-groupe $G$ des isométries de l'espace-temps (ici un
sous-groupe $\Z$ des translations), si on ne garde que les états de la
théorie initiale invariants par $G$, la fonction de partition obtenue n'est
pas invariante modulaire, et il faut ajouter des états où le champ $X$ sur
la surface d'univers n'est périodique en $\s$ qu'à un élément de $G$ près
(ici les états enroulés). Nous reverrons cela à l'\oe uvre dans le
chapitre~\ref{WZW} lorsque nous construirons les cordes sur $\RP^3$ à
partir des cordes sur $S^3$.

La dimension compactifiée admet le développement en modes suivant :
\index{developpement@développement en modes!corde fermée}
\[
X = x + \a'\frac{n}{R}\t + wR\s + i \sqrt{\frac{\a'}{2}} \sum_{n\neq0}
 \left( \frac{\a_n}{n}e^{-in\s^+} + \frac{\bar{\a}_n}{n}e^{-in\s^-}
 \right).
\]

Les modes de l'énergie-impulsion restent donnés par l'équation \eqref{Ln},
à cela près que maintenant
\[
\a_0 = \sqrt{\frac{\a'}{2}}\frac{n}{R} + \frac{wR}{\sqrt{2\a'}}
\quad\text{et}\quad
\bar{\a}_0 = \sqrt{\frac{\a'}{2}}\frac{n}{R} - \frac{wR}{\sqrt{2\a'}}
\]
d'où le spectre de masse (définie à partir de l'impulsion des dimensions
non compactes) :
\index{spectre!corde bosonique fermée}
\[
m^2 = \frac{n^2}{R^2} + \frac{w^2R^2}{\a'^2} + \frac{2}{\a'}(N+\bar{N}-2)
\]
avec la contrainte
\[
N - \bar{N} + nw = 0.
\]

Dans la limite $R\rightarrow\infty$, les états enroulés deviennent
infiniment massifs, ce qui est normal car ce sont des cordes très longues,
l'impulsion compacte tend vers un spectre continu, et on retrouve donc le
spectre d'une dimension non compacte. Dans la limite $R\rightarrow0$, comme
prévu les états d'impulsion compacte non nulle deviennent infiniment
massifs. En revanche, les états enroulés ont maintenant un spectre continu,
ressemblant beaucoup à une dimension non compacte, puisqu'il est facile
d'enrouler une corde sur un petit cercle. On aboutit donc à un spectre
ressemblant beaucoup à celui d'une dimension non compacte, là où en théorie
des champs on tendrait vers un spectre à une dimenension de moins, car il
n'y a pas de nombre d'enroulement.

\index{T-dualite@T-dualité}
C'est qu'en fait les limites $R\rightarrow\infty$ et $R\rightarrow0$ sont
\emph{identiques} ; en effet, le spectre est invariant sous
\[
R \rightarrow \frac{\a'}{R}, \quad n \leftrightarrow w.
\]
Sachant que $X$ se sépare en des parties gauche et droite :
$X(\t,\s)=X_L(\s^+)+X_R(\s^-)$, ceci revient à remplacer $X$ par
$X'(\t,\s)=X_L(\s^+)-X_R(\s^-)$, c'est-à-dire que c'est une parité sur la
partie droite de $X$. Sachant que, pour la corde fermée, les deux parties
de $X$ sont essentiellement indépendantes, cela est une symétrie de
l'ensemble de la théorie, et pas seulement du spectre, connue sous le nom
de T-dualité.

\section{Cordes ouvertes et D-branes}
Après avoir étudié les cordes fermées, nous allons passer aux cordes
ouvertes, c'est-à-dire des segments. Notre étude nous amènera à découvrir
tout un ensemble d'objets dynamiques étendus, nommés D-branes, auxquels est
consacrée une grande partie de cette thèse. Ce sujet étant inconnu de
\cite{GSW}, et que \cite{Pol} est loin d'être complet à ce propos, le
lecteur pourra se référer aux revues sur le sujet, notamment
\cite{DiV,BacBr}.

\subsection{Le spectre}
\index{spectre!corde bosonique ouverte}
Pour obtenir des cordes ouvertes, on considère une surface d'univers avec
des bords correspondants aux extrémités de la corde, situé en $\s=0$ et
$\s=\pi$.

La présence d'un bord implique des conditions aux bords. Dans le cas du
tenseur énergie-impulsion, il s'agit d'imposer qu'il n'y ait pas d'énergie
qui sorte de la corde par le bord, ce qui impose que $T=\bar{T}$ au bord.
Il en découle que $\bar{L}_n=L_n$, autrement dit il n'y a plus qu'une
algèbre de Virasoro.

\index{condition aux bords!Neumann}
\index{developpement@développement en modes!corde ouverte}
En ce qui concerne les champs $X^\mu$ de la corde, la condition aux bords,
qui découle de la variation de l'action, s'écrit $\d_\s X^\mu=0$. Il en
découle le développement en modes de $X^\mu$ :
\begin{equation}
\label{ModesOuvN}
X^\mu(\t,\s) = x^\mu + 2\a'p^\mu\t 
   + i\sqrt{2\a'} \sum_{n\neq0} \frac{\a^\mu_n}{n}e^{-in\t}\cos(n\s).
\end{equation}
Les relations de commutations sont alors les mêmes que pour la corde fermée
(à ceci près qu'il n'y a plus de $\bar{\a}$, car la condition aux bords
identifie les modes gauches et droits).

Comme pour la corde fermée, le spectre découle de l'annulation de $L_0-1$
sur les états, et on trouve
\[
m^2 = \frac{1}{\a'}(N - 1).
\]
Là aussi, il y a un tachyon qui disparaît dans la théorie des supercordes.

Le premier état excité s'écrit de façon générale $e_\mu\a^\mu_{-1}|k\>$, où
$|k\>$ est un état du vide d'impulsion $k$. L'annulation de $L_1$ implique
la condition de polarisation $k^\mu e_\mu=0$, et les états ayant
$e_\mu\propto k_\mu$ sont de norme nulle, ce qui conduit à considérer comme
équivalents des états dont la polarisation diffère d'un multiple de
$k_\mu$. On reconnaît là les excitations d'un champ de jauge abélien.

\index{champ de fond!électromagnétique}
Comme dans le cas fermé, on peut ajouter à l'action un terme pour ajouter
un champ de fond électromagnétique correspodant à une superposition
d'excitations de corde ouverte. Ce terme s'écrit comme une intégrale sur le
bord de la surface d'univers :
\[
S_{\text{ém}} = \int_{\d\Sigma} d\t A_\mu(X) \d_\t X^\mu,
\quad\text{avec } \int_{\d\Sigma} = \int_{\s=\pi} - \int_{\s=0}
\]
c'est-à-dire que la corde est un dipôle pour ce champ.

Nous verrons plus loin l'action effective à basse énergie correspondante.

\subsection{T-dualité et D-branes}
\index{T-dualite@T-dualité}
On considère une corde ouverte dans un espace plat où la dimension $X$ est
compactifiée, avec $X\sim X+2\pi R$. Comme en théorie des champs et pour la
corde fermée, l'impulsion est quantifiée ; en revanche, il n'y a pas de
nombre d'enroulement. On en conclut que, dans la limite $R\rightarrow0$, on
a simplement une dimension en moins, comme en théorie des champs.

Sachant que les cordes ouvertes sont toujours, d'après ce que nous avons
vu, accompagnées de cordes fermées, nous avons donc ensemble des cordes
ouvertes vivant à $D-1$ dimensions et des cordes fermées vivant, comme nous
l'avons vu précédemment, à $D$ dimensions parce que le rayon nul est T-dual
d'un rayon infini, ce qui paraît quelque peu étrange. C'est pourtant, en un
sens, ce qui arrive. Regardons en effet les conditions aux bords. Elles
s'écrivent, nous l'avons vu, $\d_\s X=0$, ou encore $\d_+X_L-\d_-X_R=0$.
En termes du champ $X'$, cela devient $\d_+X'_L+\d_-X'_R=0$, soit
$\d_\t X'=0$, c'est-à-dire que la valeur de $X'$ reste constante le long du
bord de la corde. Plus précisément, la coordonnée $X'$ admet le
développement en modes suivant :
\index{developpement@développement en modes!corde ouverte}
\begin{equation}
\label{ModesOuvD}
X'(\t,\s) = x'_0 + 2nR'\s 
   + \sqrt{2\a'} \sum_{n\neq0} \frac{\a_n}{n}e^{-in\t}\sin(n\s).
\end{equation}
où $R'=\a'/R$ est le rayon de compactification T-dual, et $x'_0$ n'est pas
un mode, mais une constante arbitraire liée au fait que $X'$ est défini à
une constante près. Nous voyons clairement ici que, après T-dualité, la
corde a ses deux extrémités contraintes de rester sur l'hyperplan
$X'=x'_0$.
\index{Dirichlet|see{condition aux bords}}
\index{condition aux bords!Dirichlet}
\index{D-brane!espace plat}
On est donc conduit naturellement à introduire dans la théorie un nouvel
objet, baptisé \emph{D-brane}, qui est un hyperplan où les extrémités des
cordes ouvertes sont contraintes de rester ; le nom provient des conditions
aux bords, appelées \emph{conditions de Dirichlet}, et du mot membrane. Le
nombre quantique $n$ qui intervient ici, qui était, avant de T-dualiser,
l'impulsion selon la dimension compacte, est un nombre d'enroulement, qui a
un sens bien défini ici, puisqu'on peut imaginer une corde qui, partant de
la D-brane, s'enroule $n$ fois autour de la dimension compacte avant de
revenir à la D-brane ; il n'y a pas d'impulsion selon la dimension compacte
car la symétrie par translation est brisée par la D-brane.

En compactifiant et T-dualisant plusieurs dimensions, on peut obtenir une
D-brane en dimensions inférieures ; on appelle D$p$-brane un telle membrane
avec $p$ dimensions spatiales (et donc $p+1$ dimensions en tout). Par
extension, on considère que lorsqu'on a des conditions aux bords de Neumann
sur toutes les coordonnées, on a affaire à une D25-brane remplissant tout
l'espace-temps.

\index{couplage!à une D-brane}
On peut modifier l'action pour faire en sorte que les conditions au bord de
Dirichlet découlent de l'action. Pour obtenir une D-brane localisée en
$X^i=Y^i$, il suffit d'ajouter à l'action un terme
\begin{equation}
\label{AcOuv}
S_D = \pm\frac{1}{2\pi\a'} \int_{\d\Sigma} d\t (X^i-Y^i) \d_\s X_i,
\end{equation}
le signe étant $+$ pour une surface d'univers minkowskienne, et $-$ dans le
cas euclidien. Il est alors aisé de voir que lorsqu'on varie $X^i$, le
terme de bord dans la variation de l'action qui en découle est
$(X^i-Y^i)\d_\s\delta X_i$, d'où la condition de Dirichlet. On remarque que
le terme $Y^i\d_\s X_i$ est T-dual du terme d'interaction de la corde avec
un champ électromagnétique, ce qui conduit à voir une T-dualité entre un
tel champ et la position de la D-brane :
\[
Y^i \leftrightarrow -2\pi\a' A_i.
\]
Sachant que $A_i$ peut ne pas être constant, il découle que $Y^i$ aussi,
autrement dit on peut avoir une D-brane courbe de la forme $X^i=Y^i(X^\a)$,
où les $X^i$ et les $X^\a$ sont des ensembles disjoints de coordonnées.
Cela n'a rien d'étonnant : la théorie des cordes contenant la gravité, la
métrique fluctue, et donc on imagine mal qu'une D-brane puisse rester plate
dans une géométrie fluctuante.

\index{Chan-Paton, indice de}
\index{groupe de jauge!unitaire}
Il est possible d'avoir plusieurs D-branes simultanément. Dans ce cas,
chaque extrémité de corde ouverte est attachée à une D-brane particulière%
\footnote{À l'époque où la notion de D-brane n'était pas connue, on mettait
à chaque extrémité de corde un \emph{indice de Chan-Paton} prenant $n$
valeurs, là où en langage moderne on considère qu'il y a $n$ D-branes
superposées remplissant l'espace.},
tous les couples de D-branes étant a priori possibles. Si les deux
D-branes auxquelles une corde est attachée sont séparées, ses états (autre
que l'état fondamental, qui disparaît dans la théorie des supercordes) sont
tous massifs ; si elles sont superposées, il y a des états vectoriels de
masse nulle entre eux. Dans le cas de $n$ D-branes superposées, on a donc
$n^2$ vecteurs de masse nulle, qui forment un champ de jauge $\U(n)$, la
symétrie $\U(n)$ mélangeant les D-branes entre elles. On verra qu'avec des
cordes non orientées on pourra obtenir d'autres groupes de jauge.

\subsection{Interactions}
\index{interactions!corde ouverte}
Comme précédemment, les interactions s'obtiennent par sommation sur des
surfaces. Dans le cas des cordes ouvertes, on considère des surfaces avec
un bord, et les opérateurs de vertex pour les cordes ouvertes vivent sur le
bord. Dans le cas d'une amplitude à l'ordre des arbres, le diagramme est un
disque, et l'ajout de boucles consiste à ajouter des trous (des bords
supplémentaires) dans la surface d'univers.

\begin{figure}
\begin{center}
\includegraphics[scale=0.7]{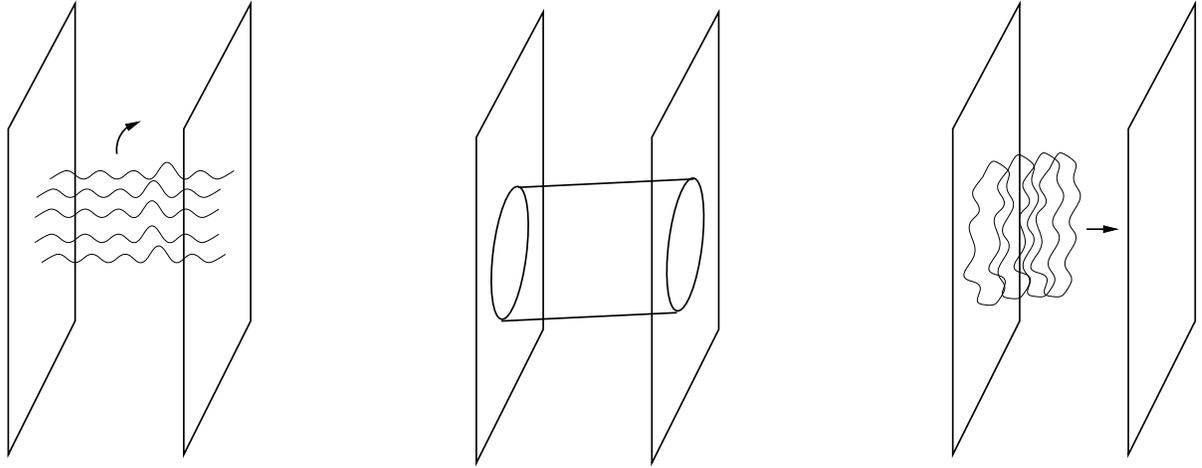}
\end{center}
\caption{Les deux interprétations du diagramme du cylindre : à gauche en
  termes de cordes ouvertes, à droite en termes de cordes fermées.}
\label{anneau}
\end{figure}

Un point important est que, alors qu'on peut avoir des cordes fermées sans
cordes ouvertes de façon cohérente, l'inverse n'est pas vrai. Il est en
effet aisé d'imaginer une interaction où les deux bouts d'une même corde
ouverte fusionnent pour donner une corde fermée. D'autre part, considérons
le diagramme du vide ayant la topologie d'un cylindre. Un tel diagramme
peut s'interpréter comme une boucle de corde ouverte, c'est-à-dire une
corde ouverte identifiée périodiquement dans le temps, et c'est ce que nous
utiliserons pour évaluer la fonction de partition. Il s'interprète aussi
comme l'émission puis l'absorption d'une corde \emph{fermée} par une
D-brane (fig. \ref{anneau}).
\index{dualite@dualité ouvert-fermé}
Il en découle qu'une théorie cohérente de cordes ouvertes doit contenir des
cordes fermées. L'interaction entre cordes fermées et cordes ouvertes
s'obtient en insérant simultanément des opérateurs de vertex sur le bord de
la surface d'univers pour les cordes ouvertes, et à l'intérieur pour les
cordes fermées. Signalons que les surfaces avec bord interviennent dans les
amplitudes ne contenant que des cordes fermées : par exemple, un disque
avec deux opérateurs de vertex à l'intérieur peut s'interpréter comme une
interaction où une corde fermée s'ouvre puis se referme.

Pour une surface orientable de genre $n$ avec $b$ bords, la caractéristique
d'Euler \eqref{Euler} vaut $2-2n-b$, ce qui conduit à ajouter un facteur
$g$ par boucle de cordes ouvertes.

\subsection{Fonction de partition et états de bord}
\index{fonction de partition!anneau}
\index{cylindre|see{fonction de partition, anneau}}
\index{anneau|see{fonction de partition}}
Revenons à l'amplitude du cylindre (qu'on appelle aussi amplitude de
l'anneau, puisqu'on peut passer de l'un à l'autre par une transformation
conforme). On peut caractériser un cylindre aux dilatations près par un
paramètre réel, qui est le rapport $t$ entre la longueur d'un bord et
l'écart entre les bords. L'amplitude s'obtient alors en considérant une
corde ouverte (de longueur $\pi$) qu'on fait évoluer pendant le temps
euclidien $\pi t$ avant de l'identifier à la corde initiale. Cela donne la
trace
\begin{equation}
\label{FoncPartOuv}
Z(t) = \Tr\bigl[ \exp(-\pi t H_o) \bigr]
     = \Tr\bigl[ \sqrt{q}^{L_0-c/24} \bigr]
\end{equation}
avec $q=e^{-2\pi t}$ (le fait de prendre $\t=it$ est lié au fait que le
cylindre peut être vu comme un tore de paramètre $it$ quotienté par une
réflexion, les bords correspondant aux points fixes de la réflexion). 

\index{etatb@état de bord}
En inversant le rôle de l'espace et du temps, ce qui est l'analogue d'une
transformation modulaire pour les cordes fermées, et en effectuant une
dilatation appropriée, on obtient une corde fermée (de longueur $2\pi$) qui
se propage entre deux D-branes pendant un temps $2\pi/t$, d'où une autre
expression de la fonction de partition :
\begin{equation}
\label{FoncPartFer}
Z(t) = \<f| \exp\!\left(\!-\frac{2\pi}{t} H_{\!f}\!\right) |i\>
     = \<f| \tilde{q}^{L_0+\bar{L}_0-c/12} |i\>
\end{equation}
où $\tilde{q}=e^{-2\pi/t}$. $|i\>$ et $|f\>$ sont des \emph{états de bord},
c'est-à-dire des états de corde fermée caractérisant les D-branes émettant
et absorbant la corde fermée. Les conditions qu'ils vérifient s'obtiennent
à partir des conditions aux bords des cordes ouvertes qui se terminent
dessus en inversant le temps et l'espace. Considérons d'abord le cas d'une
condition aux bords quelconque. Une condition nécessaire de cohérence est
qu'elle respecte l'invariance conforme. Pour cela, nous l'avons vu, il est
nécessaire que $T=\bar{T}$ au bord ; il en découle que l'état de bord
doit vérifier $(T-\bar{T})|B\>=0$, ce qui, en termes des modes de
Fourier, s'écrit $(L_n-\bar{L}_{-n})|B\>=0$. En particulier, on a
$L_0|B\>=\bar{L}_0|B\>$, ce qui conduit à simplifier la formule
précédente en $Z(t)=\<f|(\tilde{q}^2)^{L_0-c/24}|i\>$. On verra dans le
chapitre \ref{WZW} que, dans certains cas, ceci s'écrit comme une
combinaison linéaire des caractères $\chi_R(\tilde{q}^2)$.

Dans le cas plus spécifique d'une D-brane dans un espace-temps plat où les
$X^i$ ont des conditions de Dirichlet et les $X^\a$ des conditions de
Neumann, l'état de bord vérifie, à $\t=0$,
\[
(X^i - Y^i)|B\> = 0 \text{ et } \d_\t X^\a |B\> = 0.
\]
Si on développe les $X^\mu$ en modes (équation \eqref{DevModesF}), ces
conditions s'écrivent
\begin{align*}
(\a^\a_n+\bar{\a}^\a_{-n})|B\> &= 0 & p^\a|B\> &= 0 \\
(\a^i_n-\bar{\a}^i_{-n})|B\> &= 0 & (x^i-Y^i)|B\> &= 0
\end{align*}
et on peut en tirer explicitement l'état de bord :
\[
|B\> = \prod_i\left[ \delta(x^i-Y^i) \smash{\prod_{n=1}^{\infty}} 
       \exp\!\left( \frac{1}{n} \a^i_{-n} \bar{\a}^i_{-n} \right) \right]
       \prod_\a \prod_{n=1}^{\infty}
       \exp\!\left(\! -\frac{1}{n} \a^\a_{-n} \bar{\a}^\a_{-n} \right)
       |0, p^\a=0\>
\]
à une normalisation près que l'on peut déterminer en imposant l'égalité de 
\eqref{FoncPartOuv} et \eqref{FoncPartFer}.

Un exemple simple d'utilisation des états de bords est le calcul de
l'amplitude d'émission d'une corde fermée par une D-brane. Par exemple,
pour un état de masse nulle, de polarisation $\zeta_{\mu\nu}$ et
d'impulsion $k$, l'amplitude s'écrit :
\[
A = \<B| \zeta_{\mu\nu} \a^\mu_{-1} \bar{\a}^\nu_{-1} |0,k\>
  \propto (\eta^{\a\b} \zeta_{\a\b} - \eta^{ij} \zeta_{ij}).
\]

\subsection{Action effective}
\index{action effective!D-brane|see{Born-Infeld}}
\index{Born-Infeld, action de}
Nous allons maintenant donner la forme de l'action effective à basse
énergie pour une D$p$-brane et expliquer, sans la démontrer à proprement
parler, pourquoi elle est raisonnable.

On paramètre la D-brane par $p+1$ coordonnées $\xi^a$. Les champs sur la
D-brane sont alors son plongement dans l'espace-temps $X^\mu(\xi^a)$ et le
champ de jauge $A_a$. L'action, appelée \emph{action de Born-Infeld},
s'écrit alors
\begin{equation}
\label{AcBI}
S = -T_p \int d^{p+1}\xi e^{-\Phi} \sqrt{-\det(\hat{G} + \hat{B} + 2\pi\a'F)}
\end{equation}
où $T_p$ est une constante proportionnelle à la tension de la D-brane, et
$\hat{G}$ et $\hat{B}$ sont les métrique et 2-forme induites sur la
D-brane : $\hat{G}_{ab}=\d_aX^\mu\d_bX^\nu G_{\mu\nu}$, et de même pour $B$.

Dans le cas où il n'y a ni dilaton ni 2-forme, le terme $\sqrt{-\det G}$
obtenu est naturel et n'est rien d'autre que Nambu-Goto généralisé en
dimension quelconque.

La dépendance en le dilaton découle du fait que cette action vient du
diagramme du disque, dont la caractéristique d'Euler est 1, d'où un
facteur $g^{-1}$ si le dilaton est constant, et donc $e^{-\Phi}$.

La dépendance en $F$ se comprend par T-dualité. Prenons une D-brane qui
s'étend dans les directions $X^1$ et $X^2$ d'un espace plat, avec un champ
magnétique constant $F_{12}$. On peut choisir une jauge $A_2=F_{12}X^1$. Si
on prend le T-dual de cela dans la direction $X^2$, on obtient une D-brane
penchée dans le plan $(X^1,X^2)$ : $X^2=-2\pi\a'X^1F_{12}$. On obtient
alors un facteur géométrique dans l'action :
\[
\int dX^1 \sqrt{1+(\d_1X^2)^2} = \int dX^1 \sqrt{1+(2\pi\a'F_{12})^2}.
\]
Dans le cas d'un champ quelconque, on peut, en mettant le champ sous une
forme diagonale par blocs, se ramener à un produit de tels facteurs, qui
revient au déterminant de l'action \eqref{AcBI}.

Enfin, la dépendance en $B$ découle du fait suivant : la dépendance en $B$
et $F$ dans l'action sur la surface d'univers de la corde s'écrit
\[
\frac{1}{2\pi\a'} \int_\Sigma \hat{B} + \int_{\d\Sigma} A.
\]
Il en découle les invariances de jauge suivantes :
\begin{align*}
\delta B &= 2\pi\a' d\Lambda \\
\delta A &= -\hat{\Lambda} + d\lambda
\end{align*}
où $\Lambda$ est une 1-forme et $\lambda$ un scalaire. Il en découle que
$F=dA$ n'est pas invariant de jauge, mais qu'en revanche $\hat{B}+2\pi\a'F$
l'est, et c'est donc cette combinaison qui doit apparaître dans l'action.

L'action \eqref{AcBI} montre que la tension (énergie par unité de volume)
de la D-brane est $T_pe^{-\Phi_0}$. Le fait qu'elle diverge à couplage
faible montre que la D-brane est un objet \emph{non perturbatif}, mais sur
lequel on sait néanmoins dire plein de choses par un calcul perturbatif
grâce à son interprétation en termes de lieu où les extrémités des cordes
sont confinées. Le coefficient $T_p$, quant à lui, peut être calculé en
considérant le couplage de la D-brane au graviton, ce qui s'obtient avec
l'état de bord évoqué précédemment.

Avec cette action, on peut également étudier les petites fluctuations de la
D-brane ; nous verrons dans les chapitres suivants des exemples explicites
de tels calculs, où nous montrerons que les modes de fluctuations ainsi
trouvés coïncident avec le calcul exact en CFT, lorsqu'on sait le faire.
\index{metrique@métrique!de corde ouverte}
Une propriété générale est que les fluctuations de la forme de la D-brane
et du champ de jauge sont couplées par la \emph{métrique de corde ouverte},
donnée par
\[
G_o^{-1} = [(\hat{G} + \hat{B} + 2\pi\a'F)^{-1}]_S
\]
où l'indice $S$ signifie qu'on symétrise la matrice. Cela donne
\begin{equation}
\label{MetOuv}
G_o = \hat{G} - (\hat{B}+2\pi\a'F)\hat{G}^{-1}(\hat{B}+2\pi\a'F).
\end{equation}
On verra sur les exemples des chapitres suivants que dans certains cas la
géométrie vue par les cordes ouvertes est très différente de celle vue par
les cordes fermées.

\section{Cordes non orientées}
\index{orientifold|(}
Jusqu'à présent, nous avons considéré des cordes \emph{orientées},
c'est-à-dire où on peut définir un sens particulier sur la corde (par
exemple le sens des $\s$ croissants). Sachant que l'opération
$\Omega:\s\rightarrow-\s$ ($\s\rightarrow\pi-\s$ dans le cas des cordes
ouvertes) est une symétrie de la théorie, on peut obtenir une thérie de
cordes non orientées en se restreignant aux états invariants par cette
transformation, ce qui revient à identifier une corde orientée dans un sens
avec la même corde orientée dans l'autre sens.

\subsection{Cordes fermées non orientées et orientifolds}
Considérons d'abord le cas des cordes fermées dans un espace plat, avec des
champs de fond triviaux. L'action de $\Omega$ sur les opérateurs $\a^\mu_n$
découle du développement en modes \eqref{DevModesF} :
$\Omega\a^\mu_n\Omega^{-1}=\bar{\a}^\mu_n$. L'action sur le tachyon, elle,
découle de la transformation de l'opérateur de vertex ; en l'occurrence,
$\Omega=+1$. Il en découle que, parmi les états tachyonique et non massifs,
le tachyon, le graviton et le dilaton sont présents dans la théorie non
orientée, tandis que la 2-forme en est exclue.

On peut s'intéresser à ce qui se passe si on compactifie la théorie sur un
cercle et qu'on en fait tendre le rayon vers zéro. Il est alors préférable
de décrire la situation en terme de la coordonnée T-duale $X'=X_L-X_R$,
considérée comme non compactifiée. Sachant que $\Omega$ agit en échangeant
$X_L(\s^+)$ et $X_R(\s^-)$, son action sur $X'$ s'écrit :
\[
\Omega : X'(\t,\s) \rightarrow -X'(\t,-\s).
\]
Autrement dit, on obtient le produit d'une réflexion selon la coordonnée
$X'$ et d'un changement d'orientation de la corde. La théorie obtenue en se
restreignant aux états tels que $\Omega=+1$ consiste à identifier une corde
avec son image par réflexion par rapport au plan $X'=0$, après un
changement d'orientation de la corde. Une telle théorie est appelée un
orientifold. On notera que, puisque une corde est identifiée à une corde
située ailleurs, la physique locale loin du plan $X'=0$ est celle de la
corde \emph{orientée}.

Des orientifolds plus généraux, pas forcément issus d'une T-dualité,
s'obtiennent en quotientant la théorie orientée par le produit de
l'inversion d'orientation $\Omega$ de la corde et d'une isométrie
involutive $h$. En présence de champs de fond non triviaux, des conditions
additionnelles sur $h$ devront être imposées pour que $\Omega h$ soit une
symétrie de la théorie initiale : par exemple, comme nous le verrons dans
le chapitre \ref{WZW}, s'il y a un champ $B$ non trivial, il est nécessaire
que $h$ inverse l'orientation de l'espace-temps. Signalons enfin que dans
la suite, le terme orientifold désignera aussi l'ensemble, ou une partie,
des points fixes de $h$ ; il ne devrait néanmoins pas y avoir de confusion.

Une différence essentielle entre un orientifold et une D-brane est que
l'orientifold n'est pas dynamique : il n'y a en effet pas de cordes
attachées à l'orientifold qui puissent représenter ses fluctuations. Notons
que l'argument comme quoi une D-brane doit fluctuer dans une géométrie
fluctuante, utilisé pour arguer du caractère dynamique des D-branes, ne
s'applique pas ici : en effet, l'identification de l'orientifold conduit à
des conditions au bord pour le champ gravitationnel (et les autres champs)
qui empêche toute fluctuation de géométrie de l'orientifold.
\index{action effective!orientifold}
Ceci n'empêche en revanche pas l'orientifold d'émettre des cordes fermées ;
il en découle une action effective très similaire à celles des D-branes :
\[
S = -T_{\O p} \int d^{p+1}\xi e^{-\Phi} \sqrt{-\det\hat{G}}.
\]
La dépendance en $\Phi$ découle du fait que cette action vient du diagramme
du plan projectif (nous verrons plus loin qu'il faut inclure des surfaces
non orientables pour les cordes non orientées), dont la caractéristique
d'Euler est 1, comme le disque. Les conditions aux bords sur l'orientifold
empêchent qu'il y ait un couplage à $B$, et l'orientifold n'interagit pas
avec les cordes ouvertes, donc il n'y a pas de champ de jauge.
Contrairement au cas des D-branes, la tension $T_{\O p}$ peut être négative.

\subsection{Fonction de partition des cordes fermées et bouteille de Klein}
\index{Klein, bouteille de|see{fonction de partition}}
\index{fonction de partition!bouteille de Klein}
Nous nous plaçons ici dans un cadre général où le spectre de la corde
fermée orientée se décompose en paires $(R,\bar{R})$ de représentations des
algèbres de Virasoro gauche et droite. L'existence d'une projection
orientifold entraîne que si $(R,\bar{R})$ est dans le spectre, alors
$(\bar{R},R)$ aussi. Dans le cas où $R\neq\bar{R}$, la projection garde une
combinaison linéaire de $(R,\bar{R})$ et $(\bar{R},R)$ ; si $(R,R)$ est
dans le spectre initial, soit il reste, soit il est éliminé par la
projection. La fonction de partition en découle :
\begin{equation}
\label{FoncPartFNor}
Z(\t) = \frac{1}{2}\left[
  \smash[b]{\sum_{R,\bar{R}}} n_{R,\bar{R}} \chi_R(q) \chi_{\bar{R}}^*(\bar{q})
  + \smash[b]{\sum_{R}} k_R \chi_R(q\bar{q}) \right]
\end{equation}
où la deuxième somme est sur les $R$ tels que $(R,R)$ est dans le spectre,
et $k_R$ est le nombre d'états $(R,R)$ préservés par la projection moins le
nombre d'états éliminés (c'est-à-dire que chaque état $(R,R)$ est compté
$1/2$ fois dans le premier terme, et $\pm1/2$ fois dans le second).

L'interprétation en termes d'amplitude sur des surfaces d'univers bien
choisies se fait en écrivant la fonction de partition de la façon
suivante :
\begin{equation}
\begin{split}
\label{FoncPartFNor2}
Z(\t) &= \Tr\left[ \frac{1+\Omega h}{2} 
                  q^{L_0-c/24}\bar{q}^{\bar{L}_0-c/24} \right] \\
      &= \frac{1}{2} \left[ 
           \Tr\bigl[q^{L_0-c/24}\bar{q}^{\bar{L}_0-c/24}\bigr]
           +  \Tr\bigl[\Omega q^{L_0-c/24}\bar{q}^{\bar{L}_0-c/24}\bigr]
         \right] \\
      &= \frac{1}{2}(\T+\K).
\end{split}
\end{equation}
Dans la première ligne, $(1+\Omega)/2$ projette sur les états vérifiant
$\Omega=+1$, c'est-à-dire précisément les états de la corde non orientée.
Les quantités $\T$ et $\K$ de la troisième ligne sont définies comme
respectivement le premier et le deuxième terme de la deuxième ligne ; $\T$
n'est rien d'autre que l'amplitude du tore. L'identification avec
l'équation \eqref{FoncPartFNor}, et le fait que son premier terme soit
$\T$, conduit à identifier $\K$ et le second terme de \eqref{FoncPartFNor},
ce qui permettra de calculer $\K$. C'est en revanche l'équation
\eqref{FoncPartFNor2} qui permet d'interpréter $\K$ en termes de surface
d'univers : en effet, $\K$ y apparaît clairement comme s'obtenant en prenant
une corde, en la faisant évoluer pendant le temps euclidien $2\pi t$ et en
recollant les bouts après une inversion d'orientation ; le résultat obtenu
est une surface non orientable de caractéristique d'Euler nulle (comme le
tore et le cylindre), baptisée \emph{bouteille de Klein}. Plus
généralement, les amplitudes de cordes non orientées font apparaître, en
plus des surfaces orientables, les surfaces non orientables.

\index{crosscap}
La bouteille de Klein peut être définie en considérant le plan $\t,\s$ avec
les identifications suivantes :
\[
(\t,\s) \sim (\t,\s+2\pi), \quad\text{et}\quad (\t,\s) \sim (\t+2\pi t,-\s).
\]
L'expression de $\K$ ci-dessus découle du choix de domaine fondamental
$-\pi\leq\s\leq\pi$ et \mbox{$0\leq\t\leq2\pi t$}, avec les identifications
$(\t,-\pi)\sim(\t,\pi)$ et $(0,\s)\sim(2\pi t,-\s)$. Un autre choix de
domaine fondamental est $0\leq\s\leq\pi$ et $0\leq\t\leq4\pi t$, avec les
identifications $(0,\s)\sim(4\pi t,\s)$, $(\t,0)\sim(\t+2\pi t,0)$ et
$(\t,\pi)\sim(\t+2\pi t,\pi)$, c'est-à-dire qu'on peut voir la bouteille de
Klein comme un cylindre, à ceci près que sur chaque bord on identifie les
points diamétralement opposés. Le résultat ainsi obtenu est que les
extrémités ne sont plus des bords, mais ce que l'on appelle des
\emph{crosscaps} ; de façon générale, toute surface non orientée peut se
construire à partir d'une surface orientée en y insérant des crosscaps.

\index{etatc@état de crosscap}
\index{crosscap!état de}
Après avoir échangé le temps et l'espace, et une dilatation appropriée, la
bouteille de Klein s'interprète comme la propagation d'une corde fermée
pendant le temps $\pi/2t$ entre deux \emph{états de crosscap}, qui sont aux
orientifolds ce que les états de bord sont aux D-branes. Comme pour les
états de bord, la préservation de l'invariance conforme impose
$(L_n-\bar{L}_{-n})|C\>=0$. Dans le cas plus spécifique des orientifolds
plats de l'espace plat, on peut les construire explicitement en imposant
$(X(\s+\pi)-X(\s))|C\>=0$ et $(\d_\t X(\s+\pi)+\d_\t X(\s))|C\>=0$ pour les
coordonnées parallèles à l'orientifold, et les mêmes équations avec un
signe opposé pour les coordonnées orthogonales à l'orientifold.

Il en découle que l'amplitude de la bouteille de Klein peut s'écrire
\begin{equation}
\label{Klein}
\begin{split}
\K &= \<C| (\tilde{q}^{1/4})^{L_0+\bar{L}_0-c/12} |C\> \\
   &= \<C| \sqrt{\tilde{q}}^{L_0-c/24} |C\>.
\end{split}
\end{equation}
On verra dans le chapitre \ref{WZW} que, dans certains cas, ceci s'écrit
comme une combinaison linéaire des caractères $\chi_R(\sqrt{\tilde{q}})$.

\subsection{Cordes ouvertes et D-branes en présence d'un orientifold}
\index{D-brane!avec orientifold|see{orientifold}}
\index{orientifold!avec cordes ouvertes|(}
L'isométrie $h$ qui intervient dans la définition de l'orientifold
identifie une D-brane à son image par $h$, et donc identifie un état de
corde allant de la D-brane $i$ à la D-brane $j$, noté $|ij\>$, à un état
$|j'i'\>$, où $i'$ et $j'$ sont les images des D-branes $i$ et $j$.
Notamment, s'il y a une D-brane à une position donnée, il doit y avoir une
D-brane image à la position symétrique par $h$. Par souci de simplicité,
nous nous limiterons aux cas où une D-brane est soit complètement séparée
de son image, soit géométriquement confondue avec l'orientifold. Nous nous
intéresserons ici principalement aux cordes de masse nulle vivant sur une
D-brane ou un groupe de D-branes superposées, c'est-à-dire notamment le
champ de jauge.

Dans le cas de D-branes séparées de l'orientifold, la physique locale est
essentiellement la même que sans orientifold. Notamment, pour $n$ D-branes
superposées (et leurs $n$ images), les excitations de masse nulle
contiennent un champ de jauge $\U(n)$.

Considérons donc maintenant le cas de $n$ D-branes superposées contenues
dans l'orientifold. L'action de $\Omega h$ sur les opérateurs $\a^\mu_m$
découle des développements en modes \eqref{ModesOuvN} et
\eqref{ModesOuvD} : $\Omega\a^\mu_m\Omega^{-1}=(-)^m\a^\mu_m$ dans les cas
que nous considérons ici. L'action la plus générale de l'orientifold est
alors
\[
\Omega h |\psi,ij\> = (-)^N \g_{ii'} \g_{jj'}^* |\psi,j'i'\>
\]
où $N$ est le nombre d'excitations de l'état $\psi$ (d'où un signe $-$ pour
les bosons de jauge), $\g$ est une matrice unitaire, et on sous-entend une
sommation sur $i'$ et $j'$. La condition $(\Omega h)^2=1$ entraîne que $\g$
doit être symétrique ou antisymétrique.

\index{groupe de jauge!orthogonal ou symplectique}
Dans le cas où $\g$ est symétrique, par un changement de base approprié
dans l'espace des D-branes, on se ramène à $\g=\I$. Les bosons de jauge
ainsi trouvés sont alors des combinaisons antisymétriques des états
$|ij\>$, donc le groupe de jauge est $\SO(n)$.

Dans le cas où $\g$ est antisymétrique, ce qui n'est possible que si $n$
est pair, on se ramène à $\g=i\begin{pmatrix}0&\I\\-\I&0\end{pmatrix}$, et
le groupe de jauge est alors $\USp(n)$.

\subsection{Fonction de partition des cordes ouvertes et ruban de Möbius}
\index{Mobius@Möbius, ruban de|see{fonction de partition}}
\index{fonction de partition!ruban de Möbius}
On note $I$ et $J$ deux ensembles de D-branes superposées, et on
s'intéresse à la fonction de partition des cordes ouvertes entre ces deux
ensembles de D-branes. Si l'image de $I$, notée $I'$, est distincte de $J$,
alors les cordes $|IJ\>$ et leurs images $|J'I'\>$ par l'orientifold sont
distinctes, et la projection en garde une combinaison linéaire, de sorte
que la fonction de partition est simplement la moitié de l'amplitude de
l'anneau. Si $J=I'$, alors il faut y ajouter un terme pour les cordes
$|II'\>$. Le fait que $\Omega\a^\mu_m\Omega^{-1}=(-)^m\a^\mu_m$ fait
apparaître la quantité $\Tr_R[(-)^N\sqrt{q}^{L_0-c/24}]$, où $R$ est une
représentation de l'algèbre conforme apparaissant dans le spectre des
cordes $|II'\>$. Sachant que $L_0=h_R+N$, où $h_R$ est le poids des états
conformes non excités de la représentation $R$, ceci peut s'écrire
\begin{equation}
\label{hatchi}
\hat{\chi}_R(q) \equiv e^{-i\pi(h_R-c/24)} \chi_R(-\sqrt{q})
\end{equation}
et la fonction de partition s'écrit alors
\[
Z(t) = \frac{1}{2}(\A+\M), \text{ où }
\M = n \sum_R m_R \hat{\chi}_R(q).
\]
où $m_R$ indique le nombre d'états $R$ préservés par la projection moins
le nombre d'états éliminés. Dans le cas $I=I'$, où on considère donc des
cordes entre un ensemble de D-branes et lui-même, l'une de ces
représentations contient les bosons de jauge ; pour obtenir le bon nombre
de bosons de jauge ($n(n-1)/2$ pour $\SO(n)$, et $n(n+1)/2$ pour
$\USp(n)$), $m_R$ pour cette représentation doit valoir $-1$ pour $\SO(n)$,
et $+1$ pour $\USp(n)$. Ceci nous amènera à trouver des conditions sur les
groupes de jauge des différentes D-branes pour assurer la cohérence de la
théorie.
\index{groupe de jauge!orthogonal ou symplectique}

De façon analogue au cas des cordes fermées, $\M$ s'interprète comme une
amplitude obtenue en prenant une corde ouverte, en la faisant évoluer
pendant le temps $\pi t$ et en recollant les bouts après une inversion
d'orientation ; le résultat obtenu est une surface non orientable, avec un
bord, de caractéristique d'Euler nulle, appelée \emph{ruban de Möbius}.

Le ruban de Möbius peut être défini en considérant la bande
$0\leq\s\leq\pi,\t$ avec l'identification suivante :
\[
\quad (\t,\s) \sim (\t+\pi t,\pi-\s).
\]
L'expression de $\M$ ci-dessus découle du choix de domaine fondamental
$0\leq\s\leq\pi$ et \mbox{$0\leq\t\leq\pi t$}, avec l'identification
$(0,\s)\sim(\pi t,\pi-\s)$. Un autre choix de domaine fondamental est
$0\leq\s\leq\pi/2$ et $0\leq\t\leq2\pi t$, avec les identifications
$(0,\s)\sim(2\pi t,\s)$ et $(\t,\pi/2)\sim(\t+\pi t,\pi/2)$, c'est-à-dire
qu'on peut voir le ruban de Möbius comme un cylindre où l'un des bords a
été remplacé par un crosscap. Après avoir échangé le temps et l'espace et
effectué une dilatation appropriée, le ruban de Möbius s'interprète comme
la propagation d'une corde fermée pendant le temps $\pi/2t$ entre un état
de bord et un état de crosscap, d'où découle que l'amplitude du ruban de
Möbius peut s'écrire :
\begin{equation}
\label{Mobius}
\M = \<C| \sqrt{\tilde{q}}^{L_0-c/24} |B\>.
\end{equation}
On verra dans le chapitre \ref{WZW} que, dans certains cas, ceci s'écrit
comme une combinaison linéaire des caractères $\chi_R(-\sqrt{\tilde{q}})$.
\index{orientifold!avec cordes ouvertes|)}
\index{orientifold|)}

\section{Supercordes}
\index{supercorde!type II|(}
La théorie des cordes bosoniques que nous avons vue jusqu'à présent a deux
défauts majeurs :
\begin{itemize}
\item Elle comporte un tachyon, signe qu'on développe autour d'un vide
  instable.
\item Il n'y a pas de fermions, ce qui est fâcheux si on prétend décrire
  notre monde avec.
\end{itemize}
Ces deux défauts sont résolus par la théorie des supercordes. Nous allons
décrire ici le formalisme de Ramond-Neveu-Schwarz (RNS). Dans la mesure où
les détails de la construction de la supercorde n'interviendront pas dans
le reste de cette thèse, nous serons plus brefs que pour la corde
bosonique.

\subsection{Supercordes de type II}
Le formalisme RNS consiste à avoir, en plus des champs scalaires $X^\mu$,
des spineurs de Majorana $\psi^\mu$, et d'ajouter à l'action de Polyakov
\eqref{AcPol} (en fixant la métrique à $\g_{\a\b}=\eta_{\a\b}$ un terme
\[
S_\psi = \frac{i}{\pi} \int d^2\s (\psi^\mu_- \d_+ \psi_{-\mu} 
                                 + \psi^\mu_+ \d_- \psi_{+\mu})
\]
où $\psi_-$ et $\psi_+$ sont les composantes de $\psi$. Cette action admet
une supersymétrie $\N=(1,1)$ de surface d'univers, qui relie d'une part la
partie gauche de $X$ et $\psi_+$, d'autre part la partie droite de $X$ et
$\psi_-$.

Les équations du mouvement correspondantes disent alors que $\psi_\pm$ ne
dépendent que de $\s^\pm$. L'annulation de l'anomalie de Weyl nécessite que
la dimension de l'espace-temps soit 10 ; nous nous placerons dans ce cas
dans la suite.

$\psi_+$ et $\psi_-$ étant des fermions, ils sont définis à un signe près.
Il en découle que dans le cas des cordes fermées, on a $\psi_+^\mu(\s+2\pi)
= \pm\psi_+^\mu(\s)$, et de même pour $\psi_-^\mu$. Le cas $+$ est appelé
condition de périodicité de Ramond, et donne lieu à un développement en
modes avec de modes entiers, tandis que le cas $-$ est la condition de
périodicité de Neveu-Schwarz, qui donne lieu à des modes demi-entiers.
Comme on peut choisir indépendamment le signe pour $\psi_+$ et $\psi_-$ ---
le signe doit être indépendant de $\mu$ pour assurer l'invariance de
Poincaré --- on aura quatre secteurs dans le spectre, nommés NS-NS, NS-R,
R-NS et R-R.

\index{condition aux bords!supercorde}
Dans le cas des cordes ouvertes, le terme de bord dans la variation de
$S_\psi$ s'écrit
\[
\frac{i}{2\pi} \int d\t (\psi_-\delta\psi_- - \psi_+\delta\psi_+).
\]
La stationnarité de l'action impose alors l'annulation des fermions sur le
bord, d'où découle leur annulation partout à cause des équations du
mouvement. Il faut donc ajouter à chaque bord un terme
\[
S_\psi^{\mathrm{bord}} = \pm \frac{i}{2\pi} \int d\t \psi_-\psi_+
\]
où le signe $\pm$ peut dépendre du bord, ce qui donne la condition au bord
$\psi_-=\pm\psi_+$, et permet en outre de préserver une des deux
supersymétries. Il est toujours possible de prendre un signe $+$ en $\s=0$,
en redéfinissant au besoin $\psi_-\rightarrow-\psi_-$. Il reste alors un
signe en $\s=\pi$ ; le signe $+$ impliquant des modes entiers, et le signe
$-$ des modes demi-entiers, on parlera de condition au bord de Ramond dans
le premier cas, et de Neveu-Schwarz dans le deuxième cas, obtenant ainsi
deux secteurs de cordes ouvertes.

\subsubsection{Spectres NS et R}
\index{spectre!supercorde fermée}
On considère d'abord le spectre engendré par un seul ensemble de modes NS
ou R, c'est-à-dire la corde ouverte ou un seul côté (gauche ou droit) de la
corde fermée.

Puisque dans le cas NS il n'y a pas de mode zéro, le vide des excitations
fermioniques est non dégénéré. La condition à imposer sur les états
physiques est qu'ils soient annulés par $(L_0-\frac{1}{2})$, donc l'état
correspondant est un tachyon et la première excitation, engendrée par les
opérateurs $\psi_{-1/2}^\mu$, est un vecteur de masse nulle. Plus
généralement les excitations NS sont des bosons.

Dans le cas R, les opérateurs $\psi_0^\mu$, qui transforment un état
fondamental en un autre état fondamental, forment une algèbre de Dirac, de
sorte que l'état fondamental est un spineur de Dirac, dont on montre qu'il
est de masse nulle. Plus généralement, les excitations R sont des fermions.

Il existe un opérateur, baptisé \emph{nombre fermionique de surface
d'univers}, qui anticommute avec les $\psi$, et a comme valeurs propres
$\pm1$. Il est conservé multiplicativement lors des interactions, ce qui
nous permettra d'envisager des théories dont le spectre NS ou R ne contient
que des états de nombre fermionique $+1$ ou $-1$. Parmi les états NS, le
tachyon est de nombre fermionique $-1$ et le vecteur de masse nulle de
nombre fermionique $+1$. Pour les états R de masse nulle, il est donné par
la chiralité de l'état.

\subsubsection{Cordes fermées}
Dans le cas de la corde fermée, les oscillateurs gauches et droits étant
découplés, les états s'obtiennent comme produit d'un état gauche et d'un
état droit, et on a un nombre fermionique pour chaque côté. On obtient donc
des secteurs du spectre où on combine un secteur gauche et un secteur droit
parmi NS$+$, NS$-$, R$+$ ou R$-$. Les secteurs (NS,NS) et (R,R) donnent des
bosons, tandis que les secteurs (NS,R) et (R,NS) donnent des fermions.

\index{projection GSO}
Il faut ensuite imposer l'invariance modulaire. Il en découle que seules
certaines combinaisons de secteurs sont possibles. Les deux qui nous
intéresseront sont :
\begin{itemize}
\item Type IIA : (NS$+$,NS$+$) (R$+$,R$-$) (NS$+$,R$-$) (R$+$,NS$+$),
\item Type IIB : (NS$+$,NS$+$) (R$+$,R$+$) (NS$+$,R$+$) (R$+$,NS$+$).
\end{itemize}
Une telle procédure, consistant à imposer des conditions indépendantes sur
les parties gauche et droite des états, s'appelle la \emph{projection GSO},
du nom de ses inventeurs (Gliozzi, Scherk et Olive). Ces deux spectres ne
comportent pas de tachyon. Les deux autres théories invariantes modulaires
contiennent un tachyon et pas de fermion, ce qui en limite l'intérêt.

Voyons maintenant le contenu de ces spectres à basse énergie (états de
masse nulle). (NS$+$,NS$+$), produit de deux vecteurs, est identique aux
états de masse nulle de la corde bosonique, c'est-à-dire un scalaire
(dilaton $\Phi$), une 2-forme ($B_{\mu\nu}$) et un graviton. Cela nous
permettra, dans la suite (chapitres 2 et 4), d'étudier la corde bosonique
en faisant abstraction des problèmes liés au tachyon : les résultats que
nous obtiendrons seront valides dès lors que nous ne considérerons que les
états de masse nulle, et nous n'aurons pas à introduire toute la complexité
liée aux supercordes.

\index{champ de fond!Ramond-Ramond}
Le secteur (R,R), produit de deux spineurs, se décompose en formes
différentielles : dans la théorie IIA, on obtient un vecteur $C_{(1)}$ et
une 3-forme $C_{(3)}$ ; dans la théorie IIB, on obtient un scalaire
$C_{(0)}$, une 2-forme $C_{(2)}$ et une 4-forme self-duale $C_{(4)}$
(c'est-à-dire que $dC_{(4)}=*dC_{(4)}$). Il sera utile, dans la suite, de
définir des formes de degré plus élevé, avec la propriété
$dC_{(8-n)}=*dC_{(n)}$. Pour plus de détails sur cette décomposition, on se
référera au début de \cite{BacBr}.

Les secteurs (NS,R) et (R,NS) contiennent chacun un gravitino (particle de
spin 3/2) et un spineur de Majorana-Weyl ; les gravitinos de ces deux
secteurs sont de même chiralité pour la corde IIB et de chiralités
différentes pour la corde IIA.

À ce stade, il paraîtrait naturel d'écrire une généralisation de l'action
\eqref{ModSigma}. Dans le cas de fonds NS-NS, cela est possible, et on
obtient l'action \eqref{ModSigma} plus des termes fermioniques que le
lecteur pourra trouver dans la littérature. Dans le cas de fonds R-R, il
n'y a pas de telle action, ce qui rend difficile leur étude, sauf dans
certains cas où des dualités permettent de se ramener à des fonds NS-NS.
Nous en verrons un exemple dans le chapitre~3.

\index{supersymetrie@supersymétrie d'espace-temps}
De même qu'un vecteur de masse nulle ne peut interagir de façon cohérente
que s'il existe une symétrie locale avec une quantité conservée scalaire,
un spin 3/2 de masse nulle ne peut interagir que s'il existe une symétrie
locale avec une quantité conservée spinorielle, autrement dit une
supersymétrie d'espace-temps. Puisqu'il y a deux gravitinos, c'est même
d'une supersymétrie étendue qu'il s'agit, avec $\N=(1,1)$ pour la théorie
IIA, et $\N=(2,0)$ pour la théorie IIB, soit 32 supercharges dans les deux
cas. On notera que les supersymétries de surface d'univers et
d'espace-temps sont deux choses distinctes, qui ne sont reliées que de
façon indirecte.

On peut trouver étrange que la supersymétrie d'espace-temps n'ait pas été
manifeste dès le début. Il existe une formulation manifestement
supersymétrique des supercordes, dite \emph{corde de Green-Schwarz}, où, au
lieu de fermions de surface d'univers, on a des fermions d'espace-temps. Le
problème de cette formulation est qu'elle est très difficile à quantifier
sans briser l'invariance de Lorentz, c'est pourquoi on lui préfère souvent
la formulation RNS. C'est ce que nous ferons dans la suite.

\index{supercorde!heterotique@hétérotique}
Par souci de complétude, signalons l'existence d'un autre type de
supercorde, la \emph{corde hétérotique}. Elle s'obtient en mettant, en lieu
et place des fermions gauches se transformant sous le groupe de Lorentz
d'espace-temps, des champs se transformant selon une symétrie interne,
c'est-à-dire qu'on ne garde que la supersymétrie de surface d'univers
droite. On obtient alors dans le spectre de masse nulle un dilaton, une
2-forme, un graviton, un gravitino et un vecteur de groupe de jauge
$\SO(32)$ ou $E_8\times E_8$. Dans la mesure où cela n'interviendra pas
dans cette thèse, nous n'en parlerons pas plus.

\subsubsection{T-dualité}
\index{T-dualite@T-dualité!supercorde}
Comme nous l'avons vu précédemment, la symétrie de T-dualité change la
composante droite de $X$ en son opposé. La supersymétrie de surface
d'univers entraîne alors qu'il faut aussi appliquer cette transformation à
$\psi_-$. Cela a pour effet de changer la chiralité du secteur R droit, ce
qui échange les théories IIA et IIB. Il en découle également des
transformations pour les champs R-R, qui consistent essentiellement à
enlever l'indice correspondant à la dimension compactifiée lorsqu'il y
était, et à l'ajouter lorsqu'il n'y était pas. Ainsi, dans le cas où la
dimension compactifiée est la 9, on a :
\[
C \leftrightarrow C_9, \quad C_\mu \leftrightarrow C_{9\mu}, \quad
\text{etc.}
\]

\subsection{Cordes ouvertes et D-branes}
Comme dans le cas de la corde bosonique, on peut ajouter des D-branes et y
accrocher des cordes ouvertes, avec des conditions aux bords de Neumann
parallèlement à la D-brane, et Dirichlet orthogonalement à la D-brane. La
supersymétrie de surface d'univers impose que le signe dans la condition au
bord $\psi_-=\pm\psi_+$ pour une coordonnée Dirichlet soit opposé à celui
pour une coordonnée Neumann. Ceci est cohérent avec le fait qu'une
T-dualité le long d'une coordonnée la fait passer de Neumann à Dirichlet,
ou inversement.

\index{spectre!supercorde ouverte}
Les D-branes à nombre pair de dimensions spatiales, pour la corde IIA, ou
impair, pour la corde IIB, préservent la moitié des supersymétries
d'espace-temps, soit 16 supercharges. Il en découle leur stabilité : en
effet, la projection GSO appliquée aux cordes ouvertes élimine alors le
tachyon, et leur spectre de masse nulle comporte un boson de jauge, des
fluctuations de position de la D-brane (secteur NS$+$, similaire à la corde
bosonique) et un spineur (secteur R$\pm$). (Il existe aussi des D-branes
stables ne préservant pas de supersymétrie ; nous n'en parlerons pas dans
cette thèse.)

\subsubsection{Action effective}
\index{Born-Infeld, action de!avec terme de Wess-Zumino}
\index{Wess-Zumino, terme de!action effective de D-brane|see{Born-Infeld}}
Comme dans le cas bosonique, la D-brane admet une action effective. Nous
nous limiterons ici à la partie concernant les champs bosoniques.

Le couplage aux champs NS-NS est le même que pour la corde bosonique, et
est donné par l'action de Born-Infeld \eqref{AcBI}.

On montre qu'une D$p$-brane ($p\leq8$) est, en outre, chargée sous la
$(p+1)$-forme R-R $C_{(p+1)}$, d'où découle un terme $\int C_{(p+1)}$ dans
l'action. (Le cas des D9-branes est spécial, puisqu'elles n'existent que
dans le cas de la théorie de type I, que nous évoquerons plus loin.) La
supersymétrie entraîne que la charge correspondante est égale au couplage
$T_p$ de la D-brane au graviton. D'autre part, des raisonnements de
T-dualité similaires à ceux que nous avons fait dans le cas bosonique
montrent l'existence de termes, dits \emph{termes de Wess-Zumino}, couplant
les champs R-R et le champ de jauge, que l'on peut résumer par
\[
\int \exp(\hat{B} + 2\pi\a'F) \wedge \sum_q C_{(q)}
\]
étant entendu que l'on extrait de ce produit les termes qui sont des
$(p+1)$-formes. Il en découle l'action effective de
Born-Infeld-Wess-Zumino :
\begin{equation}
\label{BIWZ}
S_{\mathrm{eff}} = T_p \left(
   - \int d^{p+1}\xi\, e^{-\Phi} \sqrt{-\det(\hat{G} + \hat{B} + 2\pi\a'F)}
   + \int \exp(\hat{B} + 2\pi\a'F) \wedge \sum_q C_{(q)}
   \right).
\end{equation}

Le couplage $T_p$ peut être calculé en comparant l'amplitude de l'échange
d'un graviton entre deux D-branes calculée en théorie des cordes et celle
calculée avec la théorie des champs effective, et on trouve alors
\[
T_p = \frac{\sqrt{\pi}}{\kappa_0} (2\pi\sqrt{\a'})^{3-p}.
\]
On remarque tout d'abord la relation $T_p=2\pi\sqrt{\a'}T_{p+1}$, qui peut
se trouver par T-dualité. D'autre part, on note une dépendance en
$\kappa_0$ qui, comme nous l'avons signalé précédemment, n'est pas une
quantité physique, les quantités physiques étant $\kappa_0 e^{\Phi_0}$
(couplage gravitationnel) et $\t_p=T_p e^{-\Phi_0}$ (tension de la
D$p$-brane).

Dans le cas de la théorie IIB, on a deux objets à une dimension spatiale,
la D1-brane, de tension $\t_1$, et la corde fondamentale, de tension
$\t_F=1/2\pi\a'$. Le rapport $\t_F/\t_1$ est proportionnel au couplage
$g=e^{\Phi_0}$ ; il apparaîtra dans la suite intéressant de faire en sorte
qu'il soit égal à $g$, ce qui impose de choisir $\kappa_0=8\pi^{7/2}\a'^2$.
Le fait de fixer $\kappa_0$ fixe aussi $\Phi_0$ pour un fond donné,
autrement dit cela fixe la normalisation additive du dilaton.

\index{NS5-brane}
Puisque, pour chaque champ R-R, il existe un objet chargé électriquement et
un chargé magnétiquement sous lui (un objet chargé électriquement sous
$C_{(p)}$ est chargé magnétiquement sous $C_{(8-p)}$), on peut se demander
s'il en est de même pour le champ NS-NS $B$. Il découle de l'action
\eqref{ModSigma} que l'objet chargé électriquement sous $B$ n'est autre que
la corde fondamentale ; l'objet chargé magnétiquement sous $B$ est une
\emph{NS5-brane}, dont les fluctuations n'ont pas de description
perturbative comme celles des D-branes (qui sont les cordes ouvertes qui y
sont accrochées). Comme les D-branes, la NS5-brane préserve la moitié des
supersymétries d'espace-temps.

\subsection{Actions effectives à basse énergie et S-dualité}
\index{action effective!supercorde fermée}
Comme pour la corde bosonique, les excitations de la supercorde admettent
une action effective à basse énergie. L'absence de modèle sigma pour une
corde en présence de champs R-R n'empêche pas d'écrire une action
effective. En effet, la supersymétrie d'espace-temps est très
contraignante, et fixe complètement l'action à d'éventuelles redéfinitions
des champs près ; l'action obtenue est celle de la supergravité IIA ou IIB,
selon le cas.

Pour la corde IIA, on obtient
\[
S_{\mathrm{eff}} = \frac{1}{2\kappa_0^2} \int d^Dx \sqrt{-\det G} \left[
  e^{-2\Phi} \left( R + 4\d_\mu\Phi\d^\mu\Phi - \frac{1}{12} H^2 \right)
  - \frac{1}{8} (dC_{(1)})^2 - \frac{1}{48} (dC_{(3)})^2
\right]
\]
plus des termes cubiques et quartiques en les formes différentielles.

Dans le cas de la corde IIB, la self-dualité de $C_{(4)}$ exclut
l'existence d'une action covariante. Néanmoins, les équations du mouvement
peuvent se déduire de l'action suivante
\[
\begin{split}
S_{\mathrm{eff}} &= \frac{1}{2\kappa_0^2} \int d^Dx \sqrt{-\det G} \left[
  e^{-2\Phi} \left( R + 4\d_\mu\Phi\d^\mu\Phi - \frac{1}{12} H^2 \right)
  \right. \\ &\quad \left.
  - \frac{1}{2} (dC_{(0)})^2 - \frac{1}{12} (dC_{(2)})^2
  - \frac{1}{480} (dC_{(4)})^2 \text{ + termes cubiques et quartiques}
\right]
\end{split}
\]
en imposant la contrainte $dC_{(4)}=*dC_{(4)}$ sur les solutions.

\index{S-dualite@S-dualité}
Cette action possède une symétrie $\SL(2,\R)$ par laquelle les champs se
transforment comme suit :
\begin{align*}
\t' &= \frac{a\t+b}{c\t+d}\ , \text{ avec } \t = C_{(0)} + i e^{-\Phi} \\
\begin{pmatrix} C'_{(2)} \\ B' \end{pmatrix} &=
\begin{pmatrix} a & b \\ c & d \end{pmatrix}
\begin{pmatrix} C_{(2)} \\ B \end{pmatrix}
\end{align*}
où $ad-bc=1$, et $C_{(4)}$ et la métrique d'Einstein $G_E=e^{-\Phi/2}G$
restent inchangés.

On peut se demander si la théorie des cordes préserve tout ou partie de
cette symétrie de la supergravité. Puisque ces transformations mélangent
les 2-formes NS-NS et R-R, cela conduirait, dans le cas où $a$, $b$, $c$ et
$d$ sont quelconques, à transformer la corde fondamentale en un état de
charges quelconques sous les deux 2-formes, ce qui violerait la condition
de quantification de Dirac, sauf dans le cas où $a$, $b$, $c$ et $d$ sont
entiers. On est donc amené à conjecturer l'existence d'une symétrie
$\SL(2,\Z)$, nommée \emph{S-dualité} pour la théorie IIB. Le générateur
transformant $\t$ en $-1/\t$ échange les cordes fondamentales et les
D1-branes, les NS5- et D5-branes et laisse les D3-branes invariantes. Dans
ce dernier cas, il découle une symétrie $\SL(2,\Z)$ de la théorie des
champs effective sur la D3-brane, dont le lecteur pourra trouver
l'expression et la démonstration en \citemoi{C}. Signalons que, dans le cas
ou $C_{(0)}=0$, la transformation $\t\rightarrow-1/\t$ change $\Phi$ en
$-\Phi$, et donc le couplage en son inverse ; c'est donc une dualité non
perturbative.

\subsection{Orientifolds et cordes de type I}
\index{orientifold!supercordes}
Comme dans le cas des cordes bosoniques, il existe en théorie des
supercordes des orientifolds, obtenus en quotientant la théorie par le
produit de la parité de surface d'univers $\Omega$ et d'une isométrie
involutive de l'espace-temps.

Dans le cas des cordes IIB, $\Omega$ est une symétrie de la théorie ; les
isométries d'espace-temps changeant l'orientation ne le sont pas parce que
le spectre est chiral. Il en découle la possibilité d'orientifolds de
dimension spatiale impaire. La moitié des supersymétries d'espace-temps est
préservée.

\index{supercorde!type I}
Si l'isométrie intervenant dans le quotient est l'identité, on a affaire à
un orientifold $\O9$, c'est-à-dire à des cordes non orientées. Après
projection, il reste dans le spectre de masse nulle le graviton et le
dilaton, comme pour les cordes bosoniques, ainsi que la 2-forme R-R
$C_{(2)}$. La théorie ainsi obtenue souffre d'anomalies dont la
compensation requiert l'introduction de 32 D9-branes, donc de cordes
ouvertes, avec un groupe de jauge $\SO(32)$ ; le résultat obtenu est appelé
\emph{théorie des cordes de type~I}. Réciproquement, l'existence de
D9-branes introduit une anomalie dont la compensation requiert la présence
d'un $\O9$, et donc n'est possible que dans le cadre de la théorie de
type~I. Il peut y exister des D1- et D5-branes, chargées respectivement
électriquement et magnétiquement sous $C_{(2)}$. Comme les cordes de type~I
n'interviendront pas dans la suite de cette thèse, nous n'en parlerons pas
plus.

Dans le cas des cordes IIA, $\Omega$ n'est pas une symétrie de la théorie,
car elle est chirale sur la surface d'univers. En revanche, le produit de
$\Omega$ par une isométrie d'espace-temps changeant l'orientation l'est, et
on obtient alors un orientifold de dimension spatiale paire.
\index{supercorde!type II|)}

%% file: WZW.tex
\chapter{Cordes, D-branes et orientifolds dans un groupe compact : 
l'exemple de $S^3$ et $\pmb{\RP}^3$}
\label{WZW}
Dans le chapitre précédent, nous nous sommes intéressé principalement aux
cordes dans un espace-temps plat. On voudrait, bien évidemment, envisager
le cas d'espaces plus compliqués, mais les calculs y sont le plus souvent
très difficiles, voire impossibles. Il y a néanmoins un cas où l'on sait
calculer beaucoup de choses : c'est celui où l'espace-temps est un groupe
de Lie, ou un produit de groupes de Lie, chacun d'eux étant muni d'une
métrique invariante par l'action du groupe, c'est-à-dire, dans le cas d'un
groupe dont l'algèbre de Lie est simple, sa forme de Killing. Il découle en
effet de la structure algébrique supplémentaire qu'est la structure de
groupe une extension de la symétrie conforme permettant, au moins dans le
cas compact, l'obtention du spectre exact des cordes fermées et ouvertes.

Nous nous limiterons ici essentiellement aux cas non triviaux les plus
simples que sont la 3-sphère ($\SU(2)$), qui est simplement connexe, et
$\RP^3$ ($\SO(3)$), qui ne l'est pas, et pour lequel nous mettrons l'accent
sur les différences avec son recouvrement universel $S^3$.
\index{NS5-brane}
Outre le fait d'être un exemple simple de groupe, $S^3$ apparaît dans la
limite proche de l'horizon de plusieurs fonds intéressants, notamment celle
d'une superposition de NS5-branes \cite{CHS}, qui s'écrit
$S^3\times\R_{\Phi}\times\R^{1,5}$, où $\R_{\Phi}$ est une dimension en
laquelle le champ de dilaton est linéaire. Un autre exemple est l'état lié
formé par des NS5-branes et des cordes fondamentales, qui est $AdS_3\times
S^3\times T^4$, qui sera étudié au chapitre suivant.

\section{Cordes fermées orientées : le modèle de Wess-Zu\-mi\-no-Witten}
\index{Wess-Zumino-Witten, modèle de|(}
\index{Wess-Zumino, terme de!dans WZW|see{Wess-Zumino-Witten}}
\index{champ de fond!$B_{\mu\nu}$}
On s'intéresse ici aux cordes fermées dans un groupe de Lie compact
$G$ muni d'une métrique proportionnelle à sa forme de Killing. Par
exemple, $\SU(2)$ est la 3-sphère munie de sa métrique usuelle. Si on ne
met pas d'autre champ de fond que la métrique, le modèle sigma obtenu
n'est pas invariant conforme. On peut le rendre invariant conforme
par l'ajout d'un terme approprié, et on obtient l'action \cite{Wit,GW} :
\begin{equation}
\label{AcWZW}
S = -\frac{k}{16\pi} \int_\Sigma d\t d\s \Tr'(\d_m g\d^m g^{-1})
    + \frac{k}{24\pi} \int_M \Tr'((g^{-1}dg)\wedge(g^{-1}dg)\wedge(g^{-1}dg))
\end{equation}
où $\Tr'$ est une trace normalisée de sorte que la longueur carrée des
racines longues de l'algèbre soit 2 (pour $\SU(2)$, $\Tr'=2\Tr$ dans la
représentation fondamentale), et $M$ est une variété à trois dimensions de
bord $\Sigma$, ce qui revient à avoir un champ de fond $B_{\mu\nu}$ non
nul, le deuxième terme (appelé terme de Wess-Zumino) étant
$\int_M H = \int_\Sigma B$. Comme $k$ apparaît en facteur devant la
métrique, il est proportionnel au carré du « rayon » du groupe ; dans le
cas de la 3-sphère, le carré de son rayon vaut $k\a'$.

Cette action est ambiguë à cause du choix de M ; ce qui est important
est que $e^{iS}$ soit bien défini, ce qui est vrai si $k$, que l'on
appelle le niveau du modèle WZW, est entier.

Elle admet une symétrie bien plus large que l'invariance conforme,
puisque elle est invariante par la transformation suivante :
\begin{equation}
\label{SymWZW}
g(\t,\s) \rightarrow \Omega(\s^+) g(\t,\s) \bar\Omega^{-1}(\s^-)
\end{equation}
où $\Omega$ et $\bar\Omega$ sont deux fonctions indépendantes à
valeurs dans $G$. Il en découle les courants conservés à valeur dans
l'algèbre de Lie de $G$ (notée $\Lg$)
\begin{equation}
\label{CourKM}
J(\s^+) = kg\d_+g^{-1} \quad\text{et}\quad \bar J(\s^-) = kg^{-1}\d_-g.
\end{equation}
Leurs modes de Fourier, donnés par
\[
J(\s^+) = \sum_n J_n e^{-2in\s^+},
\]
et de même pour $\bar J$, vérifient les relations de commutation
\[
[J^a_n, J^b_m] = f^{abc} J^c_{n+m} + kn\delta^{ab}\delta_{n+m,0}
\]
où $J=J^at^a$, les $t^a$ sont des générateurs orthonormés (au sens de
la forme de Killing) et les $f^{abc}$ sont les constantes de structure
de $\Lg$. L'algèbre de Lie qu'ils engendrent, notée $\hat{\Lg}$, est
l'algèbre de Kac-Moody \index{Kac-Moody (algèbre de)} associée à
$\Lg$. Les $J_0$ forment une sous-algèbre isomorphe à $\Lg$.

Ces relations de commutation sont assez similaires à celles des
$\a^\mu_n$ dans le cas plat. Il en découle que les états sont obtenus en
agissant sur un état fondamental avec les opérateurs $J_{-n}$.
L'ensemble des états fondamentaux étant invariant par les $J_0$, ils
forment une représentation de $\Lg$, et à chaque représentation de $\Lg$
est associée une représentation de l'algèbre de Kac-Moody.

\index{Sugawara, tenseur de}
\index{tenseur énergie-impulsion!WZW}
Le tenseur énergie-impulsion associé vaut classiquement
$(1/2k)\sum_{a}J^aJ^a$. Il est toutefois renormalisé par les effets
quantiques. En imposant que les relations de commutation vérifiées par
$T$ soient correctes, on obtient la forme correcte, qu'on appelle
\emph{tenseur de Sugawara} :
\[
T = \frac{1}{2(k+h(\Lg))} \sum_{a}J^aJ^a
\]
où $h(\Lg)$ est le nombre dual de Coxeter de $\Lg$. Notamment,
$h(\mathfrak{su}(2))=2$. On montre que la charge centrale vaut alors
\[
c = \frac{k\dim(G)}{k+h(\Lg)}
\]
et que les relations de commutation entre $T$ et $J$ sont
\[
[L_n, J^a_m] = -mJ^a_{m+n},
\]
ce qui, là encore, constitue une similitude entre les $J^a$ et les
$\a^i$ du cas plat.

\index{spectre!corde bosonique fermée!WZW}
Pour chaque paire de représentations $(R,\bar{R})$ de $\Lg$, on peut
alors déduire le spectre associé. Bien sûr, considérer $G$ seul comme
espace-temps n'aurait guère de sens, et on considérera donc le produit
de $G$ par un espace de Minkowski. Il découle alors de l'annulation de
$L_0+\bar{L}_0-2$ sur les états physiques que la masse des états
(définie à partir de l'énergie-impulsion de l'espace de Minkowski) est
donnée par :
\begin{equation}
\label{SpFerWZW}
m^2 = \frac{2}{\a'} \left(
   -2 + \frac{c_R+c_{\bar{R}}}{k+h(\Lg)} +N + \bar{N} + M + \bar{M} \right)
\end{equation}
où $c_R$ est le Casimir quadratique de la représentation $R$ ($j(j+1)$
dans le cas de la représentation de spin $j$ de $\mathfrak{su}(2)$),
$N$ et $\bar{N}$ sont les nombres d'excitations gauches et droites sur
le groupe (étant entendu que $J_{-n}$ crée n excitations), et $M$ et
$\bar{M}$ sont les excitations de l'espace de Minkowski. L'annulation
de $L_0-\bar{L}_0$ impose la contrainte
\[
M + N + \frac{c_R}{k+h(\Lg)} = 
   \bar{M} + \bar{N} + \frac{c_{\bar{R}}}{k+h(\Lg)}.
\]

Il se pose alors la question de savoir quelles représentations sont
présentes dans la théorie. Un résultat remarquable est que, pour
toutes sauf un nombre fini d'entre elles, toute fonction de
corrélation contenant un opérateur se transformant sous une telle
représentation s'annule. Les représentations restantes sont dites
intégrables. Dans le cas de $\SU(2)$, ce sont les représentations de spin
inférieur ou égal à $k/2$.

Dans le cas d'un groupe simplement connexe, il y a correspondance
entre les états de cordes fermées et les opérateurs de vertex, qui
sont des fonctions sur le groupe. Sachant que l'ensemble des fonctions
sur $G$ se décompose en représentations de $G\times G$ par
\[
\mathrm{Fon}(G) = \bigoplus_{R} R \otimes R^*
\]
où $R^*$ est la représentation conjuguée de $R$ (dans le cas de $\SU(2)$,
on a toujours $R^*=R$). On s'attend alors à ce que le spectre des
cordes sur $G$ soit cet ensemble restreint aux représentations
intégrables.

\index{fonction de partition!tore}
\index{invariance modulaire}
Il faut néanmoins s'assurer de l'invariance modulaire d'un tel
spectre. Comme nous l'avons vu précédemment, la fonction de partition
s'écrit
\begin{align*}
Z(\t) &= \Tr(q^{L_0-c/24}\bar{q}^{\bar{L}_0-c/24}) \\
      &= \sum_{R,\bar{R}} n_{R,\bar{R}} \chi_R(q) \chi_{\bar{R}}^*(\bar{q})
\end{align*}
avec $q=e^{2\pi i\t}$, où $\t$ est le paramètre modulaire du tore,
$n_{R,\bar{R}}$ est la multiplicité de la représentation $(R,\bar{R})$ dans
le spectre, et
\[
\chi_R(q) = \Tr_R(q^{L_0-c/24}).
\]

\index{transformation modulaire}
\index{caractere@caractère de Kac-Moody}
Ces quantités, qu'on appelle \emph{caractères de Kac-Moody}, ont les
transformations modulaires suivantes :
\begin{align}
\chi_R(\t+1) &= \exp\!\left( 2\pi i\left( \frac{c_R}{k+h(\Lg)}
                -\frac{c}{24} \right)\!\right) \chi_R(\t) \\
\label{TransMod}
\chi_R\left(-\frac{1}{\t}\right) &= \sum_{R'} S_{RR'} \chi_{R'}(\t)
\end{align}
où $S$ est une matrice unitaire, qui dans le cas de $\SU(2)$ vaut
\begin{equation}
\label{MatS}
S_{jj'} = \sqrt{\frac{2}{k+2}} \sin\!\parfrac{\pi(2j+1)(2j'+1)}{k+2}.
\end{equation}
L'unitarité de $S$ entraîne que la fonction de partition
\[
Z(\t) = \sum_R \chi_R \chi^*_{R^*}
\]
dite diagonale, et qui n'est autre que celle découlant du spectre que
nous avons conjecturé dans le cas d'un groupe simplement connexe, est
invariante modulaire. Nous en concluons que ce spectre est le bon.

Il reste maintenant à trouver le spectre dans le cas où $G$ n'est pas
simplement connexe. On a alors $G=\tilde{G}/H$, où $\tilde{G}$ est le
recouvrement universel de $G$, et $H$ un sous-groupe discret de
$\tilde{G}$. Il est aisé de voir que les états de cordes homotopes à
un point (dits non twistés) sont ceux de $\tilde{G}$ invariants par
$H$, c'est-à-dire ceux associés aux représentations de $\tilde{G}$ où
les éléments de $H$ sont représentés par l'identité. La fonction de
partition associée à ce seul spectre n'étant pas invariante modulaire,
il faut y ajouter des états de cordes non homotopes à un point (dits
états twistés). Le spectre complet peut être déterminé complètement en
imposant l'invariance modulaire.

\index{RP@$\RP^3$}
\index{invariance modulaire}
\index{etatt@état twisté}
Le spectre des cordes fermées dans $\SO(3)=\SU(2)/\Z_2$ a été dérivé
explicitement dans \cite{GW}. Tout d'abord, il est nécessaire que $k$ soit
pair. Ceci est dû au fait que, le groupe étant de dimension 3, la variété
$M$ dans le terme de Wess-Zumino de l'action \eqref{AcWZW} est
nécessairement le groupe entier ; $\SO(3)$ étant deux fois plus petit que
$\SU(2)$, il s'ensuit que $k$ doit être pair pour que ce terme soit
correctement quantifié. En ce qui concerne le spectre, on trouve que :
\begin{itemize}
\item Les états non twistés contiennent les représentations de la
  forme $(j,j)$, avec $j$ entier inférieur à $k/2$.

\item Les états twistés contiennent les représentations de la forme
  $(j,k/2-j)$ où $j$ est entier si $k$ est multiple de 4, et
  demi-entier sinon.
\end{itemize}
\index{Wess-Zumino-Witten, modèle de|)}

\section{D-branes symétriques sur les groupes compacts}
\subsection{Description des D-branes}
\index{D-brane!sur un groupe|(}
\index{etatC@état de Cardy}
Les D-branes sur les groupes ont été initialement construites comme
des états de bord, dits états de Cardy, dans le cadre des théories
conformes rationnelles (dont fait partie le modèle de
Wess-Zumino-Witten) pour lesquelles la fonction de partition est
diagonale \cite{Cardy}.

\index{condition aux bords!WZW}
La condition au bord vérifiée par les courants $J$ et $\bar{J}$ s'écrit
$J=-\bar{J}$, d'où découle que les états de bord associés à des D-branes
symétriques, c'est-à-dire préservant l'algèbre de Kac-Moody doivent
vérifier $(J+\bar{J})|B\>=0$, ce qui, en termes de modes de Fourier,
s'écrit $(J_n+\bar{J}_{-n})|B\>=0$. On montre que pour chaque
représentation $j$ de l'algèbre de Kac-Moody il existe un unique état ne
contenant que la représentation $j$, qu'on appelle \emph{état d'Ishibashi}%
\cite{Ishi} et que nous noterons $|j\>$.

\index{fonction de partition!anneau}
\index{couplage!entre D-brane et cordes fermées}
Dans le cas où la fonction de partition est l'invariant diagonal, les
$|j\>$ engendrent l'ensemble des états présents dans le spectre, et donc la
fonction de partition du cylindre dans le secteur des cordes fermées
(équation \eqref{FoncPartFer}) s'écrit
\[
Z(t) = \sum_j \<f|j\> \<j|i\>
              \<j|\Tr\bigl[(\tilde{q}^2)^{L_0-c/24}\bigr]|j\>.
\]
La quantité $\<f|j\>$ s'interprète comme le couplage de l'état de bord
$|f\>$ avec les cordes fermées dans la représentation $j$, et nous la
noterons $D^j_{\!f}$ dans la suite. La quantité
$\<j|\Tr\bigl[(\tilde{q}^2)^{L_0-c/24}\bigr]|j\>$, quant à elle, n'est
autre que le caractère $\chi_j(\tilde{q}^2)$, d'où l'expression :
\[
Z(t) = \sum_j D^j_{\!f} D^j_i \chi_j(\tilde{q}^2).
\]

\index{dualite@dualité ouvert-fermé}
Elle doit d'autre part être égale à la fonction de partition dans le canal
ouvert (équation \eqref{FoncPartOuv}), qui s'écrit
\[
Z(t) = \sum_j n^j_{if} \chi_j(\sqrt{q})
\]
où $n^j_{if}$ est le nombre de fois où la représentation $j$ apparaît dans
le spectre des cordes entre les D-branes $i$ et $f$. Le fait que les
caractères apparaissant dans les équations précédentes sont reliés par la
relation \eqref{TransMod}, et que les $n^j_{if}$ doivent être des entiers
positifs, est une contrainte forte sur les états de bord possibles.

\index{spectre!corde bosonique ouverte!WZW}
Ces états ont été construits par Cardy \cite{Cardy}, qui a montré qu'on
peut associer à chaque représentation $r$ de l'algèbre de Kac-Moody un état
$|\tilde{r}\>$ (qu'on appellera « état de spin $r$ » dans le cas de
$\SU(2)$), et que $n^j_{\tilde{r}\tilde{s}}$ est le nombre d'apparition de
la représentation $j$ dans le produit des représentations $r^*$ et $s$
(coefficient de fusion).

Ainsi, étant données deux D-branes de spins $r$ et $s$, on connaît le
spectre des cordes qui sont entre elles, donc leur fonction de partition
dans le canal ouvert. Par une transformation modulaire, on obtient le canal
fermé, et donc les $D^j_r$. Nous utiliserons abondamment cela dans la suite
de ce chapitre.

Pour une représentation donnée de l'algèbre de Kac-Moody, la masse des
états s'obtient de façon similaire aux cordes fermées, à ceci près
qu'il n'y a qu'une algèbre de Virasoro ; avec les mêmes notations que
dans l'équation \eqref{SpFerWZW}, on trouve
\begin{equation}
\label{SpOuvWZW}
m^2 = \frac{1}{\a'} \left( -1 + \frac{c_R}{k+h(\Lg)} + N + M \right).
\end{equation}

Dans le cas de $\SU(2)$, la règle de fusion pour les représentations
intégrables de $\widehat{\mathfrak{su}(2)}$ est
\[
[j_1] \otimes [j_2] = [|j_1-j_2|] \oplus [|j_1-j_2|+1] \oplus \ldots
                      \oplus [\min(j_1+j_2,k-j_1-j_2)],
\]
ce qui donne le spectre.

\begin{figure}
\begin{center}
\includegraphics[scale=0.7]{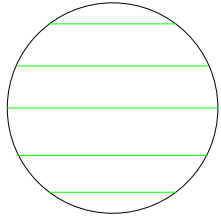}
\end{center}
\caption{Position des D-branes sur la 3-sphère (cas $k=4$).}
\label{DbranesS3}
\end{figure}

Leur géométrie a été étudiée plus tard \cite{AS,FFFS}. On montre que
les D-branes symétriques sont localisées sur des classes de
conjugaison du groupe. Dans le cas de $\SU(2)$, ce sont des 2-sphères
données, dans les coordonnées polaires usuelles, par
\begin{equation}
\label{PosDbrane}
\psi = \frac{(2j+1)\pi}{k+2}
\end{equation}
(fig. \ref{DbranesS3}).

\index{RP@$\RP^3$}
Dans le cas d'un groupe non simplement connexe, puisque la fonction de
partition du tore n'est pas un invariant modulaire diagonal, on ne
peut pas appliquer directement la théorie de Cardy. En revanche,
puisqu'un groupe non simplement connexe s'obtient par identification
de points de son recouvrement universel par un sous-groupe discret, on
pourra trouver ses D-branes par identification des D-branes du
recouvrement universel.

\index{RP@$\RP^3$}
Dans le cas de $\RP^3$, qui est traité dans \citemoi{B}, il s'agit donc
de considérer les D-branes de la 3-sphère et d'identifier celles qui
sont diamétralement opposées. Sachant que l'identification $\Z_2$ agit
par
\begin{equation}
\label{Z2}
\psi \rightarrow \pi-\psi, \quad \theta \rightarrow \pi - \theta,
\quad \phi \rightarrow \phi + \pi,
\end{equation}
il en découle que les états de Cardy d'isospins $j$ et $k/2-j$ sont
identifiés.

Pour $j<k/4$, cela identifie deux D-branes différentes, et l'on
obtient donc une D-brane sphérique, que nous baptiserons « brane de
type $j$ ».

Dans le cas $j=k/4$, c'est-à-dire sur l'« équateur », qui est un plan
projectif, on peut envisager deux types d'identification :
\begin{itemize}
\item si on identifie un point d'une D-brane équatoriale de $S^3$ avec
le point opposé sur une \emph{autre} D-brane, on obtient une D-brane
sphérique enroulée sur l'équateur, avec deux points de la D-brane sur
chaque point de l'équateur.

\item si on identifie un point d'une D-brane équatoriale de $S^3$ avec
le point opposé sur la \emph{même} D-brane, la D-brane obtenue est un
plan projectif.

\end{itemize}

Pour dériver le spectre et le groupe de jauge des D-branes
équatoriales, nous utiliserons la description suivante : on a $N$
D-branes sphériques équatoriales sur $S^3$, et l'identification $\Z_2$
agit comme une permutation involutive. Sachant qu'un état de bord peut
être une combinaison linéaire quelconque de D-branes, l'espace des
états de bord est un espace vectoriel de dimension $N$ et l'action de
$\Z_2$ sur lui est une matrice unitaire $Z$. Puisque $Z^2=1$, $Z$ est
diagonalisable avec $n_+$ valeurs propres $+1$ et $n_-$ valeurs
propres $-1$, avec $n_++n_-=N$, et on peut donc décrire les D-branes
comme $n_+$ D-branes $\RP^2$ de signe $+$, et $n_-$ D-branes $\RP^2$
de signe $-$. Nous verrons plus loin la signification de ces signes.

Dans cette description, une D-brane équatoriale sphérique correspond à
$Z=\left(\begin{smallmatrix}0&1\\1&0\end{smallmatrix}\right)$, qui a
une valeur propre $+1$ et une valeur propre $-1$, ce qui conduit à la
considérer comme la combinaison de deux D-branes $\RP^2$ de signes
opposés.

\subsection{Couplage des D-branes aux cordes fermées}
Nous nous limiterons ici à $\SU(2)$ et $\SO(3)$.

\subsubsection{Cas de SU(2)}
\index{fonction de partition!anneau!$S^3$}
La fonction de partition des cordes ouvertes entre deux D-branes de
types $r$ et $s$ est donnée par le diagramme de l'anneau :
\[
\A_{rs} = \Tr(\sqrt{q}^{L_0-c/24})
        = \sum_{j=0}^{k/2} N_{rs}^j \chi_j(\sqrt{q}).
\]
où $q=e^{-2\pi t}$.

\index{couplage!entre D-brane et cordes fermées}
Comme nous l'avons vu dans le cas plat, cette amplitude peut
s'interpréter dans le canal fermé comme l'échange d'une corde fermée
entre les deux D-branes. Pour cela, il s'agit d'exprimer l'amplitude
en termes de $\tilde{q}=e^{-2\pi/t}$ à l'aide des équations
\eqref{TransMod} et \eqref{MatS}. On trouve alors, comme on s'y attend,
\[
\A_{rs} = \sum_{j=0}^{k/2} D_r^j D_s^j \chi_j(\tilde{q}^2)
\]
où $D^j_s$ vaut
\begin{equation}
\label{CouplageSU2}
D^j_s = \sin\!\parfrac{(2j+1)(2s+1)\pi}{k+2}\sqrt{\N_j}
\end{equation}
avec
\begin{equation}
\label{norm}
(\N_j)^{-1} = \sqrt{\frac{k+2}{2}} \sin\!\parfrac{(2j+1)\pi}{k+2}.
\end{equation}

\index{groupe de jauge!unitaire}
Le groupe de jauge est, bien sûr, $\prod_{j}\U(n_j)$.

\index{RP@$\RP^3$}
\index{fonction de partition!anneau!$\RP^3$}
\subsubsection{Cas de SO(3)}
Les cordes entre deux D-branes de types $r$ et $s$ correspondent sur
la 3-sphère à des cordes entre les D-branes de types $r$ et $s$ d'une
part, et $k/2-r$ et $s$ d'autre part (les deux autres possibilités
sont identifiées à celles-ci). L'amplitude de l'anneau correspondante
est donc :
\[
\A_{rs} = \sum_{l=0}^{k/2}(N_{rs}^l + N_{k/2-r,s}^l) \chi_l(\sqrt{q}).
\]

Pour obtenir les couplages des D-branes aux cordes fermées, on
réécrit, comme précédemment, ces amplitudes en termes de la variable
$\tilde{q}$. On trouve alors que pour $j$ demi-entier, $D^j_s=0$, ce
qui est normal puisque il n'y a pas d'état non twisté dans la
représentation $(j,j)$ pour $j$ demi-entier, et pour $j$ entier
$D^j_s(\SO(3))=\sqrt{2}D^j_s(\SU(2))$.

Il convient de noter que, pour des raisons topologiques, les D-branes
sphériques ne peuvent pas être couplées aux cordes fermées twistées,
donc il s'agit là de couplages aux cordes non twistées uniquement. En
outre, comme les D-branes équatoriales sphériques sont très similaires
aux autres, on peut les considérer ici comme des D-branes de type $k/4$,
donc leurs couplages sont donnés par $D^j_{k/4}$.

Pour une corde ouverte entre une D-brane $\RP^2$ et une de type $s$,
l'anneau s'écrit
\[
\A_{rs} = \sum_{l=0}^{k/2} N_{k/4,s}^l \chi_l(\sqrt{q})
\]
où $R$ désigne la D-brane $\RP^2$. Ceci est indépendant du signe de la
D-brane $\RP^2$. On trouve alors
\begin{equation}
\label{DjR}
D^j_R = \frac{1}{2} D^j_{k/4}
\end{equation}
ce qui n'est pas surprenant puisque une D-brane équatoriale sphérique
est une combinaison de deux D-branes $\RP^2$.

L'amplitude précédente ne contenant aucune information sur le couplage
des D-branes $\RP^2$ aux cordes fermées twistées, nous considérons les
cordes entre deux D-branes $\RP^2$ de signe défini, ce qui correspond
sur la 3-sphère à des cordes entre des D-branes équatoriales. En
notant $Z$ l'action de l'identification $\Z_2$, on doit avoir
\begin{equation}
\label{Z}
Z|N,l,ij\> = (-)^l s_i s_j |N,l,ij\> \text{ pour }l = 0,1,\ldots,k/2,
\end{equation}
où $N$ est le nombre d'excitations de l'état, $l$ le spin de la
représentation de $\SO(3)$ sous laquelle il se transforme, $i$ et $j$
les indices des deux D-branes, $s_i$ et $s_j$ leurs signes. Les
valeurs indiquées pour $l$ sont celles où $N_{k/4,k/4}^l$ ne s'annule
pas et le signe $(-)^l$ est la parité du $l$-ième harmonique sphérique.
Il en découle que l'on a uniquement des cordes d'isospin pair entre
D-branes de même signe, et uniquement d'isospin impair entre D-branes
de signes différents. L'amplitude de l'anneau correspondante s'écrit
donc
\[
\A_{RR} = \frac{1}{2} \sum_{\substack{l=0\\l \text{ entier}}} \chi_l(\sqrt{q})
\pm \frac{1}{2} \sum_{\substack{l=0\\l \text{ entier}}} (-)^l \chi_l(\sqrt{q})
\]
où $\pm$ est le produit des deux signes.

\index{etatt@état twisté}
Comme le premier terme contient tous les couplages de la D-brane
$\RP^2$ avec les cordes fermées non twistées, le deuxième ne contient
que les couplages avec les cordes twistées. Après une transformation
modulaire, on trouve que la D-brane n'est couplée qu'à l'état twisté
avec $(j,\bar{j})=(k/4,k/4)$, et le couplage est
\[
D_R^{\text{twisted}} = \pm \frac{1}{2} \sqrt{\frac{k+2}{2}}
\]
où $\pm$ est le signe de la D-brane. Ce signe n'est donc rien d'autre,
à un facteur numérique près, que sa charge sous l'état twisté
$(k/4,k/4)$.

Considérons maintenant le champ de jauge. Sachant que nos D-branes ont une
géométrie $S^2\times M'$ dans $S^3\times M$ (puisque la physique sur $S^3$
seule n'aurait pas de sens), on s'intéresse au champ de jauge effectif sur
$M'$ après réduction dimensionnelle de la 2-sphère, donc à la partie du
champ de jauge indépendante des coordonnées sur la 2-sphère, qui se trouve
dans la représentation de spin 0 de l'algèbre de Kac-Moody. Sachant que, en
ce qui concerne les D-branes équatoriales, la représentation de spin 0
apparaît entre des D-branes de même charge twistée, on conclut que le
groupe est $\prod_{j<k/4}\U(n_j)\times \U(n_+)\times \U(n_-)$.

\subsection{Calculs semi-classiques et stabilisation par le flux}
Nous allons maintenant nous intéresser à la description effective des
D-branes à basse énergie, c'est-à-dire l'action de Born-Infeld. 

\index{D-brane!stabilisation par le flux}
\subsubsection{Stabilité des D-branes sphériques}
Tout d'abord, une question est de savoir d'où vient, au niveau de
l'action effective, la quantification de la position des D-branes
donnée par la théorie conforme (équation \eqref{PosDbrane}), ainsi que
leur stabilité, sachant qu'elles n'ont pas une surface minimale. On va
voir que cela est lié directement à la quantification du flux
magnétique à travers la D-brane.

Le cas de la 3-sphère a été considéré initialement en \cite{BDS}, et
le calcul a ensuite été généralisé à des groupes compacts quelconques
\cite{BRS}. Nous nous restreindrons ici à la 3-sphère.

Dans les coordonnées sphériques habituelles, la métrique et la 3-forme
de Neveu-Schwarz s'écrivent :
\index{champ de fond!$B_{\mu\nu}$}
\begin{gather*}
ds^2 = \tilde{k}\a' \left[ d\psi^2 
          + \sin^2\psi (d\theta^2 + \sin^2\theta\,d\phi^2) \right] \\
H = dB = 2\tilde{k}\a' \sin^2\psi \sin\theta\,d\psi\,d\theta\,d\phi
\end{gather*}
où $\tilde{k}\a'$ est le rayon de la 3-sphère (en l'absence de
correction quantique, $\tilde{k}$ serait le niveau de l'algèbre de
Kac-Moody). On peut choisir pour $B$ une jauge préservant une symétrie
$\SO(3)$ :
\begin{equation}
\label{BSU2}
B = \left[ \tilde{k}\a' \left( \psi - \frac{\sin2\psi}{2} \right)
           + \pi\a'n_0 \right] \sin\theta\,d\theta\,d\phi.
\end{equation}
Ceci a des singularités aux deux pôles $\psi=0$ et $\pi$. La fonction
d'onde d'une corde enroulée autour de cette singularité acquiert une
phase $\int_{S^2}B/2\pi\a'$, qui vaut $2\pi n_0$ en $\psi=0$ et
$2\pi(\tilde{k}+n_0)$ en $\psi=\pi$. Cela est acceptable si elle est
multiple de $2\pi$, donc $n_0$ et $\tilde{k}$ doivent être entiers.
Ceci est un moyen simple de voir que le rayon de la 3-sphère doit être
quantifié.

\index{champ de fond!électromagnétique}
Considérons maintenant une D2-brane sphérique à une valeur fixée de
$\psi$. On peut la munir d'un flux magnétique uniforme
\[
\label{FSU2}
F = dA = -\frac{n}{2} \sin\theta\,d\theta\,d\phi
\]
où $n$, égal à l'opposé du flux magnétique, est un entier. On notera
que $n$ n'est pas invariant de jauge, puisque la 2-forme invariante
est $\hat{B}+2\pi\a'F$, et non $F$. En revanche, $n-n_0$ l'est. Nous
prenons également en compte un fond de dilaton, supposé constant dans
un premier temps.

\index{Born-Infeld, action de}
L'énergie de Born-Infeld pour la D2-brane s'écrit
\[
\begin{split}
E &= T_{D2} \int d\theta\,d\phi \,e^{-\Phi} 
                \sqrt{\det(\hat{g}+\hat{B}+2\pi\a'F)} + \ldots \\
  &= 4\pi\tilde{k}\a'T_{D2}e^{-\Phi} \sqrt{\sin^4\psi + 
  \left( \psi - \frac{\sin2\psi}{2} - \frac{\pi(n-n_0)}{\tilde{k}} \right)^2}
  + \ldots
\end{split}
\]
Si $0<n-n_0<\tilde{k}$, ceci admet un minimum pour 
\[
\psi_n = \frac{\pi(n-n_0)}{\tilde{k}}.
\]
L'identification avec l'équation \eqref{PosDbrane} conduit aux
relations suivantes entre ces paramètres à basse énergie et ceux de
l'état de Cardy en théorie conforme :
\begin{equation}
\label{FluxType}
n - n_0 = 2j + 1
\end{equation}
\[
\tilde{k} = k + 2.
\]
Noter la renormalisation du rayon $k \rightarrow k+2$, déjà rencontrée
dans l'expression du tenseur énergie-impulsion.

Dans le cas de $\RP^3$, la stabilisation des D-branes sphériques se
fait par le même mécanisme. Les D-branes équatoriales, quant à elles,
sont stabilisées par leur topologie, qui est telle qu'elles ne peuvent
pas se déformer continûment en un point.

\index{couplage!entre D-brane et cordes fermées}
\subsubsection{Couplage aux cordes fermées}
Les couplages $D^j_s$ peuvent s'obtenir semi-classiquement à partir du
couplage linéarisé des D-branes au dilaton. Celui-ci peut se décomposer en
harmoniques hypersphériques :
\[
\Phi = \sum_j \sum_{L=0}^{2j} \sum_{M=-L}^{L}
       \Phi_{jLM} F_{jLM}(\psi) Y_{LM}(\theta, \phi).
\]
Sachant que le couplage linéarisé s'écrit, pour une D-brane située en
$\psi_n$,
\[
\int_{S^2} d\theta\,d\phi\,\Phi \sin\psi_n \sin\theta,
\]
seuls apparaissent les harmoniques $L=M=0$. Comme
\[
F_{j00}(\psi) = \sqrt{\frac{2}{\pi}} \frac{\sin(2j+1)\psi}{\sin\psi},
\]
on trouve un couplage effectif du dilaton valant
\[
D^j_{s\text{ eff}} \propto \sin\!\parfrac{(2j+1)(2s+1)\pi}{k+2}.
\]
On note tout d'abord que la dépendance en $s$ est la même que celle
donnée par la théorie conforme (équation \eqref{CouplageSU2}). En ce
qui concerne la dépendance en $j$, il convient de remarquer que dans
les calculs de théorie conforme on considérait des états propres de
$J^3$ et $\bar{J}^3$, correspondant aux spins gauche et droit de
$\mathfrak{so}(4)=\mathfrak{su}(2)_L\oplus\mathfrak{su}(2)_R$, alors
que dans ce calcul semi-classique nous avons considéré plutôt des
états propres du spin du sous-groupe diagonal $\SO(3)$, l'un et
l'autre étant reliés par
\[
|j,L=M=0\> = \frac{1}{\sqrt{2j+1}} \sum_{m=-j}^{j} |j,m,-m\>,
\]
d'où découle qu'il faut en fait comparer $D^j_s$ et
$(2j+1)^{-1/2}D^j_{s\text{ eff}}$. On voit alors que, dans la limite
de grand $k$, ce qui revient à la limite $\a'\rightarrow0$ à rayon
fixé, les deux couplages sont égaux à un facteur indépendant de $j$
près qui correspond essentiellement au couplage gravitationnel (cf.
\cite{BDS} pour une prise en compte précise de tous les facteurs). Le
fait que l'identification ne marche pas pour $k$ fini n'a rien
d'étonnant, puisque après tout l'action de Dirac-Born-Infeld n'est
valide que pour $\a'$ petit ; il est au contraire tout à fait
remarquable que, avec seulement la renormalisation $k\rightarrow k+2$,
l'action effective reproduise exactement la position des D-branes, la
dépendance en la D-brane du couplage, et, comme nous le verrons dans
la suite, le spectre des fluctuations à basse énergie.

\subsubsection{Petites fluctuations}
\index{spectre!corde bosonique ouverte!WZW}
Jusqu'à maintenant, nous avons considéré des D-branes à $\psi$
constant. Nous allons maintenant les faire fluctuer, ainsi que le champ
électromagnétique sur elles. Dans la mesure où, comme dit précédemment,
un espace-temps compact n'aurait guère de sens, nous allons ajouter
une dimension temporelle (le fait de ne pas ajouter d'autre dimension
signifie simplement que nous n'allons pas nous préoccuper des
fluctuations dans ces dimensions). En ce qui concerne le champ
électromagnétique, nous nous plaçons dans la jauge $A_0=0$. Nous avons
donc trois champs fonctions de $(t,\theta,\psi)$, que nous écrivons
comme suit :
\[
\psi = \psi_n + \delta, \quad
A_{\theta} = \frac{\tilde{k}}{2\pi}\,\a_{\theta}, \quad \text{et} \quad
A_{\phi} = \frac{n}{2}(\cos\theta-1) + \frac{\tilde{k}}{2\pi}\,\a_{\phi}.
\]
Il s'agit maintenant de développer le lagrangien de Born-Infeld
jusqu'à l'ordre quadratique en les fluctuations. On a
\[
\hat{g} + \hat{B} + 2\pi\a'F = \tilde{k}\a'
\begin{pmatrix}
-\frac{1}{\tilde{k}\a'} + (\d_t\delta)^2 &
\d_t\delta\d_\theta\delta + f_{t\theta} &
\d_t\delta\d_\phi\delta + f_{t\phi} \\
\d_t\delta\d_\theta\delta - f_{t\theta} &
\sin^2\psi + (\d_\theta\delta)^2 &
\d_\theta\delta\d_\phi\delta + \F_{\theta\phi} \\
\d_t\delta\d_\phi\delta - f_{t\phi} &
\d_\theta\delta\d_\phi\delta - \F_{\theta\phi} &
\sin^2\psi\sin^2\theta + (\d_\phi\delta)^2
\end{pmatrix}
\]
avec
\[
f_{mn} = \d_m\a_n - \d_n\a_m \quad\text{et}\quad
\F_{\theta\phi} = \left( \delta - \frac{\sin2\psi}{2} \right)
                  + f_{\theta\phi}.
\]
Après quelques calculs sans difficultés, on trouve la variation du
lagrangien.

Tout d'abord on a un terme linéaire proportionnel à $f_{\theta\phi}$.
À première vue, cela semble contredire le fait qu'on développe autour
d'une solution classique du mouvement. En fait il n'en est rien, car
l'intégrale de ce terme sur la 2-sphère est la fluctuation du flux du
champ magnétique, laquelle est nécessairement nulle puisque le flux
est quantifié. On voit donc que cette quantification est essentielle à
la stabilisation de la D-brane.

Les termes quadratiques peuvent s'écrire comme suit :
\[
\begin{split}
\delta^{(2)}S &= -T_{D2}(\tilde{k}\a')^2\sin\psi_n \,\times \\
              &\quad \times \int d\theta\,d\phi
                \sqrt{-\det\g}
\left[ \frac{1}{2}\d_m\delta\d^m\delta + \frac{\tilde{k}\a'}{4} f_{mn}f^{mn}
       + \frac{1}{\tilde{k}\a'}\delta^2 
       + \frac{2}{\tilde{k}\a'}\delta\frac{f_{\theta\phi}}{\sin\theta} \right]
\end{split}
\]
où la métrique effective $\g$, qui sert à faire monter les indices,
est donnée par
\[
ds^2 = -dt^2 + \tilde{k}\a'(d\theta^2 + \sin^2\theta\,d\phi^2).
\]
Cette métrique n'est autre que la métrique de corde ouverte de
l'équation \eqref{MetOuv}. On voit clairement sur cet exemple que les
champs sur la D-brane voient une géométrie donnée par cette métrique,
qui est différente de la géométrie vue par les cordes fermées :
notamment, les D-branes ont toutes le même rayon pour les cordes
ouvertes.

Il en découle, après quelques calculs, les équations du mouvement
suivantes :
\[
\frac{d^2}{dt^2}
    \begin{pmatrix} \delta \\ f_{\theta\phi}/\sin\theta \end{pmatrix}
  = -\frac{1}{\tilde{k}\a'}
    \begin{pmatrix} -\triangle+2 & 2 \\ -2\triangle & -\triangle \end{pmatrix}
    \begin{pmatrix} \delta \\ f_{\theta\phi}/\sin\theta \end{pmatrix}
\]
où $\triangle$ est le laplacien sur la 2-sphère de rayon unité.

Pour trouver les valeurs propres de cette matrice, on décompose
$\delta$ et $f_{\theta\phi}/\sin\theta$ en harmoniques sphériques :
\begin{equation}
\label{FlucHarmSph}
\delta = \sum_{l=0}^{\infty} \sum_{m=-l}^{l} \delta_{lm} Y_{lm}(\theta,\phi)
\quad\text{et}\quad
f_{\theta\phi}/\sin\theta = \sum_{l=1}^{\infty} \sum_{m=-l}^{l}
                             f_{lm} Y_{lm}(\theta,\phi).
\end{equation}
On notera l'absence de terme $l=0$ pour $f$ à cause de la
quantification du flux, d'où découle que pour $l=0$ la seule
fluctuation est celle de $\delta$, avec comme masse $2/(k+2)\a'$. Pour
$l>0$ la matrice de masse s'écrit
\[
\frac{1}{(k+2)\a'}
\begin{pmatrix} l(l+1)+2 & 2 \\ 2l(l+1) & l(l+1) \end{pmatrix}
\]
dont les valeurs propres sont $(l+1)(l+2)/(k+2)\a'$ et
$l(l-1)/(k+2)\a'$. Il en découle le spectre :
\begin{equation}
\label{MFluc}
m^2 = \frac{j(j+1)}{(k+2)\a'} \ \text{dans les représentations}\ 
(j-1)\oplus(j+1),
\end{equation}
où $j$ prend toutes les valeurs entières positives, étant entendu que
seule la représentation de spin 1 apparaît pour $j=0$ ; ces modes zéro
correspondent à des rotations arbitraires de la 2-sphère dans la
3-sphère. Les autres modes sont massifs, ce qui assure la stabilité de
la D-brane.

La comparaison de ce résultat avec la CFT est aisée. Les masses trouvées
sont celles de l'équation \eqref{SpOuvWZW}, avec $N+M=1$, ce qui correspond
justement aux excitations de basse énergie, le cas $N+M=0$ étant exclu par
la projection GSO. Ces états sont de la forme $J^a_{-1}|j\>$, donc se
transforment sous les représentations $(j-1)\oplus j\oplus(j+1)$. La
contrainte de Virasoro éliminant la représentation $j$ (c'est celle-là qui
est éliminée car il y a autant de contraintes que d'états dans la
représentation $j$ de $\SU(2)$, soit $2j+1$), le spectre est finalement en
accord avec Born-Infeld. Il y a toutefois une différence : c'est que la CFT
nous apprend que $j$ est borné (inférieur à $k/2$), ce que Born-Infeld ne
permet pas de voir. Là encore, cela se comprend aisément si on se rappelle
que Born-Infeld est valide dans la limite $k\rightarrow\infty$.

\index{RP@$\RP^3$}
Dans le cas de $\RP^3$, il est nécessaire d'imposer $n_0=-\tilde{k}/2$
pour que $B$ soit univaluée. Cela étant fait, les choses sont très
similaires à $S^3$, en particulier pour les D-branes sphériques où les
résultats sont identiques. Le spectre des petites fluctuations de la
D-brane sphérique équatoriale ne contient que la moitié des cordes
ouvertes légères car elle est enroulée sur un plan projectif.

Le calcul des fluctuations de la D-brane $\RP^2$ requiert d'imposer
$F=0$ afin que l'action de Born-Infeld soit bien définie. Les
fluctuations s'obtiennent de la même façon, à cela près que l'on doit
imposer l'invariance par les transformations \eqref{Z2}, c'est-à-dire
que l'on doit avoir
\begin{align*}
\delta(\pi-\theta,\phi+\pi) &= =\delta(\theta,\phi) \\
\a_\theta(\pi-\theta,\phi+\pi) &= -\a_\theta(\theta,\phi) \\
\a_\phi(\pi-\theta,\phi+\pi) &= \a_\phi(\theta,\phi),
\end{align*}
ce qui implique que, dans la décomposition \eqref{FlucHarmSph}, il
n'apparaît que des harmoniques impairs, d'où découle que dans
l'équation \eqref{MFluc} $j$ doit être pair, ce qui est conforme aux
prévisions de la théorie conforme.
\index{D-brane!sur un groupe|)}

\index{orientifold!$S^3$ et $\RP^3$|(}
\section{Orientifolds sur $S^3$ et $\RP^3$}
Cette section est consacrée à l'étude des orientifolds dans les
groupes compacts non abéliens les plus simples que sont $S^3$ et
$\RP^3$, telle que nous l'avons faite dans les articles \citemoi{A} et
\citemoi{B}. Contrairement aux approches antérieures, qui soit étaient
purement algébriques \cite{Sag,SS}, soit partaient de considérations
algébriques pour trouver la géométrie \cite{Bru,HSS}, nous sommes
parti de la géométrie, c'est-à-dire que nous avons d'abord trouvé la
position des orientifolds possibles, puis leurs couplages aux cordes
fermées par une méthode similaire à ce qui a été fait pour les
D-branes dans la section précédente. Nous traiterons ici des
orientifolds sans D-branes ; l'interaction entre les deux fera l'objet
de la section suivante.

\subsection{Description géométrique des orientifolds}
\subsubsection{SU(2)}
On peut paramétrer $\SU(2)$ comme suit :
\[
g = \begin{pmatrix} X_1+iX_2 & X_3+iX_4 \\ -X_3+iX_4 & X_1-iX_2 \end{pmatrix},
\]
où les $X_i$ sont à valeurs sur une 3-sphère de rayon 1. Outre les
coordonnées polaires usuelles :
\[
X_1 = \cos\psi, \quad X_2 = \sin\psi \cos\theta, \quad
X_3+iX_4 = \sin\psi \sin\theta\,e^{i\phi},
\]
nous serons amené à utiliser les angles d'Euler, définis par
\[
X_1+iX_2 = \sin\a\,e^{i\beta}, \quad
X_3+iX_4 = \cos\a\,e^{i\g},
\]
avec $\a\in[0;\pi/2]$ et $\beta,\g\in[0;2\pi]$.

L'orientifold, comme nous l'avons vu précédemment, consiste à combiner
une inversion d'orientation de la surface d'univers ($\Omega:
\s\rightarrow-\s$) et une isométrie involutive $h$ de l'espace-temps.
Le fait de prendre une isométrie assure l'invariance du premier terme
de l'action de WZW (équation \eqref{AcWZW}). Cela n'est, en revanche,
pas le cas pour le terme de Wess-Zumino : puisque $\Omega$ inverse
l'orientation de toute variété ayant pour bord la surface d'univers de
la corde, il faut que $h$ inverse aussi l'orientation de
l'espace-temps pour assurer l'invariance du terme de Wess-Zumino. Nous
imposerons également, comme lors de notre étude des D-branes, la
préservation de l'une des algèbres de Kac-Moody, ce qui entraîne que
$h$ doit commuter avec les rotations à $\psi$ fixé.

\begin{figure}
\begin{center}
\includegraphics[scale=0.7]{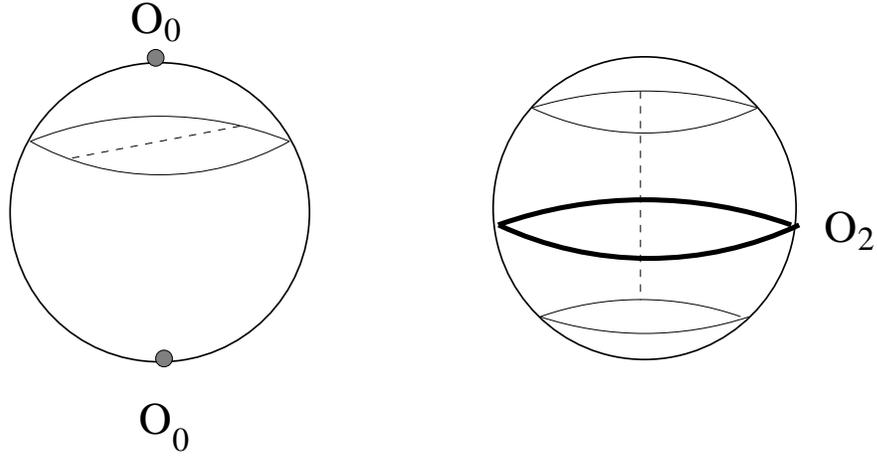}
\end{center}
\caption{Positions possibles pour les orientifolds sur la 3-sphère.}
\label{orien}
\end{figure}

Sachant que les isométries involutives de la 3-sphère s'écrivent, à
conjugaison par un élément du groupe près, $\diag(\pm1,\pm1,\pm1,\pm1)$, 
on aboutit finalement à deux choix pour $h$ (fig. \ref{orien}) :
\begin{itemize}
\item $h_0 = \diag(1,-1,-1,-1)$, c'est-à-dire $g\rightarrow g^\dagger$,
ce qui en termes de coordonnées s'écrit
\begin{alignat*}{3}
(\psi,\theta,\phi) &\rightarrow (\psi,\pi-\theta,\phi+\pi), &\quad
&\text{et} &\quad (\a,\beta,\g) &\rightarrow (\a,-\beta,\g+\pi).
\\
\intertext{Les points fixes sont deux orientifolds $\O0$ situés aux pôles de la
3-sphère ($\psi=0$ et $\pi$).
\item $h_2 = \diag(-1,1,1,1)$, c'est-à-dire $g\rightarrow -g^\dagger$,
ce qui en termes de coordonnées s'écrit}
(\psi,\theta,\phi) &\rightarrow (\pi-\psi,\theta,\phi), &\quad
&\text{et} &\quad (\a,\beta,\g) &\rightarrow (\a,\pi-\beta,\g).
\end{alignat*}
L'ensemble des points fixes est un orientifold $\O2$ situé à
l'équateur ($\psi=\pi/2$).
\end{itemize}

Remarquons en passant que ce type de transformations se généralise aisément
à n'importe quel groupe : pour tout élément $\lambda$ du centre du groupe,
$g\rightarrow\lambda g^{-1}$ définit un orientifold.

\subsubsection{SO(3)}
\index{RP@$\RP^3$}
Pour obtenir les orientifolds de $\RP^3$, on considère \textit{a priori}
une isométrie de $S^3$ avec $h^2=\pm1$ (puisque $1$ et $-1$ sont
identifiés). La nécessité que $h$ inverse l'orientation impose en fait de
se limiter à $h_0$ ou $h_2$, comme pour $\SU(2)$. Ces deux isométries étant
identifiées, elles donnent toutes deux une même configuration, avec comme
points fixes un $\O0$ au pôle et un $\O2$ plan projectif à l'équateur.

\subsection{Couplage aux cordes fermées}
\index{couplage!entre orientifold et cordes fermées}
Comme nous l'avons vu précédemment, l'amplitude de la bouteille de Klein
peut s'interpréter soit comme une boucle de corde fermée changeant
d'orientation, soit comme une émission et absorption d'une corde fermée par
un orientifold. En sachant quels états restent et lesquels sont supprimés
par la projection orientifold, on pourra trouver l'amplitude de la
bouteille de Klein et en déduire le couplage de l'orientifold aux cordes
fermées.

\subsubsection{SU(2)}
Les états de corde fermée, avant projection, sont de la forme
\[
P(J^a_n,\bar{J}^{\bar{a}}_{\bar{n}}) |j,m,\bar{m}\>
\]
où $P$ est un polynôme, $J^a_n$ et $\bar{J}^{\bar{a}}_{\bar{n}}$ sont les
générateurs de l'algèbre de Kac-Moody, et $|j,m,\bar{m}\>$ est un état
primaire (sans excitations) dans la représentation $(j,j)$ de l'algèbre.

L'action de l'orientifold sur les courants de Kac-Moody est simple : il
découle immédiatement de l'équation \eqref{CourKM} que $\Omega h_0$ et
$\Omega h_2$ échangent les courants gauche et droit.

\index{operateur@opérateur de vertex}
En ce qui concerne les primaires, on obtient l'action de l'orientifold en
considérant les opérateurs de vertex correspondants. Il s'agit
d'harmoniques sphériques qui, dans les coordonnées d'Euler, s'écrivent
\[
D^j_{m\bar{m}}(\a,\beta,\g) = e^{i(m+\bar{m})\g} e^{i(m-\bar{m})\beta}
                              P^j_{m\bar{m}}(\cos\a).
\]
La forme précise de $P^j_{m\bar{m}}$ n'interviendra pas ici : tout ce qui
importe pour nous est leur symétrie par échange de $m$ et $\bar{m}$.
L'action de l'orientifold sur les primaires en découle :
\[
\Omega h_0|j,m,\bar{m}\> = (-)^{m+\bar{m}} |j,\bar{m},m\>, \quad \text{et}
\quad \Omega h_2|j,m,\bar{m}\> = |j,\bar{m},m\>.
\]

\index{fonction de partition!bouteille de Klein!$S^3$}
Nous avons maintenant ce qu'il faut pour écrire la bouteille de Klein.
Sachant que seuls interviennent les états avec $m=\bar{m}$, et qu'on a
$2j=2m\mod 2$, on obtient :
\begin{equation}
\label{KlS3}
\K_{(0)} = \frac{1}{2} \sum_{j=0}^{k/2} (-)^{2j} \chi_j(q^2) \quad
\text{et} \quad \K_{(2)} = \frac{1}{2} \sum_{j=0}^{k/2} \chi_j(q^2).
\end{equation}

Nous avons vu d'autre part (équation \eqref{Klein}) que l'amplitude de
Möbius s'écrit aussi $\K=\<C|\sqrt{\tilde{q}}^{L_0-c/24}|C\>$. En faisant
le même raisonnement que pour l'amplitude de l'anneau avec les états
d'Ishibashi de crosscap $|j_C\>$ au lieu des états d'Ishibashi de bord, on
obtient
\[
\K = \sum_{j=0}^{k/2} (C^j)^2 \chi_j(\sqrt{\tilde{q}})
\]
où les $C^j$ sont les couplages de l'orientifold aux cordes fermées dans la
représentation $(j,j)$. Une transformation modulaire à partir de l'équation
\eqref{KlS3} permet de calculer les $C^j$, et on trouve
\begin{align}
\label{CouplageO0}
C^j_{(0)} &= \eps^j_{(0)} E_{2j+k} \sin\!\parfrac{(2j+1)\pi}{2k+4}
             \sqrt{\N_j} \\
\label{CouplageO2}
C^j_{(2)} &= \eps^j_{(2)} E_{2j} \cos\!\parfrac{(2j+1)\pi}{2k+4}
             \sqrt{\N_j}
\end{align}
où $E_n$ est le projecteur sur les nombres pairs ($E_n=(1+(-)^n)/2$),
$\N_j$ est donné par l'équation \eqref{norm} et les $\eps^j$ sont des
signes, pour l'instant inconnus puisque nous n'avons calculé que les carrés
des couplages.

\subsubsection{SO(3)}
\index{RP@$\RP^3$}
Outre les états non twistés, tous de spin entier, la bouteille de Klein
fait apparaître les états twistés dont les spins gauche et droit sont
égaux. Dans le cas de SO(3), il y en a un, avec un spin égal à $k/4$.
L'amplitude de la bouteille de Klein en découle :
\index{fonction de partition!bouteille de Klein!$\RP^3$}
\[
\K = \frac{1}{2} \biggl[ \sum_{\substack{l=0 \\ l\text{ entier}}}^{k/2} 
     \chi_l(q^2) + \zeta \chi_{k/4}(q^2) \biggr].
\]
Le signe $\zeta$ indique comment l'état twisté se transforme par
l'orientifold. Les deux signes sont \textit{a priori} cohérents, et cela
sera confirmé par l'étude de l'amplitude du ruban de Möbius.

Comme dans le cas de $\SU(2)$, on peut, par une transformation modulaire,
trouver le couplage des orientifolds aux cordes fermées. On trouve
\[
C^j = \frac{1}{\sqrt{2}}\,\eps^j E_{2j} \sqrt{\N_j} 
      \left[ \sin\!\parfrac{(2j+1)\pi}{2k+4}
      + \zeta (-)^j \cos\!\parfrac{(2j+1)\pi}{2k+4} \right].
\]
Cela ressemble fortement à la somme des couplages des deux sortes
d'orientifolds de la 3-sphère, ce qui n'a rien d'étonnant puisque justement
on a ici un $\O0$ et un $\O2$.

\subsection{Considérations semi-classiques}
Pour comprendre ces formules, notamment les facteurs $E_{2j}$ et
$E_{2j+k}$, et fixer les signes, une analyse semi-classique est utile. Nous
allons, comme pour les D-branes, nous intéresser au couplage linéarisé au
dilaton. Sachant que l'action effective de l'orientifold s'écrit
\index{action effective!orientifold}
\[
S = -T_{\O} \int e^{-\Phi} \sqrt{-\det \hat{g}},
\]
où $T_{\O}$ est la tension de l'orientifold, on peut, par des calculs
similaires au cas des D-branes, calculer un couplage effectif.

Dans le cas de l'orientifold $\O2$, on trouve
\[
C^j_{(2)\text{ eff}} \propto T_{\O} (-)^j E_{2j},
\]
ce qui confirme le facteur $E_{2j}$ et fixe le signe :
\[
\eps^j_{(2)} = (-)^j \sign(T_{\O2}).
\]
La dépendance en $j$ et $k$ de l'équation \eqref{CouplageO2} n'est pas
visible ici car ce calcul n'est valide que dans la limite où $k$ est grand,
auquel cas le cosinus vaut 1.

Le cas des orientifolds $\O0$ de la 3-sphère est plus subtil. Puisque nous
avons un orientifold à chaque pôle, leur couplage au dilaton est la somme
de deux termes :
\[
T_{\O0}^{\text{nord}} \Phi(\psi=0) + T_{\O0}^{\text{sud}} \Phi(\psi=\pi)
\]
ce qui conduit au couplage effectif
\[
C^j_{(0)\text{ eff}} \propto
   (2j+1) (T_{\O0}^{\text{nord}} + (-)^{2j} T_{\O0}^{\text{sud}}).
\]
Pour aller plus loin nous devons déterminer la nature précise des
orientifolds, qui est liée à la torsion discrète du champ $B$.

Ce qui se passe est que la présence d'un champ $B$ ajoute un facteur
$\exp\left[i\int_{\Sigma}B/2\pi\a'\right]$ à l'intégrale de chemin. Sachant que
l'orientifold transforme une 2-sphère autour d'un $\O0$ en un plan
projectif, on est amené à considérer le cas $\Sigma=\RP^2$, étant entendu
que l'intégrale est bien définie car l'orientifold change aussi le signe de
$B$. Avec les notations de l'équation \eqref{BSU2}, le facteur obtenu est
$(-)^{n_0}$ pour l'orientifold nord, et $(-)^{n_0+k}$ pour l'orientifold
sud. La tension de l'orientifold étant proportionnelle à ce facteur ---
plus précisément, elle est négative (orientifold $\O0^-$) en son absence
--- il en découle que
\[
T_{\O0}^{\text{sud}} = (-)^k T_{\O0}^{\text{nord}}
\]
 d'où
\[
C^j_{(0)\text{ eff}} \propto (2j+1) E_{2j+k} T_{\O0}^{\text{nord}}
\]
ce qui est compatible avec le calcul de la bouteille de Klein (équation
\eqref{CouplageO0}) dans la limite $k$ grand, et fixe le signe :
\[
\eps^j_{(0)} = \sign(T_{\O0}^{\text{nord}}).
\]

Le calcul semi-classique des signes $\eps^j_{(0)}$ et $\eps^j_{(2)}$ que
nous venons de faire sera confirmé, dans le cas où il y a des D-branes, par
le calcul des amplitudes de Möbius.

Dans le cas de $\RP^3$, il découle du calcul semi-classique que le couplage
est la somme d'un couplage pour le $\O0$ et d'un couplage pour le $\O2$,
c'est-à-dire ce que nous avons trouvé pour la bouteille de Klein ; on
s'attend en outre à ce que $\eps^j$ soit indépendant de $j$.

\section{D-branes en présence d'un orientifold sur $S^3$ ou $\RP^3$}
\index{orientifold!avec cordes ouvertes!sur $S^3$ et $\RP^3$|(}
Nous allons ici nous intéresser principalement à l'amplitude du vide sur le
ruban de Möbius, qui peut s'interpréter soit comme une boucle de corde
ouverte non orientée, soit comme un échange de corde fermée entre une
D-brane et un orientifold. Ceci nous fournira un test de cohérence des
calculs précédents, et nous permettra en outre de trouver les groupes de
jauge associés aux D-branes.

\index{fonction de partition!ruban de Möbius}
Nous avons vu précédemment (équation \eqref{Mobius}) que, dans le canal
fermé, l'amplitude de Möbius s'écrit en faisant intervenir les états de
bord et de crosscap : $\M=\<C|\sqrt{\tilde{q}}^{L_0-c/24}|B\>$. À l'aide
des états d'Ishibashi, ceci s'écrit
\[
\M = \sum_j \<C|j_C\> \<j|B\> \<j_C|\sqrt{\tilde{q}}^{L_0-c/24}|j\>.
\]
Sachant que les états de bord vérifient $(J_n+\bar{J}_{-n})|j\>=0$, tandis
que les états de crosscap vérifient $(J_n+(-)^n\bar{J}_{-n})|j_C\>=0$, il
découle que le terme $\<j_C|\sqrt{\tilde{q}}^{L_0-c/24}|j\>$ s'écrit
$\Tr\bigl[(-)^N\sqrt{\tilde{q}}^{L_0-c/24}\bigr]$, qui n'est autre que
$\hat{\chi}_j(\tilde{q})$, où $\hat{\chi}$ est le caractère modifié
introduit dans l'équation \eqref{hatchi}.

\index{transformation modulaire}
On montre que les $\hat{\chi}_j(\tilde{q})$ peuvent, par une transformation
modulaire, s'écrire
\[
\hat{\chi}_i(\tilde{q}) = \sum_j P_{ij} \hat{\chi}_j(q),
\]
où la matrice $P$ vaut
\[
P = T^{1/2} S T^2 S T^{1/2}.
\]
Elle a comme composantes (cf le calcul dans l'annexe de \citemoi{A})
\[
P_{ij} = \frac{2}{\sqrt{k+2}} \sin\!\parfrac{\pi(2i+1)(2j+1)}{2k+4}
         E_{2i+2j+k}.
\]

Signalons enfin que, puisque nous nous restreignons à des configurations
préservant une algèbre de Kac-Moody, nos D-branes seront parallèles au
$\O2$, ou auront les $\O0$ comme centre.

\subsection{Cas des $\O0$ sur la 3-sphère}
L'isométrie $h_0$ transformant les D-branes en elles-mêmes, l'amplitude de
Möbius fait intervenir des cordes entre une D-brane et elle-même. Sachant
que $h_0$ restreint à une D-brane est l'identification antipodale, chaque
terme sera affecté d'un signe égal à la parité de l'harmonique sphérique
correspondant, c'est-à-dire $(-)^l$. L'amplitude dans le canal ouvert en
découle :
\index{fonction de partition!ruban de Möbius!$S^3$}
\[
\M_{(0)r} = \eps'_r \sum_{\substack{l=0 \\ l\text{ entier}}}^{\min(2r,k-2r)}
            (-)^l \hat{\chi}_l(q) ;
\]
le signe global $\eps'_r$ détermine le type de projection, et donc le
groupe de jauge, pour les branes de type $r$.

À l'aide de la matrice $P$, on peut écrire l'amplitude dans le canal fermé.
Après quelques calculs, on trouve
\[
\M_{(0)r} = \eps'_r (-)^{2r} \sum_j |C^j_{(0)}| D^j_r \hat{\chi}_j(\tilde{q}).
\]
Or il faudrait trouver ce résultat sans valeur absolue ni signe
supplémentaire. Il en découle d'une part que le signe $\eps^j_{(0)}$ ne
dépend pas de $j$, comme l'analyse semi-classique nous la montré, et
d'autre part que le type de projection alterne avec le type de D-brane :
\[
\eps'_r = \sign(T_{O0}^{\text{nord}}) (-)^{2r},
\]
ce qui donne comme groupe de jauge un produit alterné de groupes
orthogonaux et symplectiques. Ceci implique en particulier que les D-branes
de type entier dans le cas d'un $\O0^+$ au nord, demi-entier dans le cas
d'un $\O0^-$ au nord, sont en nombre pair.


\index{groupe de jauge!orthogonal ou symplectique}
On pourrait, naïvement, faire le raisonnement suivant : puisque
l'intégralité de la D-brane est hors de l'orientifold, la physique sur elle
est localement celle de la corde orientée, donc le groupe de jauge doit
être $\U(n)$. En fait, ce n'est pas correct car, comme auparavant, on regarde
le champ de jauge effectif après réduction dimensionnelle sur la 2-sphère ;
ce champ a bien $n(n-1)/2$ ou $n(n+1)/2$ composantes, d'où un groupe de
jauge $\SO(n)$ ou $\USp(n)$.

\subsection{Cas des $\O2$ sur la 3-sphère}
L'isométrie $h_2$ transforme une D-brane de type $r$ en une D-brane de type
$k/2-r$. Les D-branes viennent donc nécessairement par paires, sauf les
branes équatoriales, de type $k/4$, qui n'existent que si $k$ est pair, et
l'amplitude de Möbius fait intervenir des cordes entre une D-brane et son
image. Sachant que les harmoniques sphériques sont invariants par $h_2$, il
n'y aura pas de signe dans le canal ouvert comme pour le $\O0$. L'amplitude
de Möbius s'écrit donc :
\index{fonction de partition!ruban de Möbius!$S^3$}
\[
\M_{(2)r} = \eps'_r \sum_{\substack{l=k/2-2r \\ l\text{ entier}}}^{k/2}
            \hat{\chi}_l(q).
\]
Après une transformation modulaire par la matrice $P$, on obtient le canal
fermé, et, par un raisonnement similaire au cas précédent, il en découle
que $\eps_{(2)}^j$ est bien celui que nous avons trouvé par un calcul
semi-classique, et que $\eps'_r=\sign(T_{\O2})$.

Le groupe de jauge est sans surprise : il est unitaire pour les D-branes
non équatoriales, et orthogonal ou symplectique pour les D-branes
équatoriales.

\subsection{Cas de $\RP^3$}
\index{RP@$\RP^3$|(}
\subsubsection{Branes non équatoriales}
Les cordes entre deux D-branes non équatoriales de même type $r$ sur
$\RP^3$ correspondent à deux sortes de cordes sur la 3-sphère : des cordes
entre deux branes de type $r$, et des cordes entre une brane de type $r$ et
une de type $k/2-r$. L'action de $\Omega h$ sur de telles cordes est celle
que nous avons vue dans le cas de la 3-sphère, donc l'amplitude de Möbius
s'écrit :
\index{fonction de partition!ruban de Möbius!$\RP^3$}
\[
\M_r = \eps'_r \sum_{\substack{l=0 \\ l\text{ entier}}}^{2r}
       ((-)^l \hat{\chi}_l(q) + \eps''_r \hat{\chi}_{k/2-l}(q))
\]
où $\eps'_r$ et $\eps''_r$ sont des signes. Une transformation modulaire
confirme que $\eps^j$ est une constante $\eps$, et donne les signes
$\eps'_r=(-)^{2r}\eps$ et $\eps''_r=(-)^{2r}\zeta$.

\index{groupe de jauge!orthogonal ou symplectique}
Le groupe de jauge est donné par le signe devant $\chi_0$, c'est-à-dire
$\eps'_r$. Donc, comme pour le $\O0$ de la 3-sphère, le groupe de jauge
pour les D-branes non équatoriales est une alternance de groupes
orthogonaux et symplectiques.

\subsubsection{Branes équatoriales}
Le cas des D-branes équatoriales est plus subtil car, comme nous l'avons
vu, il y en a deux sortes, et elles peuvent être mélangées par
l'orientifold. L'action la plus générale de l'orientifold sur les états
est, comme vu précédemment
\[
\Omega h |N,l,ij\> = (-)^N \g_{ii'} \g^*_{jj'} |N,l,j'i'\>
\]
où $\g$ est une matrice unitaire, symétrique ou antisymétrique. Sachant que
les états de $\RP^3$ doivent être invariants par la transformation $Z$ de
l'équation \eqref{Z}, on doit imposer que l'action de $\Omega h$ préserve
l'invariance par $Z$, d'où découle que $\g$ doit commuter ou anticommuter
avec la matrice $Z$.

Si $\g$ commute avec $Z$, alors les branes de charges différentes ne sont
pas mélangées par l'orientifold, et le groupe de jauge sur ces branes sera
$\SO(n_+)\times\SO(n_-)$ ou $\USp(n_+)\times\USp(n_-)$, où $n_+$ et $n_-$
sont les nombres de chaque sorte de D-brane. Les états apparaissant dans
l'amplitude de Möbius sont alors $|++\>$ et $|--\>$, qui sont d'isospin
pair, d'où l'expression
\index{fonction de partition!ruban de Möbius!$\RP^3$}
\begin{align*}
\M_R &= (n_+ + n_-) \eps'_R \sum_{\substack{l=0\\l\text{ pair}}}^{k/2}
       \hat{\chi}_l(q) \\
     &= \frac{1}{2}(n_+ + n_-) \eps'_R 
        \sum_{\substack{l=0\\l\text{ entier}}}^{k/2}
        \left( (-)^l \hat{\chi}_l(q) + \hat{\chi}_{k/2-l}(q) \right).
\end{align*}

Si $\g$ anticommute avec $Z$, ce qui est possible seulement si
$n_+=n_-\equiv n$, alors, par un changement de base sur les D-branes $+$ et
$-$ séparément, $\g$ peut être mise sous la forme
\[
\g = \begin{pmatrix} 0 & \pm\I \\ \I & 0 \end{pmatrix},
\]
où les premières lignes et colonnes correspondent aux D-branes $+$. L'effet
de $\Omega h$ sur les cordes ouvertes est donc :
\begin{align*}
|++\> &\leftrightarrow |--\> \\
|+-\> &\leftrightarrow \pm|+-\> \\
|-+\> &\leftrightarrow \pm|-+\>
\end{align*}
Avant de faire l'opération d'orientifold, on avait un groupe de jauge
$\U(n)$ correspondant aux états $|++\>$, et un autre pour les états
$|--\>$. Puisqu'ils sont identifiés par l'orientifold, il reste un groupe
de jauge $\U(n)$. On notera que les D-branes $+$ et $-$ ne peuvent pas
exister séparément ici, et que la matrice $\g$ définit un appariement entre
elles, de sorte qu'il est ici pertinent de considérer que nous avons $n$
D-branes équatoriales sphériques, le groupe $\U(n)$ s'interprétant alors
comme les changements de base arbitraires dans l'espace de ces D-branes.
Les états apparaissant dans l'amplitude de Möbius sont alors $|++\>$ et
$|--\>$, qui sont d'isospin pair, d'où l'expression
\index{fonction de partition!ruban de Möbius!$\RP^3$}
\begin{align*}
\M_R &= 2n \eps'_R \sum_{\substack{l=0\\l\text{ impair}}}^{k/2}
        \hat{\chi}_l(q) \\
     &= -n \eps'_R
        \sum_{\substack{l=0\\l\text{ entier}}}^{k/2}
        \left( (-)^l \hat{\chi}_l(q) - \hat{\chi}_{k/2-l}(q) \right).
\end{align*}

\index{groupe de jauge!unitaire}
\index{groupe de jauge!orthogonal ou symplectique}
Dans les deux cas, à un facteur $1/2$ près, qui est aussi présent dans les
couplages $D^j_R$ \eqref{DjR}, ces amplitudes s'écrivent comme pour les
D-branes non équatoriales, avec $\eps''_R=+1$ dans le premier cas, et $-1$
dans le second cas. La condition de cohérence s'écrit donc, comme
précédemment, $\eps'_R=\eps''_R(-)^{k/2}\eps$ et $\zeta=\eps''_R(-)^{k/2}$.
Ainsi :
\begin{itemize}
\item Si $\zeta=(-)^{k/2}$, alors le groupe de jauge est
  $\SO(n_+)\times\SO(n_-)$ ou $\USp(n_+)\times\USp(n_-)$, selon le signe
  $\eps$.
\item Si $\zeta=-(-)^{k/2}$, alors on doit avoir $n_+=n_-\equiv n$, et le
 groupe de jauge est $\U(n)$.
\end{itemize}
\index{RP@$\RP^3$|)}
\index{orientifold!avec cordes ouvertes!sur $S^3$ et $\RP^3$|)}
\index{orientifold!$S^3$ et $\RP^3$|)}

%% file: AdS.tex
\chapter{Univers branaires et D-branes anti-de~Sitter}
Comme nous l'avons vu, la théorie des supercordes n'est cohérente qu'à dix
dimensions. Si on veut avoir une chance de décrire notre univers avec, il
faut s'arranger pour obtenir une théorie effective qui soit à quatre
dimensions, au moins aux échelles d'énergie actuellement accessibles
expérimentalement.

Avant qu'on connaisse les D-branes, le seul moyen pour cela était de
prendre un espace cible de la forme $\R^4\times M$, où $M$ est une variété
compacte de dimension six ; on obtient alors une théorie effective à quatre
dimensions à des échelles très grandes devant la taille de $M$. Ceci a
donné lieu à d'importants développements dans les années 1980, notamment
concernant les variétés de Calabi-Yau, qui ont comme propriété de préserver
une partie de la supersymétrie. Le lecteur pourra se référer à
\cite{GSW,Pol} pour plus de détails.

Avec l'émergence des D-branes, et des cordes ouvertes contraintes de rester
à leur voisinage, est venue l'idée que notre Univers pourrait être une
D3-brane (un \emph{univers branaire}), et que la matière serait des cordes
ouvertes vivant sur cette D-brane. Malheureusement, cela ne semble a priori
pas marcher pour la gravité, puisque ses excitations apparaissent en
théorie des cordes comme des cordes fermées, qui peuvent se déplacer dans
tout l'espace-temps. Il faut donc, de toute façon, compactifier six
dimensions. Toutefois, la loi de Newton de la gravitation n'étant vérifiée
expérimentalement que jusqu'à des échelles de l'ordre du dixième de
millimètre, on peut envisager des dimensions compactes bien plus grandes
que sans D-branes, ce qui ouvre d'autant plus de possibilités. Tout cela
est traité en détail dans la revue récente \cite{Kir}.

Par la suite, l'article fondateur de Randall et Sundrum \cite{RS} a montré
que, en présence d'une brane massive, il peut exister un mode du champ
gravitationnel qui est confiné au voisinage de la brane, et la force
gravitationnelle qui en découle est, à basse énergie, celle de Newton.
Cette localisation de la gravité permet de réaliser le principe de
l'univers branaire sans être obligé de compactifier six dimensions. Il
apparaît alors que l'obtention d'une brane plate nécessite un ajustement
fin de la tension de la brane par rapport à la constante cosmologique de
l'espace ambiant, ajustement sans lequel la brane aura une géométrie
de~Sitter ou anti-de~Sitter. On est donc amené à étudier ces géométries,
plus particulièrement dans leur limite de grand rayon où on tend vers un
espace plat.

Le but de ce chapitre est d'étudier certains aspects des tentatives de
réalisations cordistes de l'univers branaire. Puisque, comme nous l'avons
dit, ces réalisations font apparaître des espaces anti-de~Sitter, nous
rappellerons d'abord ce que sont ces espaces.

\section{Espaces anti-de~Sitter}
\index{anti-de Sitter}
\subsubsection{Définition}
Considérons un espace de Minkowski à $n+1$ dimensions et avec deux temps,
c'est-à-dire ayant pour métrique
\[
ds^2 = - dX_0^2 + dX_1^2 + \ldots + dX_{n-1}^2 - dX_n^2.
\]
On peut, au sein de cet espace, obtenir un hyperboloïde ayant comme groupe
d'isométries $\OO(2,n-1)$ en prenant le sous-ensemble d'équation
\begin{equation}
\label{AdSdef}
- X_0^2 + X_1^2 + \ldots + X_{n-1}^2 - X_n^2 = -L^2
\end{equation}
où $L$ est un réel positif, que nous appellerons \emph{rayon} de
l'hyperboloïde, qui a un seul temps. Cet espace ayant des courbes fermées
de genre temps, on définit l'espace anti-de~Sitter $AdS_n$ comme le
recouvrement universel de l'hyperboloïde.

On définit de la même façon l'espace de de~Sitter $dS_n$ en remplaçant,
dans les équations précédentes, les signes $-$ devant $X_n$ et $L^2$ par
des $+$, et il n'y a alors pas besoin de prendre le recouvrement universel.
L'espace ainsi obtenu a comme groupe d'isométries $\OO(1,n)$.

L'espace anti-de~Sitter est solution des équations d'Einstein du vide avec
une constante cosmologique négative $-3/L^2$, tandis que de~Sitter l'est
avec une constante cosmologique $3/L^2$.

\subsubsection{Systèmes de coordonnées pour $AdS_n$}
Divers systèmes de coordonnées sont utilisés pour $AdS_n$, selon les
sous-espaces et les sous-groupes des isométries auxquels on s'intéresse.

Les coordonnées dites anti-de~Sitter s'obtiennent en posant
$X_{n-1}=L\sinh\psi$. Les sous-espaces $\psi=\text{constante}$ sont alors
des espaces $AdS_{n-1}$ de rayon $L\cosh\psi$, d'où la métrique
\begin{equation}
\label{AdSAdS}
ds^2 = L^2 \left[ d\psi^2 + \cosh^2\psi\,ds^2(AdS_{n-1}) \right]
\end{equation}
étant entendu que $ds^2(AdS_{n-1})$ désigne la métrique d'un espace $AdS$
de rayon unité.

Les coordonnées dites de~Sitter s'obtiennent en posant $X_n=L\cosh\psi$.
Les sous-espaces $\psi=\text{constante}$ sont alors des espaces $dS_{n-1}$
de rayon $L\sinh\psi$, d'où la métrique
\begin{equation}
\label{AdSdS}
ds^2 = L^2 \left[ d\psi^2 + \sinh^2\psi\,ds^2(dS_{n-1}) \right].
\end{equation}
Ces coordonnées ne couvrent pas l'hyperboloïde en entier.

Les coordonnées Poincaré s'obtiennent en posant $X_n+X_{n-1}=L^2u$ et
$X_i=Lux_i$ pour $0\leq i\leq n-2$. L'équation \eqref{AdSdef} entraîne
alors $X_n-X_{n-1}=u^{-1}+u(-x_0^2+\ldots+x_{n-2}^2)$, d'où la
métrique :
\begin{equation}
\label{AdSPoin}
ds^2 = L^2 \left[ \frac{du^2}{u^2} + u^2 (-dx_0^2 + \ldots + dx_{n-2}^2)
       \right].
\end{equation}

\section{Localisation de la gravité sur une brane}
\subsection{Cas de la gravité pure}
\index{localisation de la gravité|(}
On considère un espace à 5 dimensions avec une 3-brane massive. L'action
s'écrit :
\begin{equation}
\label{AcLoc}
S \propto \int d^5x \sqrt{-\det g} \left[ -\frac{1}{4}R - \Lambda_5 \right]
          - \lambda \int d^4x \sqrt{-\det \hat{g}}
\end{equation}
où $\Lambda_5=-3/L^2$ est la constante cosmologique, prise négative,
$\lambda$ est la tension de la brane et $\hat{g}$ est la métrique induite
sur la brane. On se place ici dans l'approximation de la brane fine
(d'épaisseur nulle), qui facilite grandement la résolution des équations du
mouvement.

Comme on cherche à avoir une brane maximalement symétrique (c'est-à-dire
avec un groupe d'isométries le plus grand possible), donc Minkowski ou
(anti-)de~Sitter, on va chercher des solutions ayant les symétries de
$\R^{1,3}$ ou $(A)dS_4$. On considère donc des solutions de la forme
\begin{equation}
\label{MetLoc}
ds^2 = A(r) \bar{g}_{ij} dx^i dx^j + B(r) dr^2
\end{equation}
où $r$ est une coordonnée telle que la brane est en $r=0$, et $\bar{g}$ est
une métrique Minkowski, de~Sitter ou anti-de~Sitter.

Un moyen pour trouver la solution est de remplacer la métrique par l'ansatz
\eqref{MetLoc} dans l'action \eqref{AcLoc}, et d'intégrer sur les quatre
dimensions $x_i$, de façon à obtenir une action à une dimension. On peut
alors varier l'action par rapport à $A(r)$ et $B(r)$, ce qui donne deux
équations pour $A$ après avoir imposé $B(r)=1$ (ce qui est toujours
possible par redéfinition de $r$) et il n'est alors pas très difficile de
trouver les solutions. On pourra trouver le détail des calculs dans
l'annexe de \cite{Kal}.

Ayant choisi $A(0)=1$ (par choix de normalisation pour $\bar{g}$), on peut
alors écrire la solution \cite{KR} :
\begin{align*}
dS_4 : A &\propto \sinh^2 \frac{c-|r|}{L} &
       L_{dS_4} &= L \sinh \frac{c}{L} &
       \lambda &= \frac{3}{L}\coth\frac{c}{L} \\
M_4 : A &= \exp\!\left(\! -\frac{2|r|}{L} \right) & &&
      \lambda &= \frac{3}{L} \\
AdS_4 : A &\propto \cosh^2 \frac{c-|r|}{L} &
        L_{AdS_4} &= L \cosh \frac{c}{L} &
        \lambda &= \frac{3}{L}\tanh\frac{c}{L} \\
\end{align*}
où $c$ est un paramètre réel positif. On voit que pour une tension faible,
la brane est anti-de Sitter, qu'elle devient Minkowski à la tension
critique $\lambda=3/L$, et enfin de Sitter pour une tension élevée.
L'espace à 5 dimensions, quant à lui, est constitué de deux morceaux
d'espace $AdS_5$ de rayon $L$ collés au niveau de la brane, la métrique
\eqref{MetLoc} pour $r>0$ ou $r<0$ n'étant rien d'autre que celle d'$AdS_5$
dans l'un des systèmes de coordonnées \eqref{AdSAdS}, \eqref{AdSdS},
\eqref{AdSPoin}. (Dans le cas Poincaré, il faut poser
$u=L^{-1}\exp(-|r|/L)$.)

Pour montrer la localisation de la gravité sur une telle brane, on regarde
les fluctuations linéarisées de la métrique autour de la solution
ci-dessus. Dans le cas des branes de Sitter et Minkowski, on trouve (cf
\cite{KR}) qu'il y a un état lié de masse nulle confiné sur la brane ; dans
le cas anti-de~Sitter, ce n'est pas le cas, mais le spectre des
fluctuations est discret, et, pour $\lambda$ suffisamment proche de $3/L$
(c'est-à-dire une courbure faible pour la brane), l'un des modes, qualifié
de mode presque zéro, est très léger devant les autres, et est aussi
confiné sur la brane. Dans tous les cas, le mode (presque) zéro reproduit
la gravité à quatre dimensions à basse énergie.

Le fait qu'un mode presque zéro puisse reproduire la gravité peut
surprendre, dans la mesure où un graviton massif à quatre dimensions a cinq
polarisations, alors qu'un graviton de masse nulle en a deux, et il en
découle que la limite $m\rightarrow0$ d'un graviton massif n'est pas un
graviton de masse nulle (discontinuité de Van~Dam-Veltman-Zakharov
\cite{VDVZ}). On montre néanmoins que, dans ce cas précis, il est possible
d'éliminer les modes en trop, de sorte que le problème ne se pose pas (cf
\cite{KKR}).
\index{localisation de la gravité|)}

\subsection{Réalisation cordiste}
\subsubsection{Obtention d'$AdS_5$}
\index{anti-de Sitter!$AdS_5$}
Considérons $N$ D3-branes plates superposées en théorie IIB, vérifiant
$x_3=x_4=x_5=x_7=x_8=x_9=0$. Comme nous l'avons vu précédemment, les
D3-branes sont chargées sous le champ de Ramond-Ramond $C_{(4)}$. On peut
explicitement trouver la solution de supergravité correspondante :
\begin{align*}
ds^2 &= f^{-1/2} (-dx_0^2 + dx_1^2 + dx_2^2 + dx_6^2)
        + f^{1/2} (dr^2 + r^2 ds^2(S^5)) \\
F_{(5)} &= (1 + *)\, dx_0\,dx_1\,dx_2\,dx_6\,df^{-1} \\
f &= 1 + \frac{L^4}{r^4} \quad\text{avec}\quad L^4 \equiv 4\pi g_s \a'^2 N
\end{align*}
où $r^2=x_3^2+x_4^2+x_5^2+x_7^2+x_8^2+x_9^2$.

Dans la limite, dite proche de l'horizon, où $r$ tend vers zéro, le terme
constant de $f$ disparaît et on obtient
\[
ds^2 = \frac{r^2}{L^2} (- dx_0^2 + dx_1^2 + dx_2^2 + dx_6^2)
       + L^2\, \frac{dr^2}{r^2} + L^2 ds^2(S^5).
\]
En posant $u=L^{-2}r$, on voit qu'il s'agit de $AdS_5\times S^5$, la
métrique de $AdS_5$ étant exprimée dans les coordonnées de Poincaré
\eqref{AdSPoin}, et où $AdS_5$ et $S^5$ ont comme rayon $L$.

Il résulte de ce fait une conjecture de dualité entre la théorie des cordes
sur $AdS_5\times S^5$ et la théorie de Yang-Mills maximalement
supersymétrique à quatre dimensions, qui est la théorie des champs
effective vivant sur la D3-brane ; cette dernière étant invariante conforme,
la conjecture est nommée « AdS/CFT ». Le lecteur pourra se reporter à
\cite{AdSCFT} pour plus de détails.

\subsubsection{$AdS_4$ dans $AdS_5$}
\index{D-brane!dans $AdS_5\times S^5$}
Dans la géométrie ci-dessus, l'hypersurface $x_6=0$ est un $AdS_4$. Pour
obtenir cela à partir de D-branes, il faut donc ajouter aux D3-branes
ci-dessus une D-brane dont l'intersection avec les D3-branes s'étend selon
les dimensions 0, 1 et 2, et vérifie $x_6=0$. D'autre part, pour assurer la
stabilité de cette configuration, on veut préserver une partie de la
supersymétrie ; on montre que pour cela, dans le cas de D-branes
perpendiculaires, il faut que le nombre de dimensions vérifiant une
condition de Neumann pour l'une des D-branes et Dirichlet pour l'autre soit
multiple de 4. Il est alors aisé de voir que cela est réalisé si on ajoute
aux D3-branes des D5-branes données par $x_6=x_7=x_8=x_9=0$. Si on néglige
l'effet des D5-branes sur la géométrie de l'espace ambiant, alors, dans la
limite proche de l'horizon de la D3-brane, on obtient des D-branes
$AdS_4\times S^2$ dans $AdS_5\times S^5$.

Bien sûr, pour obtenir la localisation de la gravité au voisinage des
D5-branes, il ne faut pas négliger leur action sur la géométrie. Une étude
précise de cela nécessite d'écrire la solution de la supergravité en
présence des D3- et D5-branes. Malheureusement, cette solution n'est pas
connue : on ne connaît que des solutions où soit les D3, soit les D5 sont
délocalisées \cite{D3D5}. Néanmoins, comme expliqué dans \cite{KR2}, on a
de bonnes raisons de penser que la gravité est localisée sur les D5-branes,
au moins au voisinage des D3-branes, de façon similaire au cas en gravité
pure que nous avons vu, même si l'approximation de brane fine risque de ne
plus marcher lorsque la partie $AdS_4$ devient plate.

Puisqu'on ne sait pas prendre en compte la rétroaction des D5-branes sur
la géométrie, nous allons, dans la suite de ce chapitre, nous placer dans
l'approximation où il y a très peu de D5-branes (plus précisément
$g_sM\ll1$, où $M$ est le nombre de D5-branes), de sorte que nous pourrons
considérer que la géométrie est celle découlant des D3-branes. 

En utilisant l'action de Born-Infeld, on peut étudier semi-classiquement
les D-branes $AdS_4\times S^2$ dans $AdS_5\times S^5$. Tout d'abord, il est
aisé de voir que, outre les D-branes $x_6=0$, les D-branes données par
$x_6=C/u$, où $C$ est une constante, sont également des $AdS_4\times S^2$,
avec un rayon anti-de~Sitter égal à $L\sqrt{1+C^2}$, et un rayon de la
sphère maximal (égal à $L$). Ces rayons sont stabilisés par le flux
magnétique à travers la 2-sphère \cite{KR3} ; les calculs étant assez
similaires à ceux du chapitre précédent, nous ne les reproduisons pas ici.
On peut aussi calculer le spectre des fluctuations quadratiques de la
D-brane \cite{DFO}.

\index{champ de fond!Ramond-Ramond}
\index{S-dualite@S-dualité}
On voudrait, bien sûr, pouvoir aller au-delà des calculs semi-classiques.
Le problème est que nous avons ici un champ de fond R-R ($F_{(5)}$ en
l'occurrence) et que l'on ne sait pas écrire de modèle sigma sur la surface
d'univers de la corde comme on sait le faire pour des champs NS-NS. Un cas
intéressant à étudier de ce point de vue là est le cas similaire des
D-branes $AdS_2\times S^2$ dans $AdS_3\times S^3$. En effet, une telle
configuration peut s'obtenir avec des champs de fond R-R ($F_{(3)}$)
similaires à ce que nous avons vu précédemment. Comme la S-dualité que nous
avons vue dans le premier chapitre échange $F_{(3)}=dC_{(2)}$ et $H=dB$,
cette configuration est S-duale à une configuration ne comportant que des
fonds NS-NS qui est un modèle de WZW, où on sait que les calculs
semi-classiques donnent des résultats exacts.

Nous sommes donc amené à considérer des D-branes $AdS_2\times S^2$ dans
$AdS_3\times S^3$ avec des champs de fond NS-NS. Si on cherche à obtenir
quelque chose qui ressemble à un univers branaire, il est intéressant de
savoir quelle est la géométrie effective vue par les champs vivant sur la
D-brane, c'est-à-dire les cordes ouvertes. Nous verrons que, comme montré
dans \cite{BP}, les rayons effectifs des parties anti-de~Sitter et
sphérique sont égaux, même si les rayons dans la géométrie ambiante sont
très différents. 

Nous étudierons ensuite cette même configuration, mais avec des fonds R-R
au lieu de NS-NS. Nous verrons que le calcul des petites fluctuations à
l'aide de l'action de Born-Infeld donne les mêmes fluctuations que dans les
fonds NS-NS, ce qui nous permet de vérifier la S-dualité dans ce cas
particulier. Ce calcul permet également de montrer que, comme dans le cas
NS-NS, on a une géométrie effective pour les cordes ouvertes, et que les
deux rayons sont aussi égaux. Nous en discuterons les conséquences.

\section{D-branes dans $AdS_3\times S^3$ avec des champs de fond NS-NS}
\index{anti-de Sitter!$AdS_3$}
\subsection{Obtention d'$AdS_3\times S^3$}
La géométrie $AdS_3\times S^3$ s'obtient à partir d'un état lié de $Q_5$
NS5-branes enroulées sur un 4-tore et de $Q_1$ cordes fondamentales le long
de la dimension non compacte des NS5-branes. On montre que la géométrie
proche de l'horizon d'une telle configuration est $AdS_3\times S^3\times
T^4$, où les rayons d'$AdS_3$ et de $S^3$ sont égaux à
$L\equiv\sqrt{Q_5\a'}$. Dans la suite, nous considérerons le $T^4$ comme
spectateur, et nous ne nous en occuperons pas.

\index{Wess-Zumino-Witten, modèle de}
$AdS_3$ et $S^3$ étant des groupes, respectivement le recouvrement
universel de $\SL(2,\R)$ et $\SU(2)$, et puisque nous avons ici une
configuration sans charge R-R, les champs de fond sont donnés par le modèle
de Wess-Zumino-Witten vu au chapitre précédent. Dans des coordonnées
appropriées,
\index{champ de fond!$B_{\mu\nu}$}
\begin{align*}
ds^2 &= L^2 \left[ d\psi^2 + \cosh^2\psi\,(d\omega^2 - \cosh^2\omega\,d\tau^2)
        + d\chi^2 + \sin^2\chi\,(d\theta^2 + \sin^2\theta\,d\phi^2) \right] \\
B &= L^2 \left[ \left( \psi + \frac{\sinh 2\psi}{2} \right)
               \cosh\omega\,d\omega \wedge d\tau
             + \left( \chi - \frac{\sin 2\chi}{2} \right)
               \sin\theta\,d\theta \wedge d\phi \right]
\end{align*}
et le dilaton est une constante $\Phi=\Phi_{\text{NS}}$.

En ce qui concerne le spectre, ce que nous avons vu dans le cas de groupes
compacts s'applique ici. Il y a tout de même quelques différences : d'une
part, il y a un nombre infini de représentations de $\SL(2,\R)$ qui
interviennent, et d'autre part il y a des subtilités à cause du fait que le
groupes des transformations \eqref{SymWZW} n'est pas connexe. Le lecteur
pourra se référer à \cite{ThSR,AdS3} pour plus de détails.

\subsection{D-branes dans $AdS_3$}
\index{D-brane!dans $AdS_3\times S^3$|(}
Les D-branes dans $AdS_3\times S^3$ s'obtiennent comme produit d'une
D-brane dans $AdS_3$ et d'une D-brane dans $S^3$. Comme nous avons déjà vu
les D-branes dans $S^3$ dans le chapitre précédent, nous allons maintenant
étudier les D-branes dans $AdS_3$.

\index{condition aux bords!WZW}
Comme dans le cas de la 3-sphère, la condition aux bords $J=-\bar{J}$ est
réalisée par des classes de conjugaison de $\SL(2,\R)$. Dans les
coordonnées \eqref{AdSdef}, elles sont données par $X_0=\text{constante}$,
et sont des espaces hyperboliques (pour $X_0<L$) ou de Sitter (pour
$X_0>L$). Dans le premier cas, il s'agit de surfaces entièrement de genre
espace, donc de D-instantons (localisés dans le temps), dont
l'interprétation n'est pas claire. Dans le deuxième cas, le champ
électrique qu'il faut mettre sur la D-brane pour qu'elle minimise le
lagrangien de Born-Infeld \eqref{AcBI} est tel que l'expression sous la
racine carrée est négative, ce qui rend ces D-branes non physiques
\cite{BP}.

Au lieu de la condition aux bords $J=-\bar{J}$, on peut prendre
$J=-\Omega(\bar{J})$, où $\Omega$ est un automorphisme de $\SL(2,\R)$. S'il
s'agit d'un automorphisme intérieur, c'est à dire de la forme
$\Omega(g)=h_0gh_0^{-1}$, avec $h_0\in\SL(2,\R)$, on obtient
essentiellement les mêmes D-branes qu'auparavant après une rotation sur le
groupe. Dans le cas de $\SU(2)$, tous les automorphismes sont intérieurs,
et c'est pour cela que nous n'avions pas évoqué ce point. En revanche,
$\SL(2,\R)$ admet des automorphismes extérieurs qui, modulo un
automorphisme intérieur, s'écrivent
\[
\Omega(g) = \begin{pmatrix} 0 & 1 \\ 1 & 0 \end{pmatrix} g
            \begin{pmatrix} 0 & 1 \\ 1 & 0 \end{pmatrix}.
\]
Les D-branes qui en découlent sont des classes de $\Omega$-conjugaison, la
classe d'un élément $g$ étant définie par $\{hg\Omega(h)^{-1}, h\in G\}$.
Ici, il s'agit de surfaces $AdS_2$ données, dans les coordonnées
\eqref{AdSAdS}, par $\psi=\psi_0$. Le champ électromagnétique
nécessaire à leur stabilisation s'écrit
\index{champ de fond!électromagnétique}
\[
F = -\frac{L^2}{2\pi\a'} \psi_0 \cosh\omega\, d\omega \wedge d\tau.
\]

Sachant qu'on a affaire ici à des D-branes non compactes, on pourrait
croire que, contrairement aux 2-sphères dans la 3-sphère, les positions de
ces D-branes ne sont pas quantifiées. En fait il n'en est rien : en effet,
de manière générale, la charge sous $B$ d'une corde (pas forcément
fondamentale) vaut
\[
q = 2\pi\a' \frac{\delta S}{\delta B_{\omega\tau}}
\]
où $S$ est l'action effective de la D-corde \eqref{BIWZ}. Elle doit être
multiple de la charge d'une corde fondamentale, c'est-à-dire, avec la
normalisation choisie, entière. Dans le cas qui nous concerne, on a
$q=-e^{-\Phi}\sinh\psi_0$, d'où une quantification de la position de la
D-brane. D'autre part, cette charge non nulle signifie qu'on a en fait
affaire à un état lié d'une D1-brane et $q$ cordes fondamentales, baptisé
corde $(1,q)$. Il convient de noter que cette quantification est de nature
différente de celle des 2-sphères dans la 3-sphère : notamment, elle est
invisible si on fait tendre le couplage vers zéro, donc on ne verra rien
dans la théorie conforme.

\subsection{$AdS_2\times S^2$ dans $AdS_3\times S^3$}
Les D-branes symétriques dans $AdS_3\times S^3$ sont, d'après ce qui
précède, données par $\psi=\psi_0$ et $\chi=\chi_0$. Le champ
électromagnétique est la somme des deux champs stabilisant respectivement
$AdS_2$ et $S^2$, donc
\[
F = -\frac{L^2}{2\pi\a'} (\psi_0 \cosh\omega\, d\omega \wedge d\tau
                           + \chi_0 \sin\theta\, d\theta \wedge d\phi).
\]

Pour une D3-brane de topologie $M\times S^2$, la charge sous $B$
s'écrit :
\[
q = 2\pi\a' \int_{S^2} d\theta\,d\phi \frac{\delta S}{\delta B_{\omega\tau}}
  \equiv \frac{1}{2\pi} \int_{S^2} d\theta\,d\phi \tilde{F}_{\theta\phi}.
\]
Le champ $\tilde{F}$ est, de manière générale, donné par
\[
\begin{split}
2\pi\a'\tilde{F}_{mn} &= \frac{1}{2}\,e^{-\Phi}\sqrt{-\det(\hat{g}+B+2\pi\a'F)}
                         [(\hat{g}+B+2\pi\a'F)^{-1}]^{[pq]} \eps_{mnpq} \\
                &\quad   + \hat{C}_{mn} + C_{(0)} (B_{mn} + 2\pi\a'F_{mn})
\end{split}
\]
où $\eps$ est le tenseur antisymétrique avec $\eps_{0123}=+1$. Pour les
branes qui nous intéressent ici, on a
$q=-\frac{L^2}{\pi\a'}e^{-\Phi}\sin\chi_0\sinh\psi_0$.

D'autre part, nous avons vu précédemment que $p=-\frac{1}{2\pi}\int_{S^2}
d\theta\,d\phi F_{\theta\phi}$ est quantifié. Ici,
$p=\frac{L^2}{\pi\a'}\chi_0$, d'où une quantification de $\chi_0$. On peut
montrer que les D-branes $S^2$ dans $S^3$ peuvent s'interpréter comme des
états liés de $p$ D-branes avec deux dimensions de moins ; on en déduit que
nos D-branes sont des cordes $(p,q)$.

On peut calculer le spectre des petites fluctuations de la D-brane avec
l'action de Born-Infeld, comme on l'a fait dans le chapitre précédent. 
Nous ne reproduirons pas ici les calculs, et laissons le lecteur se référer
à \cite{PR}, où il est également montré que le spectre trouvé est en accord
avec ce que l'on sait faire en CFT, à cela près que, comme dans le chapitre
précédent, le calcul semi-classique ne donne pas de borne supérieure pour
le spin des représentations de $\SU(2)$ permises.

\index{metrique@métrique!de corde ouverte}
Considérons maintenant la géométrie effective, c'est-à-dire vue par les
cordes ouvertes. Cette géométrie est, comme nous l'avons vu dans le premier
chapitre, donnée par la métrique de corde ouverte \eqref{MetOuv}. Il est
aisé de voir que la géométrie effective de notre D-brane est
$AdS_2\times S^2$, comme sa géométrie induite, mais avec comme rayon $L$
aussi bien pour $AdS_2$ que pour $S^2$, alors que les rayons induits sont
respectivement $L\sin\chi_0$ et $L\cosh\psi_0$. Ainsi, alors que, du point
de vue de la géométrie induite, il est possible d'avoir un $AdS_2$ presque
plat et une 2-sphère très petite, cela n'est en fait pas possible dans la
géométrie vue par les cordes ouvertes. Cela a évidemment une grande
importance si on veut obtenir finalement un univers branaire ; nous y
reviendrons lors de l'étude avec des fonds R-R.

\section{D-branes dans $AdS_3\times S^3$ avec des champs de fond R-R}
\subsection{Champs de fond}
\index{champ de fond!Ramond-Ramond}
Le fond R-R s'obtient à partir du fond NS-NS par la transformation de
S-dualité $C_{(2)}\rightarrow B$, $B\rightarrow -C_{(2)}$,
$F\rightarrow\tilde{F}$ et $\t\rightarrow -\frac{1}{\t}$, où
$\t=C_{(0)}+ie^{-\Phi}$. Il s'agit d'un état lié de $Q_5$ D5-branes et
$Q_1$ D1-branes. On montre alors (cf l'appendice de \citemoi{C}) que
$\tilde{F}$ est transformé en $-F$, ce qui garantit que la dualité préserve
l'équation du mouvement pour $F$, puisqu'elle s'écrit $d\tilde{F}=0$.
Ainsi, notre corde $(p,q)$ devient une corde $(-q,p)$, ce qui signifie que,
à un signe près, les charges de D-corde et de corde fondamentale sont
échangés.

Les champs ainsi obtenus sont
\begin{align*}
\Phi &= -\Phi_{\text{NS}} \\
C_{(2)} &= B_{\text{NS}} \\
F &= \frac{e^{\Phi}L_{\text{NS}}^2}{2\pi\a'}
     (\cos\chi_0 \cosh\psi_0 \cosh\omega\,d\omega\wedge d\tau
     + \sin\chi_0 \sinh\psi_0 \sin\theta\,d\theta\wedge d\phi) \\
ds^2 &= e^{\Phi} ds^2_{\text{NS}}.
\end{align*}
Vu la forme de la métrique, dans la suite, nous redéfinissons
$L^2\equiv L^2_{\text{RR}}=e^{\Phi}L^2_{\text{NS}}$.

\subsection{Petites fluctuations et métrique effective}
\index{Born-Infeld, action de!avec terme de Wess-Zumino}
Puisque la S-dualité est une symétrie des équations du mouvement découlant
de l'action effective à basse énergie, on s'attend à ce que les
représentations de $\SL(2,\R)\times\SU(2)$ présentes dans le spectre soient
les mêmes. Il est néanmoins utile de vérifier que la S-dualité fonctionne
bien. Pour cela, comme précédemment, on considère les fluctuations
quadratiques de l'action effective \eqref{BIWZ}. Il apparaît alors qu'après
un changement de variables approprié :
\begin{align*}
\delta\tilde{\psi} &= \sinh\psi_0 \cos\chi_0 \,\delta\psi
                     - \cosh\psi_0 \sin\chi_0 \,\delta\chi \\
\delta\tilde{\chi} &= \cosh\psi_0 \sin\chi_0 \,\delta\psi
                      + \sinh\psi_0 \cos\chi_0 \,\delta\chi
\end{align*}
les termes quadratiques sont essentiellement identiques au cas NS-NS, en
remplaçant $\delta\psi$ par $\delta\tilde{\psi}$ et $\delta\chi$ par
$\delta\tilde{\chi}$. Il en découle que le spectre est bien le même.

La S-dualité nous donne des informations supplémentaires qui ne découlent
pas de l'action effective. En effet, nous avons vu que, dans le cas NS-NS,
le spin maximal des représentations présentes est égal à la moitié de la
charge magnétique $p$ de la D-brane sous le champ $B$. Il en découle que,
si la S-dualité est vraie, dans le cas RR le spin maximal est la moitié de
$p$, qui est maintenant la charge électrique sous $B$. Un tel résultat ne
pouvait être trouvé directement car il n'y a pas de modèle sigma connu pour
la surface d'univers des cordes en présence d'un fond RR non nul.

\index{metrique@métrique!de corde ouverte}
Pour ce qui concerne la géométrie effective, il se pose le problème qu'a
priori, la formule \eqref{MetOuv} n'a pas de raison d'être valide en
présence de champs R-R. Il se pose donc la question de savoir s'il y a une
notion de métrique effective dans un tel cas. Dans le développement en
fluctuations de l'action effective apparaît naturellement une métrique
$AdS_2\times S^2$ avec des rayons égaux. Mais ceci ne dit rien sur la
normalisation de cette métrique, donc sur la valeur du rayon commun. Pour
le trouver, on va, comme on avait fait dans le cas de la 3-sphère, ajouter
une dimension supplémentaire $x$, qui sera ici spatiale puisqu'il y a une
dimension temporelle dans $AdS_2$, et considérer une D-brane qui s'étend le
long de cette dimension (c'est-à-dire qu'on fait une T-dualité le long
d'une dimension orthogonale à $AdS_3\times S^3$). L'action effective
s'exprime alors naturellement en termes de la métrique
\[
ds^2_{\text{eff}} = dx^2 + L^2(\sin^2\chi_0 + \sinh^2\psi_0)
   (d\omega^2 - \cosh^2\omega\,d\tau^2 + d\theta^2 + \sin^2\theta\,d\phi^2).
\]
Sachant que la géométrie effective de la direction $x$ doit être la même
que sa géométrie réelle (car il n'y a aucun champ de fond qui en dépende),
on conclut que cette expression donne la métrique effective avec la bonne
normalisation, d'où le rayon effectif
\[
R^2 = L^2(\sin^2\chi_0 + \sinh^2\psi_0).
\]
On peut vérifier aisément que c'est ce que l'on trouve en utilisant la
formule \eqref{MetOuv} sans aucune correction. La question de savoir si
cela se généralise reste ouverte.

Ainsi, la situation est différente du cas NS-NS, puisque le rayon dépend de
la position de la D-brane, et peut prendre n'importe quelle valeur. Cela ne
contredit pas la S-dualité, puisqu'il s'agit d'une symétrie des équations
du mouvement, et non de l'action.

\index{supersymetrie@supersymétrie d'espace-temps}
L'égalité des rayons des parties anti-de Sitter et sphérique de la
géométrie posent problème dans le cadre d'un univers branaire à la
Randall-Sundrum : ce que l'on voudrait, en effet, c'est un grand rayon
(c'est-à-dire une échelle cosmologique) pour la partie anti-de Sitter, afin
qu'elle soit presque plate, conformément aux observations, et un petit
rayon pour la partie sphérique, c'est-à-dire plus petit que l'échelle où la
loi de Newton est vérifiée expérimentalement (environ 0,1~mm). L'égalité
des rayons est reliée à la supersymétrie : en effet, l'existence d'une
théorie des champs supersymétrique dans une géométrie requiert l'existence
d'un spineur covariantement constant, ce qui n'est possible ici que si les
deux rayons sont égaux. Plus précisément, c'est de supersymétrie étendue
qu'il s'agit puisque, dans la théorie des champs obtenue en supprimant la
sphère, les isométries de la sphère deviennent des R-symétries. Par
exemple, dans le cas de la D-brane $AdS_4\times S^2$ considérée
initialement, on a une R-symétrie $\OO(4)\cong\OO(3)\times \OO(3)$, le
premier $\OO(3)$ provenant de la 3-sphère et le deuxième des dimensions
extérieures à la D-brane, d'où une supersymétrie $\N=4$ à 4 dimensions (16
supercharges). Il semblerait donc que l'obtention d'univers branaires
réalistes requière une supersymétrie $\N\leq1$. Il faudrait pouvoir faire
le calcul sans négliger la réaction de la D-brane sur la géométrie ambiante
pour voir ce qui se passe dans ce cas.

Remarquons toutefois que puisque, dans le cas de $AdS_2\times S^2$ avec des
fonds R-R, on peut obtenir un rayon très grand, on a là un moyen d'obtenir
un espace plat à quatre dimensions (cf la remarque à la fin de
\cite{AdSCFT}). La question de savoir si cela peut donner lieu à un univers
branaire réaliste mérite certainement d'être étudiée.
\index{D-brane!dans $AdS_3\times S^3$|)}

%% file: pulse.tex
\chapter{Cordes ouvertes dans une onde plane}

Jusqu'à présent, nous nous sommes intéressé à des fonds indépendants du
temps. Les fonds dépendants du temps posent en effet des problèmes à la
fois techniques et conceptuels, qui viennent notamment du fait que la
théorie des cordes perturbative est formulée comme une théorie de
matrice~$S$, alors qu'on ne peut pas toujours définir d'états asymptotiques
dans une géométrie dépendant du temps.

Parmi les fonds les plus étudiés, il y a les géométries obtenues en
quotientant l'espace plat par un boost, éventuellement de genre lumière
\cite{TDOrb}, l'intérêt étant que l'espace est plat, ce qui simplifie
grandement la quantification de la corde. Ces fonds sont des modèles jouets
pour un « rebond cosmologique » (un Big Crunch suivi d'un Big Bang), et
peuvent contenir une singularité de genre espace ou lumière ; la résolution
de telles singularités en théorie des cordes reste un problème ouvert.

\begin{figure}
\begin{center}
\scalebox{0.7}{\input{ciseaux.pstex_t}}
\end{center}
\caption{Ciseaux de genre lumière régularisés.}
\label{ciseaux}
\end{figure}
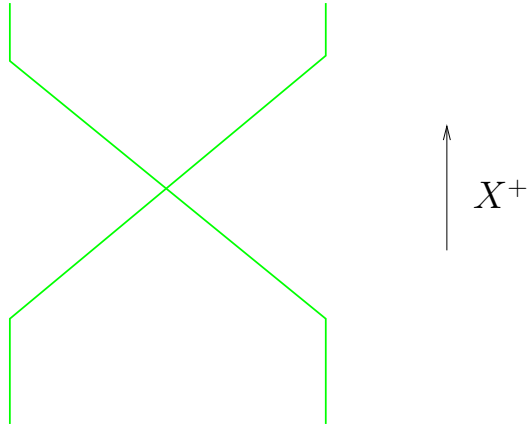

On peut aborder ces problèmes en considérant un analogue en termes de
cordes ouvertes \cite{BH}. Il s'agit d'une configuration, dite
\emph{ciseaux de genre lumière}, où deux D-branes qui se coupent bougent
l'une par rapport à l'autre de sorte que les points d'intersection vont à
la vitesse de la lumière, le tout étant plongé dans un espace de Minkowski
ordinaire. L'intérêt d'une telle étude est que l'on peut isoler les effets
cordistes de ceux dûs à une forte gravité, ces derniers étant absents dans
le cas ouvert, et qu'on peut régulariser le comportement à grands temps en
rendant les D-branes parallèles dans le passé et le futur lointains (fig.
\ref{ciseaux}) --- cela est analogue à stopper l'expansion de l'Univers
dans le cas du quotient ci-dessus, et permet d'avoir des états
asymptotiques.

Les ciseaux de genre lumière régularisés sont un cas particulier de
D-branes portant une onde plane, qui font l'objet de ce chapitre. L'intérêt
de telles D-branes est que, moyennant un choix de jauge particulier, la
\emph{jauge du cône de lumière}, la théorie des champs sur la surface
d'univers est libre avec des sources, ce qui permet de calculer diverses
quantités explicitement. On pourra donc, en particulier, envisager, dans le
cas des ciseaux de genre lumière régularisés décrits ci-dessus, de voir
s'il y a une divergence infrarouge dans l'émission de cordes ouvertes de
masse nulle par une corde tendue entre les deux D-branes, ce qui serait le
signe d'une forte rétroaction à tout couplage. On s'attend à ce que ce soit
le cas au moins dans la limite où la distance entre les D-branes tend vers
l'infini, car alors l'énergie d'une corde tendue entre elles tendrait aussi
vers l'infini, et il faut bien qu'elle prenne cette énergie quelque part,
d'où une rétroaction sur les D-branes. Un objectif de ce chapitre sera donc
de voir comment aborder ce genre de calcul.

\section{Quantification de la corde dans la jauge du cône de lumière}
\index{quantification!dans la jauge du cône de lumière}
Au début du chapitre 1, nous avons vu qu'après fixation de la métrique dans
l'action de Polyakov \eqref{AcPol}, il reste des symétries, qui sont les
transformations conformes des coordonnées. Nous allons voir que, moyennant
un abandon de la covariance manifeste de Lorentz de l'espace-temps cible,
on peut fixer complètement la jauge et imposer simplement la contrainte
\eqref{contrainte}.

La forme générale des transformations conformes est
\[
\s^+ \rightarrow \tilde{\s}^+(\s^+),\ \s^- \rightarrow \tilde{\s}^-(\s^-)
\]
avec $\s^\pm \equiv \t\pm\s$. Il en découle que $\t$ et $\s$ se
transforment en
\begin{align*}
\tilde{\t} &= \frac{1}{2}[\tilde{\s}^+(\t+\s)+\tilde{\s}^-(\t-\s)] \\
\tilde{\s} &= \frac{1}{2}[\tilde{\s}^+(\t+\s)-\tilde{\s}^-(\t-\s)].
\end{align*}
Ceci équivaut à demander que $\tilde{\t}$ soit une solution quelconque de
l'équation $\Box\tilde{\tau} = 0$, et $\tilde\s$ est alors fixé à une
constante additive près.

Dans le cas d'une corde ouverte, le fait que $\tilde{\s}$ soit constant aux
bords entraîne la condition aux bords $\d_{\s}\tilde\t$. Ces deux
conditions sont, par les équations du mouvement, vérifiées par toute
combinaison linéaire des coordonnées $X^\mu$ ; pour simplifier l'expression
de la contrainte \eqref{contrainte}, on prend $\tilde\t = X^+/(2\a'p^+)$,
la normalisation étant nécessaire pour que $\tilde\s$ ait la bonne
normalisation (c'est-à-dire un écart de $\pi$ entre les bords), ce qui
revient à considérer qu'on prend
\[
X^+ = 2\a' p^+ \t,
\]
où
\[
X^\pm = \frac{X^0 \pm X^1}{\sqrt{2}}
\]
sont les coordonnées du cône de lumière.

La contrainte \eqref{contrainte} s'écrit alors
\[
\d_\pm X^- = \frac{1}{p^+} (\d_\pm X^i)^2,
\]
$i$ parcourant les indices $2$ à $D-1$, de sorte que $X^-$ est complètement
déterminé à une constante additive près. Les variables dynamiques sont donc
le mode zéro de $X^-$ et les $D-2$ champs $X^i$.

Il découle de l'équation précédente que l'impulsion $p^-$ s'écrit :
\[
p^- = \frac{1}{2p^+} \left[ (p^i)^2 + \frac{1}{\a'} \left(
            \sum_{i=2}^{D-1}\sum_{m>0} \a_{-m}^i \a_m^i - a \right) \right]
\]
où, comme au premier chapitre, nous avons introduit une constante $a$ à
cause de l'ordre normal. Il en découle que la masse des états vaut
\[
m^2 = 2p^+p^- - (p^i)^2 = \frac{1}{\a'} (N-a)
\]
où $N$ est le nombre d'excitations selon les coordonnées 2 à $D-1$.

Il se pose alors la question de l'invariance de Lorentz de cette théorie.
Puisque nous l'avons obtenue à partir d'une théorie invariante de Lorentz,
il est nécessaire à la cohérence qu'elle le soit. On peut calculer
explicitement les générateurs du groupe de Lorentz, et on constate alors
que le commutateur $[J^{i-},J^{j-}]$, qui devrait être nul, ne l'est
effectivement que si $D=26$ et $a=1$, c'est-à-dire précisément ce que nous
avons trouvé lors de la quantification covariante, et le spectre est donc
le même. Notons toutefois qu'il n'y a pas à imposer de conditions sur les
états comme lors de la quantification covariante, et donc on obtient les
états simplement en faisant agir les $\a^i_{-n}$ sur un état non excité.

Dans le cas de la corde fermée, les choses sont similaires, avec des
parties gauche et droite indépendantes, à ceci près qu'on doit imposer la
condition $N=\tilde{N}$. Nous ne détaillerons pas plus.

\index{D-brane!ondulante|(}
\section{Corde libre sur une D-corde ondulante}
\subsection{Description du fond}
\begin{figure}
\begin{center}
\includegraphics[scale=0.7]{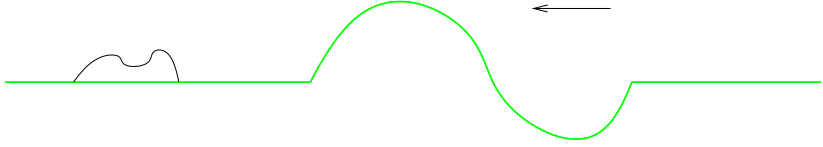}
\end{center}
\caption{Corde fondamentale accrochée à une D-corde ondulante.}
\label{pulse}
\end{figure}

La configuration qui nous intéresse ici est une D1-brane, paramétrée par
les coordonnées du cône de lumière $X^+$ et $X^-$. Les autres coordonnées
sont données par $X^i=Y^i(X^+)$, où les $Y^i$ sont des fonctions continues
à support compact (on s'attend à ce que les résultats s'étendent au cas de
fonctions décroissant « suffisamment vite »). Ceci décrit une impulsion
d'onde plane de déformation allant à la vitesse de la lumière (fig.
\ref{pulse}).
Le support compact des fonctions $Y^i$ permet, pour les cordes ouvertes sur
cette D-brane, des états asymptotiques à temps suffisamment petit ou grand
qui sont ceux d'une corde ouverte sur une D-brane plate donnée par $X^i=0$.
Le couplage de la corde avec l'onde sur la D-brane est donné, comme le
montre l'action \eqref{AcOuv}, par
\index{couplage!à une D-brane}
\[
S_I = \pm \frac{1}{2\pi\a'} \int d\t\, Y^i(X^+) \d_\s X_i.
\]
Cette action n'ayant pas de dépendance en $X^-$, l'impulsion $p^+$ est
conservée, et $X^+$ vérifie les mêmes équations du mouvement et conditions
aux bords qu'en l'absence d'onde. On peut donc passer dans la jauge du cône
de lumière comme expliqué précédemment et, du point de vue de la surface
d'univers, on a affaire à des champs libres $X^i$ couplés à des sources
dépendant du temps $Y^i(2\a'p^+\t)$. Pour alléger les notations, on prendra
dans la suite $2\a'=1$.

Par T-dualité, une telle onde est équivalente à une onde électromagnétique
plane $A_i(X^+)$ sur le volume d'univers d'une D-brane en dimension plus
élevée. Une grande partie de ce que nous allons faire ici pourra s'y
transposer facilement. Néanmoins, dans le cas électromagnétique, les
conditions aux bords de Neumann (modifiées par le champ) conduisent à
l'existence de modes zéros pour la corde qui compliqueraient quelque peu la
discussion ; aussi considérerons-nous le cas d'une onde de déformation
d'une D1-brane dans ce qui suit.

Par souci de simplicité, nous ne travaillerons ici que dans le cadre de la
corde bosonique. Signalons toutefois que dans le cas de la supercorde, le
terme d'interaction fermionique s'écrit uniquement en termes de $\psi^+$
(le partenaire supersymétrique de $X^+$), qui s'annule dans la jauge du
cône de lumière, de sorte que le terme d'interaction est in fine le même
que pour la corde bosonique. Il en découle que beaucoup de résultats de la
corde bosonique devraient pouvoir se transposer simplement à la supercorde.

Enfin, dans la mesure où les $X^i$ sont découplés entre eux, nous n'en
considérerons qu'un, et omettrons l'indice $i$.

\subsection{Fonctions de corrélation}
Nous considérons dans un premier temps une corde « libre », au sens où il
n'y a pas de coupure ou de fusion de cordes. Malgré cette absence
d'interaction entre cordes, la présence d'une source rend la matrice $S$
non triviale.

Pour calculer les amplitudes, on est amené à considérer des fonctions de
corrélation telles que
\begin{equation}
\<0_{out}| X(\t_1,\s_1) \ldots X(\t_n,\s_n) |0_{in}\>_S
= \<0_{out}| X(\t_1,\s_1) \ldots X(\t_n,\s_n) \exp(iS_I) |0_{in}\>,
\label{corgen}
\end{equation}
les états $|0_{in}\>$ et $|0_{out}\>$ étant en présence de la source dans
le premier cas, et en son absence dans le second. On a ici
$|0_{in}\>=|0_{out}\>$ (en l'absence de source), mais nous les distinguons
afin de pouvoir plus loin appliquer ce formalisme au cas des interactions.

On va se placer dans le cas d'une source générique, c'est-à-dire non
localisée au bord, et donc
\[
S_I = \int d\t\,d\s\, J(\t,\s) X(\t,\s)
\]
où $J$ est une fonction quelconque. L'équation du mouvement s'écrit alors
$\Box X=-\pi J$. Il en découle que
$\<0_{out}|X(\t,\s)|0_{in}\>_S/\<0_{out}|0_{in}\>_S$ doit également
vérifier cette équation. Cette condition, et le fait que cette quantité
doit s'annuler au bord (si $J$ ne contient pas de $\delta$ ou dérivée de
$\delta$ au bord) et ne contenir que des termes en $e^{-in\t}$ (resp.
$e^{+in\t}$), $n>0$, quand $\t\to+\infty$ (resp. $-\infty$), impose que
\index{propagateur}
\begin{equation}
\frac{\<0_{out}|X(\t,\s)|0_{in}\>_S}{\<0_{out}|0_{in}\>_S}
= i\int d\t' d\s' N(\t,\s;\t',\s') J(\t',\s')
\label{corJ}
\end{equation}
où $N$, le propagateur, doit vérifier
$\Box N(\t,\s;\t',\s')=i\pi\delta(\t-\t')\delta(\s-\s')$, s'annuler au bord
et vérifier les mêmes conditions que $\<X\>$ quand $\t\to\pm\infty$. Nous
l'expliciterons plus loin.

La fonction de corrélation $\<0_{out}|\exp(iS_I)|0_{in}\>$ vérifie
l'équation différentielle suivante :
\[
\begin{split}
\frac{\delta}{\delta J(\t,\s)} \<0_{out}| \exp(iS_I) |0_{in}\>
&= i\<0_{out}| X(\t,\s) \exp(iS_I) |0_{in}\> \\
&= -\int d\t' d\s' N(\t,\s;\t',\s') J(\t',\s')
                   \<0_{out}| \exp(iS_I) |0_{in}\>
\end{split}
\]
où la deuxième égalité découle de l'équation \eqref{corJ}. Cette équation
se résout aisément, et on trouve
\begin{equation}
\<0_{out}| \exp(iS_I) |0_{in}\> = C \exp\!\left[
   -\frac{1}{2} \int d\t'\,d\s'\,d\t''\,d\s''\,
                    N(\t',\s';\t'',\s'') J(\t',\s') J(\t'',\s'')
   \right]
\label{corexp}
\end{equation}
où $C$ est une constante indépendante de $J$. Bien sûr, dans le cas d'une
corde libre, ${C=\<0_{out}|0_{in}\>=1}$. À partir de cette amplitude, on
peut obtenir des fonctions de corrélations générales comme dans
l'équation~\eqref{corgen} en dérivant plusieurs fois \eqref{corexp} par
rapport à la source $J$. On peut notamment vérifier que
$\<0_{out}|X(\t,\s)X(\t',\s')|0_{in}\>=N(\t,\s;\t',\s')$.

Dans le cas qui nous intéresse ici, on a
\[
J(\t,\s) = -\frac{1}{\pi} (\delta'(\s) - \delta'(\s-\pi)) Y(p^+\t)
\]
d'où
\begin{equation}
\<0_{out}|0_{in}\>_S = C \exp\!\left[
   -\frac{1}{2\pi^2} \oint d\t' \oint d\t''\,
         \d_{\s'}\d_{\s''}N(\t',\s';\t'',\s'') Y(p^+\t') Y(p^+\t'')
   \right]
\label{corvide}
\end{equation}
où les intégrales sont sur le bord de la surface d'univers, et
\[
\frac{\<0_{out}| X(\t,\s) |0_{in}\>_S}{\<0_{out}|0_{in}\>_S}
= -\frac{i}{\pi} \oint d\t'\, \d_{\s'}N(\t,\s;\t',\s') Y(p^+\t').
\]

Jusqu'à présent, ce que nous avons dit ici est essentiellement indépendant
de la topologie de la surface d'univers, et nous l'utiliserons quand nous
étudierons le cas des interactions entre cordes en présence de l'onde. Le
propagateur, en revanche, en dépend. Dans le cas libre, il s'exprime
simplement comme la série suivante :
\index{propagateur!ruban}
\[
N(\t,\s;\t',\s') = \sum_{n>0} \frac{1}{n} e^{-(i+\eps)n|\t-\t'|}
                                          \sin(n\s) \sin(n\s')
\]
où $\eps$ est un petit réel strictement positif servant à faire converger
la somme, similaire à celui qu'on met habituellement dans le propagateur de
Feynman. Il est aisé de vérifier que ce propagateur satisfait toutes les
propriétés requises.

\subsection{Matrice $S$}
La présence de l'ondulation permet des transitions entre différents états
de la corde. On est donc amené à calculer des quantités du genre
\[
\<0_{out}| a^{out}_{k_1} \ldots a^{out}_{k_m}
           a^{in}_{-l_1} \ldots a^{in}_{-l_n} |0_{in}\>
\]
où les $a^{in}$ (resp. $a^{out}$) sont les opérateurs de création et
d'annihilation pour les états asymptotiques entrants (resp. sortants).

Sachant que $X(\t,\s)$ admet le développement en modes suivant :
\[
X(\t,\s) = \sum_{n\neq0} \frac{a_n}{n} e^{-in\t} \sin(n\s),
\]
on vérifie aisément que
\[
a_n = \lim_{\t\to\pm\infty}
       \frac{e^{in\t}}{\pi} \int d\s\, \sin(n\s) (nX + i\d_\t X)
\]
où la limite est en $+\infty$ pour les états sortants, et $-\infty$ pour
les états entrants. Ainsi, on peut obtenir un élément quelconque de la
matrice $S$ à partir de corrélateurs du type de l'équation~\eqref{corgen}.

Le calcul de la quantité \eqref{corvide}, dont le module au carré donne la
probabilité de non-excitation pour une corde initialement dans son état
fondamental, est fait dans \cite{BG} (dans le cas T-dual), et nous laissons
le lecteur s'y référer. On trouve, à une phase près (sans signification
physique)
\[
\<0_{out}|0_{in}\>_S = \sum_{\nposodd} \frac{8n}{(p^+)^2}
                 \left| \tilde{Y}\parfrac{n}{\,p^+} \right|^2,
\]
où $\tilde{Y}$ est la transformée de Fourier de $Y$,
\[
Y(x^+) = \int dp^-\, \tilde{Y}(p^-) e^{ip^-x^+}.
\]
Il s'agit du résultat que \cite{pulse} obtient en considérant directement
l'opérateur d'évolution $e^{i\int d\t\,H_I(\t)}$.

À partir de là, le calcul d'amplitudes entre des états quelconques est
aisé, et on peut vérifier que ce sont celles données par la matrice $S$
calculée en \cite{pulse} :
\[
S = \<0_{out}|0_{in}\>_S \ordnorm{\exp\left[ \frac{4i}{p^+}
    \sum_{n\text{ impair}} \tilde{Y}\parfrac{n}{\,p^+} a_n
    \right]}.
\]

\subsection{Calcul covariant en termes d'état de bord}
\index{etatb@état de bord!D-brane ondulante}
Il est possible de construire un état de bord correspondant à la D-corde
ondulante \cite{HTT,BG} : en effet, en partant de l'état de bord $|\D1\>$,
correspondant à la D-corde sans onde, il est aisé de voir que l'état
\begin{equation}
|B\> = \exp\biggl( \frac{1}{\pi}
   \int\limits_{\t_F=0} d\s_F\, Y^i(X^+) \d_{\t_F} X^i \biggr) |\D1\>,
\label{bordpulse}
\end{equation}
où $\t_F$ et $\s_F$ sont les coordonnées sur la corde fermée, vérifie les
conditions aux bords
\[
(X^i - Y^i(X^+)) \bigr|_{\t_F=0} |B\> = 0.
\]
(Noter que l'état de bord a un sens pour une surface d'univers euclidienne,
et qu'il faut donc, pour vérifier l'équation ci-dessus, utiliser les
relations de commutation euclidiennes
$\bigl[X(\s_F),\d_{\t_F} X(\s_F')\bigr]=\pi\delta(\s_F-\s_F')$.)

À partir du moment où on dispose d'un tel état de bord, on peut s'en servir
pour calculer des amplitudes en insérant des opérateurs de vertex entre
l'état de bord et le vide de cordes fermées. Ainsi, l'amplitude pour un
état tachyonique d'y rester est donnée par
\index{operateur!opérateur de vertex}
\begin{equation}
S_{0\to0} \propto \<0| \ordnorm{e^{-ip_\mu X^\mu(0,0)}}
                       \ordnorm{e^{ip_\mu X^\mu(0,\pi)}} |B\>
\label{amplcov}
\end{equation}
où, s'agissant de cordes fermées, la coordonnée $\s_F$ est périodique de
période $2\pi$. Nous avons utilisé le fait que le disque a une symétrie
conforme $\SL(2,\R)$ pour fixer la position des opérateurs de vertex. Quant
à la normalisation, elle sera fixée en imposant que $S_{0\to0}=1$ en
l'absence d'onde. Enfin, l'impulsion d'une corde ouverte étant
nécessairement longitudinale à la D-brane, l'indice $\mu$ ne prendra ici
que les valeurs $+$ et $-$.

Il faut tout d'abord préciser ce que l'on entend par le vide de cordes
fermées $\<0|$. Puisque nous considérons des amplitudes sans émission de
corde fermée, la prescription appropriée est naturellement que les
opérateurs d'impulsion, et notamment $p^-$ (conjugué du mode zéro de $X^+$,
noté $x^+$), s'annulent. Il en découle que le vide s'écrit
\begin{equation}
\<0| \propto \int dx^+\, \<x^+|
\label{videfer}
\end{equation}
où $\<x^+|$ est un état propre de $x^+$.

Dans la suite du calcul, on est amené à séparer les fréquences positives et
négatives de $X^\mu$. On pose donc $X^\mu(0,\s_F)=X^\mu_>+X^\mu_<+x^\mu$,
avec
\[
X^\mu_> = \frac{i}{2} \sum_{n>0} \frac{1}{n}
                    (a^\mu_n e^{-in\s_F} + \bar{a}^\mu_n e^{in\s_F})
\]
et une expression similaire pour $X^\mu_<$. De la même façon, nous serons
amené à introduire $\d_{\t_F} X^\mu_>$ et $\d_{\t_F} X^\mu_<$.

Sachant que $\<0|$ est annulé par $X^\mu_<$, et que $|\D1\>$ est annulé par
$X^\mu_>-X^\mu_<$ (pour $\mu\in\{+,-\}$), la bonne prescription pour
l'ordre normal est de mettre $X^\mu_<$ à gauche et $X^\mu_>-X^\mu_<$ à
droite. Pour cela, on écrit
\begin{equation}
\ordnorm{e^{ip_\mu X^\mu}} =
       e^{2ip_\mu X^\mu_<} e^{ip_\mu x^\mu} e^{ip_\mu(X^\mu_>-X^\mu_<)}.
\label{pulseordnorm}
\end{equation}
Dans l'expression \eqref{amplcov}, les facteurs en $X^\mu_<$ s'éliminent,
ainsi que ceux en $X^+_>-X^+_<$ parce que $(X^+_>-X^+_<)|B\>=0$. Les
facteurs avec les modes zéros, quant à eux, s'éliminent parce que les états
entrant et sortant ont même $p^+$ (parce que $p^+$ est conservé) et même
$p^-$ (parce que ce sont deux tachyons ; sinon il y aurait un facteur
$e^{ix^+\triangle p^-}$). On obtient alors
\[
S_{0\to0} \propto \<0| e^{ip^+(X^-_> - X^-_<)}|_{\s_F=0} \,
                       e^{-ip^+(X^-_> - X^-_<)}|_{\s_F=\pi} |B\>.
\]

Il reste alors à faire passer le facteur en $X^+_>-X^+_<$ à droite de
l'exponentielle de l'équation \eqref{bordpulse}. Pour cela, nous allons
utiliser la formule
\[
e^C g(B) e^{-C} = g(B + [C,B])
\]
où $[C,B]$ doit commuter avec $C$ et $B$, et $g$ est une fonction
quelconque (en principe analytique), et le commutateur
\[
[X_>(\s_F), X_<(\s_F')] = -\frac{1}{4}
    \ln\!\left( 4\sin^2\parfrac{\s_F-\s_F'}{2} \!\right),
\]
qui n'est autre que le propagateur restreint au bord $N_F(0,\s_F;0,\s'_F)$.
En appliquant ceci une fois pour chaque opérateur de vertex, et après
quelques calculs (cf \cite{BG}), on obtient
\begin{equation}
S_{0\to0} \propto \<0| \exp\!\left[ \frac{1}{\pi} \int d\s_F\,
   Y^i\!\left( X^+(\s_F) + ip^+ \ln\!\left|\tan\frac{\s_F}{2}\right| \right)
   \d_{\t_F} X^i \right] |\D1\>.
\label{eqS}
\end{equation}
(Le fait que l'on prolonge ici analytiquement une fonction qui n'a pas de
raison de être analytique ---~les fonctions analytiques à support borné
sont rares~--- peut paraître fumeux. Néanmoins, le résultat final sera
identique à celui de la jauge du cône de lumière (équation
\eqref{corvide}), ce qui laisse supposer qu'il doit y avoir un moyen de
rendre cela rigoureux.) On peut en outre remplacer $X^+(\s_F)$ par $x^+$
car, en l'absence de termes en $X^-$, les modes non zéros de $X^+$ peuvent
se déplacer librement à gauche où à droite, et s'annulent sur $\<0|$ ou
$|\D1\>$.

Pour aller plus loin, on écrit 
\[
\d_{\t_F} X^i = 2\d_{\t_F} X^i_< + (\d_{\t_F} X^i_> - \d_{\t_F} X^i_<).
\]
Le premier terme s'annule sur $\<0|$, et les conditions de Dirichlet sur
la D-corde sans onde entraînent l'annulation du deuxième terme sur $|\D1\>$.
On est donc amené à séparer l'exponentielle dans \eqref{eqS} en utilisant
la formule de Baker-Campbell-Hausdorff
\[
e^{B+C} = e^B\, e^{\frac{1}{2}[C,B]}\, e^C,
\]
applicable si $[C,B]$ commute avec $B$ et $C$, où $B$ et $C$ sont
l'exponentielle dans \eqref{eqS} en remplaçant $\d_{\t_F}X^i$ par
respectivement le premier et le deuxième terme de l'équation précédente. En
remarquant d'autre part que le terme en $\ln|\tan(\s_F/2)|$ apparaît dans
la transformation conforme qui transforme le demi-cylindre infini
($\t_F\geq0$) en ruban euclidien :
\[
-\t_E + i\s = \ln\!\left( \tan\!\parfrac{\s_F+i\t_F}{2} \!\right),
\]
on obtient 
\[
\begin{split}
S_{0\to0} = \<0| \exp\biggl[& \frac{1}{\pi^2}
                  \int d\s_F'\, Y^i(x^+ - ip^+ \t_E(\s_F'))
                  \int d\s_F''\, Y^i(x^+ - ip^+ \t_E(\s_F'')) \times \\
&\times \d_{\t_F'} \d_{\t_F''} N_F(0,\s_F';0,\s_F'') \biggr] |\D1\>
\end{split}
\]

Tout naturellement, on fait alors le changement de variable vers les
coordonnées $(\t_E,\s)$. La dérivée du propagateur se transforme de façon
triviale (noter que ce n'est pas le cas du propagateur lui-même), à ceci
près que le propagateur ouvert sur la bande a un facteur 2 par rapport au
propagateur fermé à cause du bord, donc nous devrons ajouter un facteur
$1/2$ ; pour plus de détails à ce propos, on se réfèrera à l'annexe de
\cite{BG}. Enfin, une rotation de Wick ($\t_E=i\t$) permet de revenir à un
argument réel pour $Y^i$ et transforme le propagateur euclidien en
propagateur minkowskien.

Il faut encore éliminer les $x^+$. Pour cela, on peut absorber $x^+$ dans
la variable d'intégration. Il reste alors un facteur $\int dx^+\<x^+|\D1\>$
absorbé dans la normalisation, et on aboutit à l'expression
\eqref{corvide}.

Notons que la remarque que nous avons faite à la fin de la section
\ref{eucl} est explicitement vérifiée ici : en partant de l'état de bord,
donc dans un formalisme covariant avec une surface d'univers euclidienne,
nous arrivons finalement à une expression faisant apparaître une surface
d'univers minkowskienne (puisque le propagateur dans l'expression finale
est minkowskien) dans la jauge du cône de lumière, ce qui confirme que
c'est bien une métrique minkowskienne qu'il faut considérer dans cette
jauge.

Il n'est pas difficile de considérer le processus tachyon~$\to$~photon dans
ce formalisme. Il s'agit alors de calculer l'amplitude
\[
\<0| \ordnorm{e^{-ip_\mu X^\mu(0,0)} \d_\s X^i}
     \ordnorm{e^{ip_\mu X^\mu(0,\pi)}} |B\>.
\]
On aboutit alors, après la rotation de Wick, à
\index{operateur!opérateur de vertex}
\[
\<0| \frac{i}{\pi} \oint d\t e^{i\t} Y^i(x^+ + p^+\t) \exp\Big[\ldots\Big]
     e^{ix^+/p^+} |B\>
\]
où $\exp[\ldots]$ est l'exponentielle dans $S_{0\to0}$, et le facteur
$e^{ix^+/p^+}$ vient des modes zéros dans \eqref{pulseordnorm}. Ce facteur
permet de n'avoir qu'une dépendance en $x^++p^+\t$, et on peut donc, comme
précédemment, absorber $x^+$ dans la variable d'intégration, et on obtient
finalement la même chose que dans la jauge du cône de lumière.

Les opérateurs de vertex devenant rapidement assez compliqués pour les
états excités suivants, la vérification du fait que les deux formalismes
donnent le même résultat n'a pas été faite au-delà. Néanmoins, on s'attend
à ce que la dualité ouvert-fermé soit très générale, et donc à ce que cela
marche.

\section{Interactions sur la D-corde ondulante}
Après avoir considéré des processus où une corde ouverte donne une autre
corde ouverte, éventuellement dans un état différent, nous allons envisager
des interactions. Nous nous limiterons au cas d'une corde ouverte se
cassant en deux. Plus spécifiquement, on s'intéressera au processus où une
corde dans son état fondamental émet une excitation de masse nulle et
d'impulsion tendant vers zéro. Ces excitations correspondant aux
déformations de la D-brane, une divergence infrarouge serait un signe
d'instabilité de la configuration ; à l'inverse, l'absence de divergence
signifierait que la rétroaction sur la D-corde est contrôlable,
c'est-à-dire peut être rendue négligeable avec un couplage suffisamment
petit.

\subsection{Surface d'univers dans la jauge du cône de lumière}
\begin{figure}
\begin{center}
\scalebox{0.7}{\input{ruban.pstex_t}}
\end{center}
\caption{Surface d'univers d'une corde ouverte se cassant en deux.}
\label{ruban}
\end{figure}
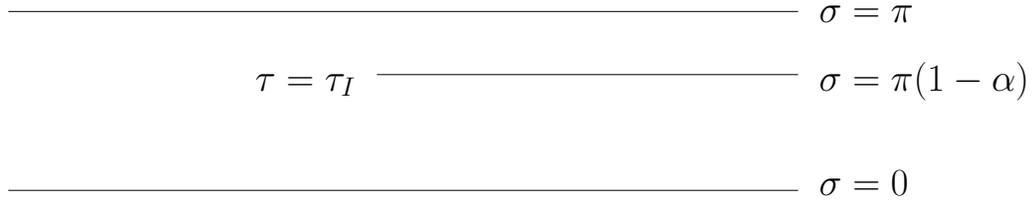

Pour décrire une corde ouverte se cassant en deux, nous considérons comme
surface d'univers une bande de largeur $\pi$ avec une coupure parallèle à
ses bords partant du temps d'interaction $\t_I$, c'est-à-dire que la
coupure est donnée par $\t\geq\t_I$ et $\s=(1-\a)\pi$ (fig. \ref{ruban}).
Dans la jauge du cône de lumière, où on prend $X^+=p^+\t$, où $p^+$ est
l'impulsion totale, qui est conservée, l'impulsion de chacun des deux
morceaux après séparation est proportionnelle à leur largeur, et vaut donc
respectivement $(1-\a)p^+$ et $\a p^+$. Nous considérerons le cas où
l'excitation de masse nulle émise est dans la partie
$\s\in[(1-\a)\pi;\pi]$ ; ainsi, la limite d'impulsion nulle pour cette
excitation correspondra à $\a\to0$.

\index{Schwarz-Christoffel, transformation de}
Le formalisme que nous développé dans le chapitre précédent reste en grande
partie valide pour des cordes en interaction. Un point essentiel qui va
différer est la forme du propagateur. Celui-ci peut s'exprimer assez
simplement dans le cas euclidien. En effet, dans ce cas-là, la
transformation conforme suivante, dite \emph{transformation de
Schwarz-Christoffel}, amène le demi-plan complexe supérieur vers la bande
coupée avec $\t_I=0$ :
\begin{equation}
\t_E + i\s = \ln(z-1) - \a\ln z + \a\ln\a + (1-\a)\ln(1-\a).
\label{TransSC}
\end{equation}
Le temps d'interaction correspond au point stationnaire de cette
transformation ; le lecteur pourra vérifier aisément qu'on a bien $\t_I=0$
(le terme constant a été ajusté dans ce but).

En termes de la coordonnée $z$, le propagateur s'écrit
\index{propagateur!ruban fendu}
\[
N_E(z;z') = -\frac{1}{2} \ln|z-z'| + \frac{1}{2} \ln|\bar{z}-z'|.
\]
En l'absence d'expression explicite pour l'inverse de la transformation
\eqref{TransSC}, nous devrons travailler dans ces coordonnées, ce qui nous
impose de travailler avec une surface d'univers euclidienne, et donc de
faire un prolongement analytique à partir du cas minkowskien.

Pour déterminer comment on doit faire ce prolongement, il faut examiner le
lien précis entre les propagateurs euclidien et minkowskien. Pour cela, on
peut écrire le propagateur euclidien dans les coordonnées $\t_E,\s$ sous
forme d'une série dépendant des parties de la surface d'univers (corde
entrante ou l'une des cordes sortantes), notées $r$ et $r'$, où se trouvent
les points $(\t_E,\s)$ et $(\t_E',\s')$ :
\[
\begin{split}
N_E(\t_E,\s;\t_E',\s') &=
\delta_{rr'} \sum_{n>0} \frac{1}{n} 
  e^{-n|\t_E-\t_E'|/|\a_r|} \sin(n\s_r) \sin(n\s'_{r'}) \\
&\quad + \sum_{n,n'>0} N_{nn'}^{rr'}
e^{n\t_E/\a_r} e^{n'\t_E'/\a_r'} \sin(n\s_r) \sin(n'\s'_{r'})
\end{split}
\]
où $\s_r$ et $\s'_{r'}$ sont des coordonnées spatiales adaptées à chaque
partie de la surface d'univers, c'est-à-dire des fonctions affines de $\s$
valant $0$ ou $\pi$ aux bords, et $\a_r$ vaut $1$ sur la corde entrante (où
$\t\leq0$) et $-(1-\a)$ et $-\a$ sur les cordes sortantes (où $\t\geq0$).
Pour plus de détails sur cette expression du propagateur, on pourra se
référer au chapitre 11 de \cite{GSW}, où on trouvera également des
expressions explicites pour les coefficients $N_{nn'}^{rr'}$, dont nous
n'aurons pas besoin ici.

À partir de cette expression, on peut obtenir une expression pour le
propagateur minkowskien :
\[
\begin{split}
N(\t,\s;\t',\s') &=
\delta_{rr'} \sum_{n>0} \frac{1}{n} 
  e^{-in|\t-\t'|/|\a_r|} \sin(n\s_r) \sin(n\s'_{r'}) \\
&\quad + \sum_{n,n'>0} N_{nn'}^{rr'}
e^{in\t/\a_r} e^{in'\t'/\a_r'} \sin(n\s_r) \sin(n'\s'_{r'}).
\end{split}
\]
Un point non trivial à vérifier est que les expressions sur les différents
morceaux de la surface d'univers se recollent bien en $\t=0$. On pourra
vérifier que si c'est le cas pour $N_E$ (ce qui est vrai puisqu'on l'a
obtenue à partir d'une expression en $z$ qui n'a pas de discontinuité),
alors c'est vrai aussi pour $N$. Il en découle le lien suivant entre les
propagateurs :
\[
N(\t,\s;\t',\s') = N_E(i\t,\s;i\t',\s').
\]

\subsection{Fonctions de corrélation et amplitudes}
L'essentiel de ce que nous avons vu sur les fonctions de corrélation et
amplitudes dans le cas libre s'applique ici ; bien sûr, puisqu'il y a deux
cordes sortantes, $\<0_{out}|$ doit être compris comme le produit tensoriel
des vides de chaque corde. Notamment, il découle l'expression suivante,
pour un temps d'interaction $\t_I$ :
\[
\begin{split}
\<0_{out}|0_{in}\>_S = C \exp\biggl[&
   -\frac{1}{2\pi^2} \oint d\t' \oint d\t''\,
         \d_{\s'}\d_{\s''}N(\t',\s';\t'',\s'') \times \\
       & \times Y(p^+(\t_I+\t')) Y(p^+(\t_I+\t''))
   \biggr].
\end{split}
\]
Après une rotation de Wick ($\t'_E=i\t'$ et $\t''_E=i\t''$), on obtient
\[
\begin{split}
\<0_{out}|0_{in}\>_S = C \exp\biggl[&
   \frac{1}{2\pi^2} \oint d\t'_E \oint d\t''_E\,
         \d_{\s'}\d_{\s''}N_E(\t'_E,\s';\t''_E,\s'') \times \\
       & \times Y(p^+(\t_I-i\t'_E)) Y(p^+(\t_I-i\t''_E))
   \biggr].
\end{split}
\]
A priori $C$ peut dépendre de $\a$. Puisqu'il ne dépend pas de l'onde, on
peut utiliser les résultats déjà connus en l'absence d'onde \cite{GSW}, qui
disent que $C$ est une constante, que l'on peut absorber dans la constante
de couplage, et que nous omettrons donc dans la suite.

À partir de là, un changement de variables vers $z'$ et $z''$ donne
\[
\<0_{out}|0_{in}\>_S = \exp\!\left[ \frac{1}{2\pi^2} \int dz'\,dz''\,
  \frac{Y(p^+(\t_I-i\t_E(z'))) \,Y(p^+(\t_I-i\t_E(z'')))}{(z'-z'')^2} \right]
\]
où $\t_E(z)$ est donné par l'équation \eqref{TransSC}. Ceci diverge quand
$z'\to z''$ ; il faut donc comprendre cette intégrale comme étant une
partie principale.

Pour calculer l'amplitude d'émission d'une excitation de masse nulle par un
tachyon, il faut calculer l'amplitude $\<0_{out}|a_1|0_{in}\>_S$, où $a_1$
est l'opérateur créant une excitation de masse nulle sortante :
\[
a_1 = \lim_{\t\to+\infty}
   \frac{e^{i\t/\a}}{\pi} \int_{(1-\a)\pi}^\pi d\s\,
   \sin\parfrac{\pi-\s}{\a} \left(\frac{X}{\a} + i\d_\t X \right).
\]
Pour obtenir une expression explicite de l'amplitude, là encore les
expressions trouvées dans le cas libre s'appliquent, et comme pour le
calcul de $\<0_{out}|0_{in}\>_S$ il faut effectuer une rotation de Wick
pour passer à la coordonnée $z$ (on peut vérifier que la limite
$\t\to+\infty$ revient à la limite $\t_E\to+\infty$ sur l'expression
euclidienne). À partir de là, le calcul est plus fastidieux que difficile :
comme la limite $\t_E\to+\infty$ correspond à $z\to0$, l'intégrale sur $\s$
devient une intégrale de contour autour de $z=0$, que l'on peut calculer
explicitement, et on trouve
\[
\frac{\<0_{out}|a_1|0_{in}\>_S}{\<0_{out}|0_{in}\>_S} =
\frac{1}{\pi} \int dz\, \frac{Y(p^+(\t_I - i\t_E(z)))}{z^2}.
\]

Ceci nous donne une amplitude à $\t_I$ fixé. Pour obtenir l'amplitude
physique qui nous intéresse, il faut intégrer sur les temps d'interaction
possibles, l'amplitude s'écrivant alors
\begin{equation}
A = \int d\t_I\, \<0_{out}|a_1|0_{in}\>_S \bigr|_{\t_I}
                 e^{i\t_I p^+ \triangle p^-}
\label{inttI}
\end{equation}
où le facteur $p^+\triangle p^-$ dans l'exponentielle n'est autre que la
variation de hamiltonien entre les états initial et final, la présence de
cette exponentielle étant un résultat classique de la théorie quantique des
perturbations dépendantes du temps.

Le moins qu'on puisse dire est que cette expression est peu sympathique.
Pour commencer, il faut donner un sens à l'expression
$Y(p^+(\t_I-i\t_E(z)))$. On peut le faire à l'aide de la transformée de
Fourier de $Y$ :
\[
Y(p^+(\t_I - i\t_E(z))) = \frac{1}{p^+} \int d\w\, \tilde{Y}\parfrac{\w}{p^+}
                          e^{i\w\t_I} e^{\w\t_E(z)}.
\]
Ainsi, pour calculer $\<0_{out}|0_{in}\>_S$, on sera amené à considérer
l'intégrale
\[
\int dz'\,dz''\, \frac{e^{\w'\t_E(z')} e^{\w''\t_E(z'')}}{(z'-z'')^2}
\]
qui contient une grande partie de la dépendance en $\a$ (pas tout puisque
$\triangle p^-$ en dépend aussi), ce qui est intéressant puisqu'on
s'intéresse à la limite $\a\to0$. L'expression analogue dans le cas de
l'amplitude à deux cordes (où l'équation pour $\<0_{out}|0_{in}\>_S$ est la
même en prenant $\t_E(z)=\ln z$) peut être calculée exactement après une
rotation de Wick $\w\to i\w_E$ (qui améliore la convergence des
intégrales), et on retrouve le résultat déjà trouvé par d'autres moyens.
Dans le cas à trois cordes, on ne sait pas faire le calcul exact ; on peut
néanmoins espérer trouver des résultats concernant la limite $\a\to0$ qui
nous intéresse.

\subsection{Calcul covariant}
\index{etatb@état de bord!D-brane ondulante}
\index{operateur!opérateur de vertex}
Pour calculer de façon covariante une amplitude à trois cordes, le calcul
est similaire au cas à deux cordes (cas libre), à ceci près qu'il y a trois
opérateurs de vertex.

Dans le cas qui nous intéresse, l'amplitude à trois tachyons sera donc
donnée, après des calculs identiques à ceux du cas libre, par
\[
S_{0\to0} \propto \<0| e^{i\a p^+(X^-_> - X^-_<)}|_{\s_F=0} \,
                       e^{i(1-\a)p^+(X^-_> - X^-_<)}|_{\s_F=\pi}
                       e^{-ip^+(X^-_> - X^-_<)}|_{\s_F=\pi/2} |B\>.
\]
Les positions des opérateurs de vertex peuvent être fixées arbitrairement
grâce à la symétrie conforme $\SL(2,\R)$. Quelques calculs plus tard, on
aboutit à
\[
S_{0\to0} \propto \<0| \exp\!\left[ \frac{1}{\pi} \int\! d\s_F\,
  Y^i\biggl(\! X^+(\s_F) - ip^+\!\left( \ln\!\left|\tan\frac{\s_F}{2}-1\right|
               -\a \ln\!\left|\tan\frac{\s_F}{2}\right| \right) \!\biggr)
   \d_{\t_F} X^i \right] |\D1\>.
\]

En utilisant la transformation conforme
\[
z = \tan\frac{\s_F + i\t_F}{2}
\]
et des calculs similaires à ceux de l'amplitude à deux cordes, on retrouve
l'expression obtenue en jauge du cône de lumière ; l'intégrale sur le temps
d'interaction (équation \eqref{inttI}) découle de l'intégrale
\eqref{videfer} et des modes zéros dans \eqref{pulseordnorm}, avec
$x^+=p^+\t_I$.

\subsection{Remarques finales}
Nos résultats restent pour l'instant très partiels. Nous espérons néanmoins
les compléter dans un avenir raisonnablement proche. On peut d'ores et déjà
avoir une idée de ce à quoi s'attendre. Dans le cas, que nous avons traité
ici, de cordes dont les deux extrémités sont sur une même D-brane, on n'a
pas de raison, a priori, de s'attendre à une divergence infrarouge. Dans le
cas à deux D-branes, en revanche, les choses sont différentes, et on
s'attend à des divergences, au moins dans la limite où leur distance finale
(à des temps asymptotiquement grands) tend vers l'infini, comme expliqué au
début de ce chapitre. Il faudra donc faire ce calcul pour en savoir plus.
\index{D-brane!ondulante|)}

%% file: ciseaux.pstex_t
\begin{picture}(0,0)%
\includegraphics{ciseaux.pstex}%
\end{picture}%
\setlength{\unitlength}{4144sp}%
\begingroup\makeatletter\ifx\SetFigFont\undefined%
\gdef\SetFigFont#1#2#3#4#5{%
  \reset@font\fontsize{#1}{#2pt}%
  \fontfamily{#3}\fontseries{#4}\fontshape{#5}%
  \selectfont}%
\fi\endgroup%
\begin{picture}(4766,3644)(3129,-3683)
\put(7111,-1816){\makebox(0,0)[lb]{\smash{{\SetFigFont{20}{24.0}{\rmdefault}{\mddefault}{\updefault}{\color[rgb]{0,0,0}$X^+$}%
}}}}
\end{picture}%

%% file: ruban.pstex_t
\begin{picture}(0,0)%
\includegraphics{ruban.pstex}%
\end{picture}%
\setlength{\unitlength}{4144sp}%
\begingroup\makeatletter\ifx\SetFigFont\undefined%
\gdef\SetFigFont#1#2#3#4#5{%
  \reset@font\fontsize{#1}{#2pt}%
  \fontfamily{#3}\fontseries{#4}\fontshape{#5}%
  \selectfont}%
\fi\endgroup%
\begin{picture}(6942,1773)(1789,-1528)
\put(8731, 29){\makebox(0,0)[lb]{\smash{{\SetFigFont{20}{24.0}{\rmdefault}{\mddefault}{\updefault}{\color[rgb]{0,0,0}$\s=\pi$}%
}}}}
\put(8731,-556){\makebox(0,0)[lb]{\smash{{\SetFigFont{20}{24.0}{\rmdefault}{\mddefault}{\updefault}{\color[rgb]{0,0,0}$\s=\pi(1-\a)$}%
}}}}
\put(4771,-556){\makebox(0,0)[rb]{\smash{{\SetFigFont{20}{24.0}{\rmdefault}{\mddefault}{\updefault}{\color[rgb]{0,0,0}$\t=\t_I$}%
}}}}
\put(8731,-1456){\makebox(0,0)[lb]{\smash{{\SetFigFont{20}{24.0}{\rmdefault}{\mddefault}{\updefault}{\color[rgb]{0,0,0}$\s=0$}%
}}}}
\end{picture}%

%% file: concl.tex
\chapter*{Conclusion}
\subsection*{Résultats obtenus}
Cette thèse avait pour objectif de mieux comprendre la théorie des cordes
en espaces-temps courbes ou dépendant du temps. Pour cela, nous en avons
étudié divers aspects et obtenu des résultats que nous allons rappeler ici.

Tout d'abord, nous avons construit, par une approche géométrique, les
orientifolds sur $S^3$ et $\RP^3$. Nous avons montré que, dans chaque cas,
on a deux façons différentes d'effectuer la projection orientifold sur les
cordes fermées : dans le cas de la 3-sphère, cela correspond à des
géométries différentes pour les orientifolds, tandis que pour $\RP^3$ les
orientifolds sont au même endroit dans les deux cas. Nous avons également
calculé les interactions de ces orientifolds avec les D-branes, et en avons
déduit les groupes de jauge possibles.

L'étude des univers branaires nous a ensuite amené à nous intéresser aux
D-branes dans des espaces anti-de~Sitter en présence de champs de fond
Ramond-Ramond. Dans le cas où l'espace est $AdS_3\times S^3$, les D-branes
sont des D3-branes, laissées invariantes par la S-dualité qui transforme
les fonds R-R en fonds Neveu-Schwarz, ce qui nous a permis de les aborder
en utilisant les résultats connus dans ce dernier cas. Nous avons vérifié
explicitement que ce que l'on sait calculer semi-classiquement est
compatible avec la S-dualité.

Enfin, nous avons étudié un des fonds dépendants du temps les plus simples,
à savoir D-brane portant une onde plane et plongée dans un espace-temps
plat. Nous n'avons obtenu que des résultats très partiels ; néanmoins, nous
avons introduit les outils qui devraient permettre, dans un temps
raisonnablement court, d'aboutir à des résultats intéressants concernant
notamment la rétroaction sur un tel fond.

\subsection*{Perspectives}
Comme indiqué, les résultats du chapitre~4 restent partiels, et il faudrait
les compléter, notamment en considérant le cas d'une corde entre deux
D-branes différentes. On pourra également considérer l'émission de cordes
fermées par une telle D-brane, et ainsi étudier l'émission d'ondes
gravitationnelles par un défaut topologique, ce qui pourrait avoir des
incidences importantes sur l'histoire de l'Univers primitif.

Au-delà, il est clair que beaucoup reste à faire en ce qui concerne les
fonds dépendant du temps, notamment cosmologiques. Les propriétés de
stabilité de tels fonds et la façon de les aborder lorsqu'on ne peut pas
définir de matrice S restent largement mystérieuses à l'heure actuelle. Il
y a également la question de savoir si la théorie des cordes peut résoudre
les singularités telles que celle qui a vraisemblablement eu lieu lors du
Big Bang, et si la façon dont cela se résout a des conséquences
observationnelles, visibles par exemple dans le fond micro-onde
cosmologique observé par des expériences telles que WMAP (Wilkinson
Microwave Anisotropy Probe).

On s'attend à ce que l'étude des univers branaires, évoqués dans le
chapitre~3, ait un impact significatif sur ces questions. Notamment, il
pourrait être intéressant de voir si on peut réaliser l'inflation
primordiale par un univers branaire « réaliste ».

En résumé, la recherche en théorie des cordes a encore de longues années
devant elle.